\documentclass[3p, preprint]{elsarticle}

\makeatletter
\def\ps@pprintTitle{%
    \let\@oddhead\@empty
    \let\@evenhead\@empty
    \let\@evenfoot\@oddfoot
    }
\makeatother
\usepackage{tabularx}
\usepackage{amsmath}
\usepackage{amssymb}
\usepackage{amsthm}
\usepackage{mathtools}
\usepackage{bm}
\usepackage{graphicx}
\usepackage[final]{hyperref}
\usepackage[ruled,vlined]{algorithm2e}
\usepackage{enumitem}
\usepackage{array}
\usepackage{multirow}
\usepackage{makecell}
\usepackage[title]{appendix}
\usepackage{empheq}
\usepackage{cleveref}
\usepackage{cases}
\usepackage{textcomp}
\usepackage{gensymb}
\usepackage{subcaption}
\hypersetup{
	colorlinks=true,       % false: boxed links; true: colored links
	linkcolor=blue,        % color of internal links
	citecolor=blue,        % color of links to bibliography
	filecolor=magenta,     % color of file links
	urlcolor=blue         
}
\usepackage{float}
\usepackage{afterpage}
\usepackage{regexpatch}

\newcommand{\R}{\mathbb{R}}

\newcommand{\N}{\mathbb{N}}

\newcommand{\norm}[1]{\left\lVert #1 \right\rVert}

\newcommand{\lap}{\Delta}
\newcommand{\grad}{\nabla}

\DeclareMathOperator*{\argmin}{arg\,min}

\DeclareMathOperator{\atantwo}{atan2}
\DeclarePairedDelimiter\ceil{\lceil}{\rceil}
\DeclarePairedDelimiter\floor{\lfloor}{\rfloor}

\newtheorem{remark}{Remark}
\newcommand{\x}{\boldsymbol{x}}
\newcommand{\bx}{\boldsymbol{X}}
\newcommand{\y}{\boldsymbol{y}}
\newcommand{\by}{\boldsymbol{Y}}

\newcommand{\bz}{\boldsymbol{Z}}
\newcommand{\data}{\boldsymbol{d}}
\newcommand{\bd}{\boldsymbol{D}}
\newcommand{\w}{\boldsymbol{w}}
\newcommand{\bw}{\boldsymbol{W}}

\newcommand{\erf}{\textrm{erf}}
\newcommand{\bdmc}[1]{\boldsymbol{\mathcal{#1}}}

\newcommand{\aaps}{\boldsymbol{\mathcal{P}}_r}

\newcommand{\expect}[2]{\mathbb{E}_{#1}\left[#2 \right]}
\newcounter{example}[section]
\newenvironment{experiment}[1][]{\refstepcounter{example}\par\medskip
   \noindent \textbf{Experiment~\theexample. #1} \rmfamily}{\medskip}

\newcolumntype{C}{ >{\centering\arraybackslash} m{0.055\linewidth} }
\newcolumntype{D}{ >{\centering\arraybackslash} m{0.095\linewidth} }
\newcolumntype{E}{ >{\centering\arraybackslash} m{0.024\linewidth} }
\newcommand{\nn}{\widetilde{\bdmc{S}}}

\newcolumntype{F}{ >{\centering\arraybackslash} m{0.69\linewidth} }
\newcolumntype{G}{ >{\centering\arraybackslash} m{0.3\linewidth} }
\usepackage{lineno}

\begin{document}
\begin{frontmatter}
\title{Bayesian Model Calibration for Diblock Copolymer Thin Film Self-Assembly Using Power Spectrum of Microscopy Data and Machine Learning Surrogate}\tnoteref{t1}
\tnotetext[t1]{This work is dedicated to Professor Thomas J.R. Hughes in recognition of his lifelong seminal contributions to the field of computational science and engineering.}
\author{Lianghao Cao\corref{cor1}\fnref{fn1}}
\ead{lianghao@oden.utexas.edu}
\author{Keyi Wu\fnref{fn1,fn2}}
\ead{keyiwu@utexas.edu}
\author{J. Tinsley Oden\fnref{fn1,fn3}}
\ead{oden@oden.utexas.edu}
\author{Peng Chen\fnref{fn4}}
\ead{pchen402@gatech.edu}
\author{Omar Ghattas\fnref{fn1,fn5}}
\ead{omar@oden.utexas.edu}

\cortext[cor1]{Corresponding author}

\fntext[fn1]{Oden Institute for Computational Engineering and Sciences, The University of Texas at Austin, 201 E. 24th Street, C0200, Austin, TX 78712, USA.} 
\fntext[fn2]{Department of Mathematics, The University of Texas at Austin, 2515 Speedway, C1200, Austin, TX 78712, USA.}
\fntext[fn3]{Department of Aerospace Engineering and Engineering Mechanics, The University of Texas at Austin, 2617 Wichita Street, C0600, Austin, TX 78712, USA.}
\fntext[fn4]{School of Computational Science and Engineering, The Georgia Institute of Technology, 756 W Peachtree St NW, Atlanta, GA 30308, USA.}
\fntext[fn5]{Walker Department of Mechanical Engineering, The University of Texas at Austin, 204 E. Dean Keaton Street, C2200, Austin, TX 78712, USA.}

\begin{abstract}
Identifying parameters of computational models from experimental data, or \textit{model calibration}, is fundamental for assessing and improving the predictability and reliability of computer simulations. In this work, we propose a method for Bayesian calibration of models that predict morphological patterns of diblock copolymer (Di-BCP) thin film self-assembly while accounting for various sources of uncertainties in pattern formation and data acquisition. This method extracts the azimuthally-averaged power spectrum (AAPS) of the top-down microscopy characterization of Di-BCP thin film patterns as summary statistics for Bayesian inference of model parameters via the pseudo-marginal method. We derive the analytical and approximate form of a conditional likelihood for the AAPS of image data. We demonstrate that AAPS-based image data reduction retains the mutual information, particularly on important length scales, between image data and model parameters while being relatively agnostic to the aleatoric uncertainties associated with the random long-range disorder of Di-BCP patterns. Additionally, we propose a phase-informed prior distribution for Bayesian model calibration. Furthermore, reducing image data to AAPS enables us to efficiently build surrogate models to accelerate the proposed Bayesian model calibration procedure. We present the formulation and training of two multi-layer perceptrons for approximating the parameter-to-spectrum map, which enables fast integrated likelihood evaluations. We validate the proposed Bayesian model calibration method through numerical examples, for which the neural network surrogate delivers a fivefold reduction of the number of model simulations performed for a single calibration task.
\end{abstract}


\end{frontmatter}

% \linenumbers
\section{Introduction}\label{sec:intro}

\subsection{Motivation}\label{subsec:motivation}
Block copolymer (BCP) melts are a large collection of linear polymers composed of blocks of distinct monomers. Upon thermal or solvent annealing, the thermodynamic incompatibility between the blocks drives them to spontaneously segregate and form periodic structures whose periodicity lengths are between 5 and 50 nanometers~\cite{Hamley1998, Bates1999}. In particular, for diblock copolymers (Di-BCPs), BCPs with two blocks, their observed periodic structures, or phases, include lamellae, perforated layers, gyroids, cylinders, and spheres~\cite{Khandpur1995}. The \textit{self-assembly} of BCP thin films can be further controlled by external guidance, such as chemically patterned substrates~\cite{Kim2003} and directional crystallization~\cite{Berry2007}, to generate desired morphological patterns over large surface areas accurately. This technology is called the \textit{directed self-assembly} (DSA) of BCPs, and it has been employed in commercial applications such as patterning for nanolithography \cite{Mansky1995, Park2003, Bates2014,Ji2016} and fabricating nano-scale devices \cite{Black2001, Xiao2005, Stoykovich2007}.

Models underlying computer simulations of BCP self-assembly are mainly derived from statistical polymer theory~\cite{Grosberg1994, Fredrickson2006}. Commonly-used ones include Monte Carlo simulation~\cite{Wang1999, Binder2012}, theoretical-informed coarse grain (TICG) simulation ~\cite{Detcheverry2009}, self-consistent field theory (SCFT)~\cite{Matsen2005,Muller2005}, and density functional theory (DFT)~\cite{Uneyama2005, Fraaije1993}. These models impact how we understand the phenomenon of BCP self-assembly and how we apply it to surmount challenges in engineering problems. Most notably, the inverse design of BCP thin film DSA~\cite{Hannon2013, Qin2013, Hannon2014, Luo2023}, a simulation-based optimization framework for designing DSA that optimally produces complex target patterns, is often recognized as the critical pathway towards transitioning DSA technology from the laboratory into commercial manufacturing of 10-nanometer class integrated circuit devices~\cite{itrs2013}.

The growing impact of computer simulations of BCP self-assembly drives us to examine the predictability and reliability of the underlying models. Identifying model parameters to fit the outputs of computer models to experimental or observational data is known as \textit{model calibration}, which belongs to the larger class of \textit{inverse problems} \cite{GhattasWillcox21}. Model calibration is often regarded as the first step for producing reliable model predictions~\cite{Oden2017foundations}. Furthermore, the model calibration and subsequent predictions should accommodate uncertainties in the data and model. Bayes' rule provides a unified way for assimilating data into models while naturally taking into full account inherent uncertainties~\cite{Jaynes2003, Oden2017, Oden2018}. This work focuses on Bayesian model calibration for Di-BCP thin film self-assembly and contributes to the methodological foundation for improving the predictability and reliability of Di-BCP self-assembly simulations.

Besides uncertainties developed during signal and image formation via microscopy or X-ray scattering, we also address \textit{aleatoric uncertainties} intrinsic to the self-assembly phenomenon. Due to uncontrolled nucleation and domain growth during phase separation, self-assembled BCP films typically exhibit surface patterns with random long-range disorders. They are often referred to as \textit{metastable states}~\cite{Cheng2008}. In BCP thin film DSA, \textit{random pattern defects} are formed due to insufficient or imprecise control by DSA guidance~\cite{Li2015}. The effective annihilation of pattern defectivity is regarded as a major hurdle in the nanolithography application of DSA because of its implication on product quality and yield~\cite{Bates2014}. To find the optimal trade-off between the control precision and the viability or manufacturability of DSA guidance, many inquiries have been made to theoretically and computationally characterize the influences of the aleatoric uncertainties on BCP self-assembly~\cite{Fredrickson1989, Nagpal2012, Rottler2020, Schneider2021}. In this work, we seek to incorporate them into the Bayesian model calibration procedure, which empowers us to assess the predictability and reliability of computer simulations of Di-BCP self-assembly using experimental data in the presence of such uncertainties.

\subsection{Related works and our proposed Bayesian model calibration method}
{\color{red}
\begin{figure}[!ht]
    \centering
    \includegraphics[width = \linewidth]{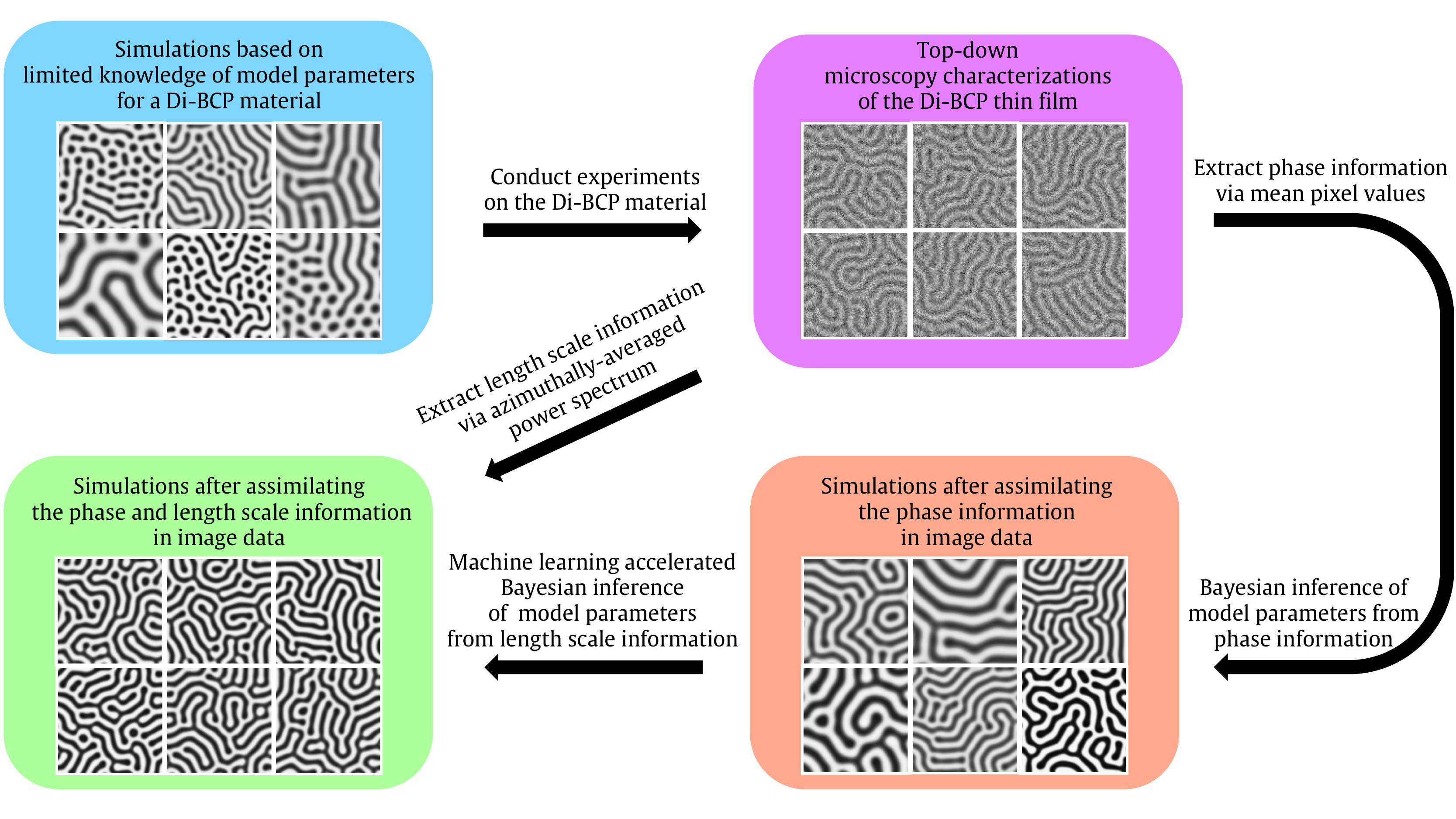}
    \caption{A high-level summary of the proposed Bayesian model calibration procedure. The simulated patterns are taken from the numerical results in Section \ref{subsec:surrogate_uncertain}}
    \label{fig:summary}
\end{figure}
}

In a recent work by Baptista et al.~\cite{Baptista2022}, a generic formulation of Bayesian model calibration for BCP self-assembly that includes the aforementioned aleatoric uncertainties is proposed. Unlike the deterministic approach employed by Khadilkar et\thinspace al.~\cite{Khaira2017} and Hannon et al.~\cite{Hannon2018}, this formulation integrates both uncertainties induced by noise in microscopy or X-ray scattering imaging and the aleatoric uncertainties into the model calibration procedure via the likelihood function in Bayes' rule. Baptista et al.~demonstrate that (i) the aleatoric uncertainties lead to \textit{integrated likelihood functions} that are generally intractable to evaluate for unguided or weakly-guided self-assembly, and (ii) high-dimensional image data lead to an ineffective inference. To overcome these challenges, Baptista et al.~advocate (i) using a \textit{likelihood-free inference approach via measure transport}~\cite{Baptista2020, Marzouk2017}, and (ii) constructing informative low-dimensional \textit{summary statistics}~\cite{Najm2017} of image data. Several summary statistics based on the \textit{azimuthally-averaged power spectrum} (AAPS) of image data were introduced in a case study of Bayesian model calibration using top-down microscopy characterization of Di-BCP thin films. These summary statistics extract from image data information about the important \textit{length scales} of the latent Di-BCP patterns, such as periodicity length and feature size (e.g., the radius of spots in spot patterns). They are often used in automated analysis of microscopy data of BCP thin film~\cite{Murphy2015}. The utility of summary statistics extracted from AAPS for model calibration is affirmed by numerical evidence~\cite[Section 8.3]{Baptista2022}.

This work is largely based on the Bayesian model calibration framework presented by Baptista et al., and inherits the same challenges in performing the calibration task. Here we pursue a likelihood-free inference approach via the \textit{pseudo-marginal method}~\cite{Andrieu2009} in the specific context of unguided Di-BCP thin film self-assembly. Using the AAPS of top-down microscopy image data for inferring model parameters is further explored. In particular, we derive analytical and approximate forms of probability distributions of the AAPS in response to image noise. They define \textit{conditional likelihood functions}\footnote{Note that analytical forms of the conditional likelihoods are considered, in a general setting, unavailable by Baptista et al.~\cite[Section 4.2]{Baptista2022}.} \thinspace based on the AAPS of image data as functions of both model and nuisance parameters. We numerically demonstrate that the conditional likelihoods based on AAPS are relatively insensitive to the nuisance parameters representing the aleatoric uncertainties. This property is well-suited for the pseudo-marginal method, for which the posterior sampling is efficient when such sensitivity is low~\cite{Sherlock2015,Warne2020}. Additionally, we show that informative prior distributions can be constructed based on the mean pixel values of image data. They compensate for the information lost due to the reduction of image data to AAPS by assimilating information about the \textit{phase} of the latent Di-BCP patterns in the image.

The data dimension reduction from images to their AAPS enables us to readily and effectively build surrogate models for accelerating Bayesian model calibration. We train small-sized multi-layer perceptrons (MLPs)~\cite{Goodfellow2016} to construct a surrogate of the \textit{parameters-to-spectrum} map. It can be used to estimate the integrated likelihood functions and transfer the computational burden of repeated model solves for each calibration task to that of a one-time sample generation and surrogate building. We numerically demonstrate that the naive neural network surrogate can achieve reasonable accuracy in posterior sampling with decreased uncertainty reduction, which can be compensated by efficiently assimilating more data into the model. 

We provide a high-level summary of the proposed Bayesian model calibration procedure in Figure \ref{fig:summary} and below:
\begin{enumerate}
    \item Implement a computational model of interest that predicts Di-BCP thin film patterns for a range of parameter values corresponding to various Di-BCP materials and experimental scenarios.
    \item Collect top-down microscopy images of a Di-BCP thin film pattern of interest.
    \item Extract the mean pixel values and AAPS of image data.
    \item Use the mean pixel values to construct a phase-informed prior distribution of model parameters.
    \item Use the AAPS entries and either (i) model simulations with the pseudo-marginal method or (ii) machine learning surrogate accelerated predictions to sample from the length scale and phase--informed posterior distribution of model parameters.
\end{enumerate}

\subsection{Contributions with broader impact}
Regarding contributions to DSA research, a comprehensive Bayesian approach that quantifies various uncertainties in computational models of the self-assembly and their predictions has not been reported beyond \cite{Baptista2022}. As discussed in Section \ref{subsec:motivation}, understanding and controlling uncertainties, particularly those associated with defect formation, are fundamental challenges for advancing BCP DSA technology. The proposed method bridges the computational models and microscopy characterizations of Di-BCP thin film self-assembly through concepts deeply embedded in the theory of Di-BCP microphase separation. For example, the phase-informed prior and the length scale--informed likelihood are proposed based on our understanding that Di-BCP material properties highly correlate with the length scales (such as periodicity length and interface length) and phase of Di-BCP equilibrium structures \cite{Leibler1980, Khandpur1995}. More importantly, the method rigorously (through probabilistic modeling) and efficiently (through a machine learning surrogate) encodes (through Bayes' rule) uncertainties from data acquisition and pattern formation in the calibrated model parameters. When deployed to search for optimal designs of DSA procedures, a calibrated computational model with quantified uncertainties often leads to much more reliable and robust designs than those obtained by a deterministically calibrated computational model; see examples of optimal design under uncertainties in \cite{chen2021optimal,alghamdi2022optimal, Luo2023}.

We believe that the proposed Bayesian model calibration method can be applied to scientific and engineering problems beyond Di-BCP thin film self-assembly. Carefully designed summary statistics and machine learning surrogates based on the power spectrum of experimental and simulated image data may help enable uncertainty quantification of models for a wide range of random small-scale pattern and structure formation; see, e.g., \cite{zhao2021image, acar2021recent, gu2021uncertainty, yoshinaga2022bayesian}.

\subsection{Layout of the paper}
The layout of the paper is as follows. Section~\ref{sec:aaps} introduces a probabilistic model of the AAPS of image data induced by the image noise for a given Di-BCP material state. In Section~\ref{sec:forward_model}, we introduce a generic forward operator for diblock copolymer self-assembly, i.e., the map from model parameters to model-predicted Di-BCP material states, and a specific DFT model called the \textit{Ohta--Kawasaki model} used in our numerical studies. In Section~\ref{sec:bayesian_model_calibration}, we define length-scale informed conditional and integrated likelihoods based on the AAPS of image data. Their utility and efficiency for performing the calibration task via the pseudo-marginal method are analyzed. Then, we introduce phase-informed prior distributions conditioned on the mean pixel value of the image data. Section~\ref{sec:nn_surrogate} describes a neural network surrogate of the parameter-to-spectrum map. In Section~\ref{sec:numerical_results}, we provide numerical examples demonstrating the capability of the proposed formulation and trained neural network surrogate to infer parameters in the Ohta--Kawasaki model for unguided diblock copolymer thin film self-assembly when considering various sources of uncertainties. Conclusions are given in Section \ref{sec:conclusion}.

\section{A probabilistic model of the azimuthally-averaged power spectrum of image data}\label{sec:aaps}

This section introduces a probabilistic model for the azimuthally-averaged power spectrum (AAPS) of top-down microscopy images that visualize the top surfaces of phase-separated Di-BCP thin films. We first define a model for the characterization process that maps a latent Di-BCP material state to an image. Then the procedure for computing AAPS from the image data is specified, which is used to derive the analytical and approximate form of the probability density for AAPS induced by image noise.

\subsection{A model of top-down microscopy characterization}
\label{subsec:characterization_model}
Let $\Omega = [0, L_1]\times[0,L_2]$ denote the surface area of the Di-BCP thin film captured by a top-down microscope image, i.e., the field of view. On the meso- to continuum scale, the latent top-surface morphology depicted by the image can be specified by an \textit{order parameter} $u:\Omega\to[-1, 1]$. It is given by the difference between normalized segment number densities of the two monomers species, labeled A and B. That is,
\begin{equation}\label{eq:order_parameter}
    u(\x) = u_A(\x)-u_B(\x)\in[-1, 1]\,.
\end{equation}
When the order parameter takes on the extreme values of $\pm 1$, it implies a locally pure monomer A or B composition. A locally mixed composition is implied when the order parameter takes on values between $\pm 1$. We denote the set of possible order parameters by $V^u(\Omega)$.

\begin{figure}[!ht]
\centering
\addtolength{\tabcolsep}{-5pt}
\scalebox{0.95}{\footnotesize
\begin{tabular}{|c c|c c c|c|c|}\hline

    \multicolumn{2}{|c|}{$\displaystyle u\in V^u([0, L]^2)$} & \multicolumn{3}{c|}{Pixelization: $\bdmc{D}(u)$} & \multirow{3}{*}{\includegraphics[width = 0.26\linewidth]{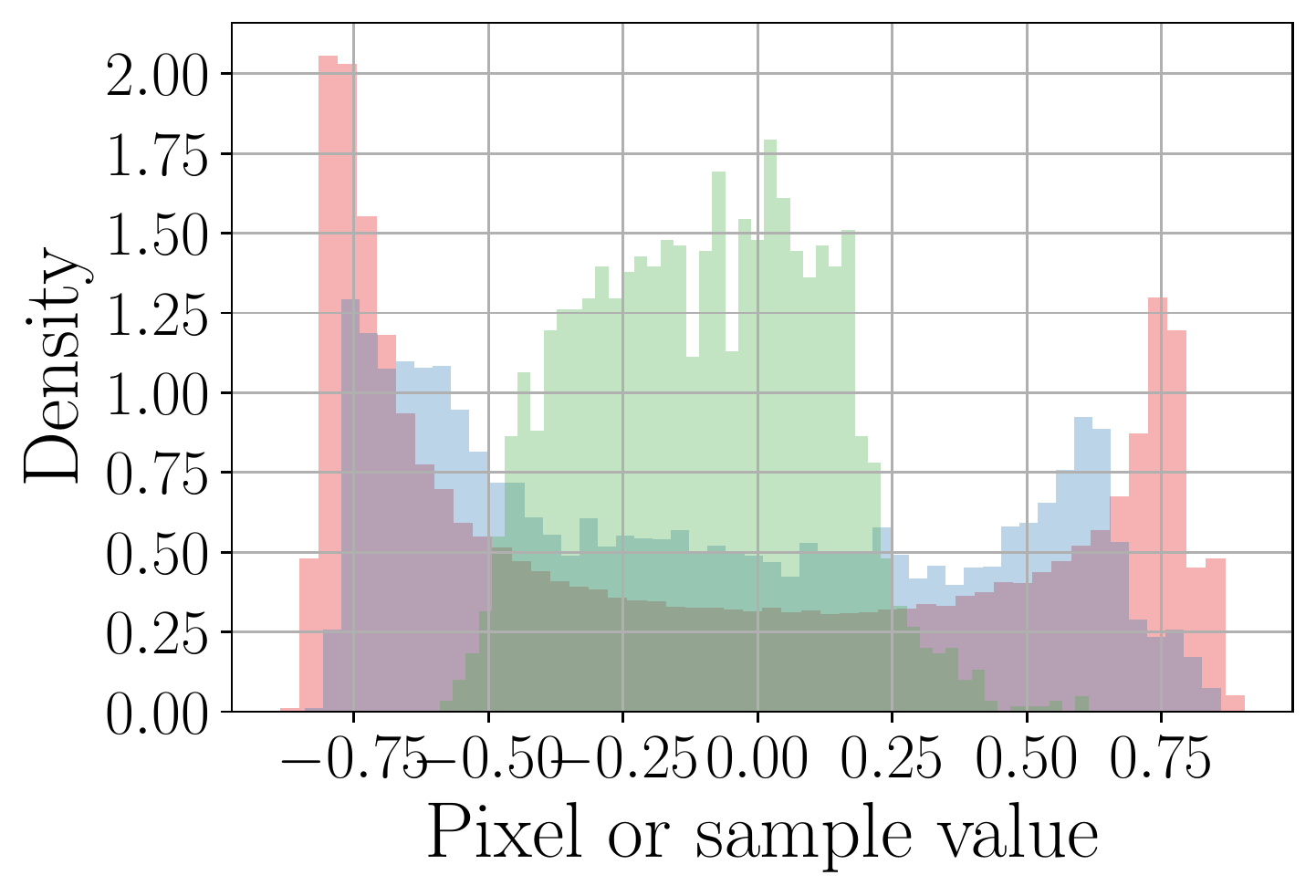}} & \multirow{3}{*}{\includegraphics[width = 0.26\linewidth]{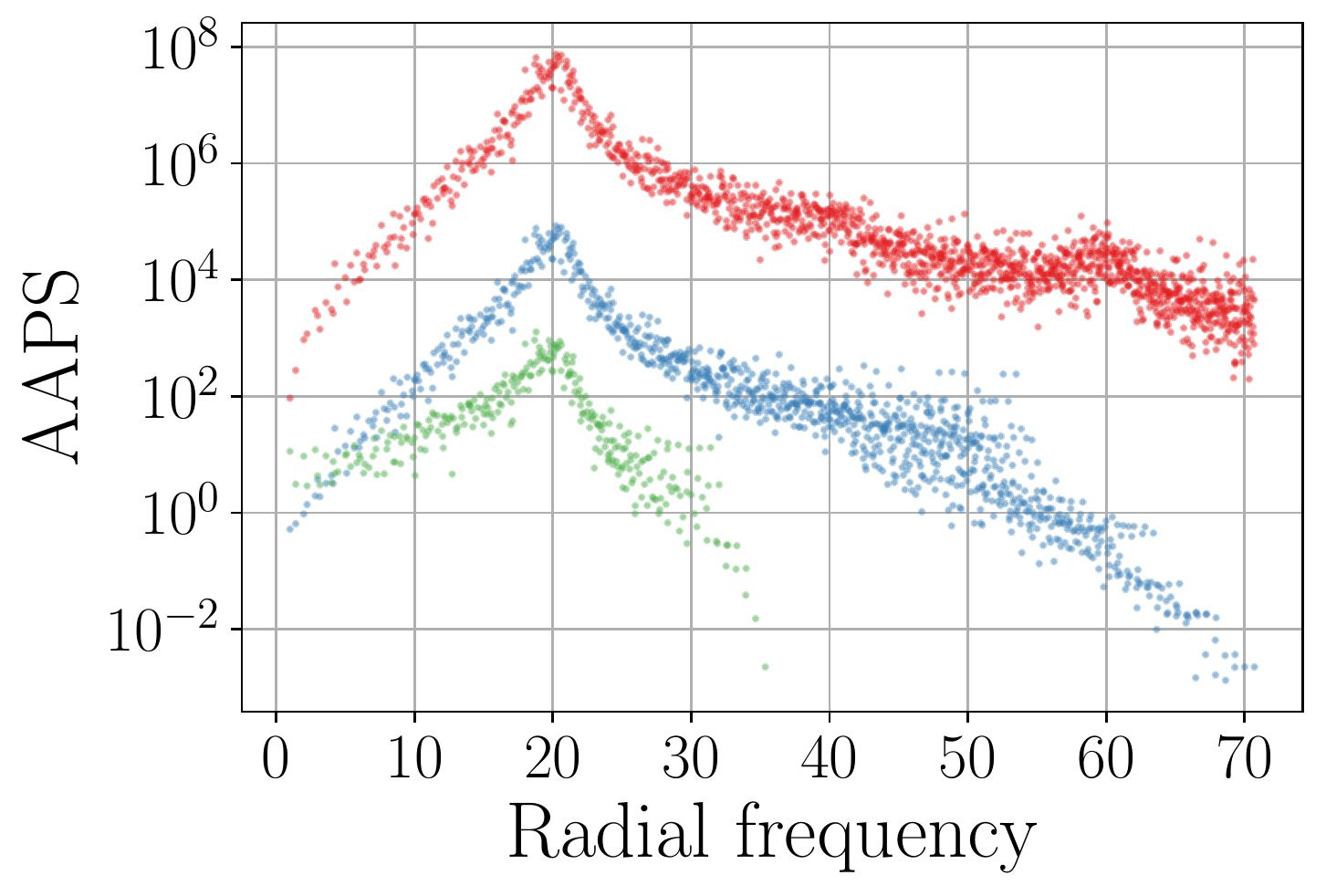}}\\\cline{1-5}
     & & $\displaystyle M = 100$ & $\displaystyle M = 50$ & & &\\
    \includegraphics[width = 0.13\linewidth]{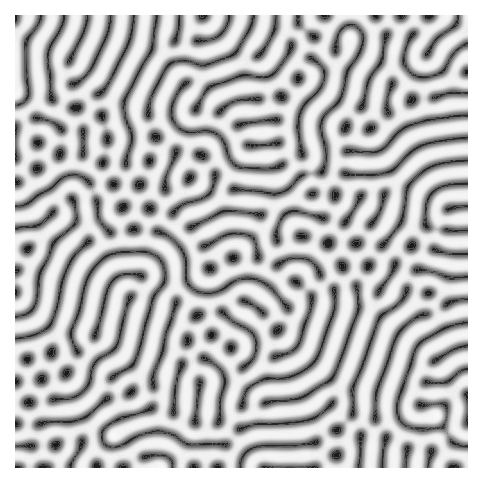}\llap{\raisebox{0.12\linewidth}{\includegraphics[width = 0.03\linewidth]{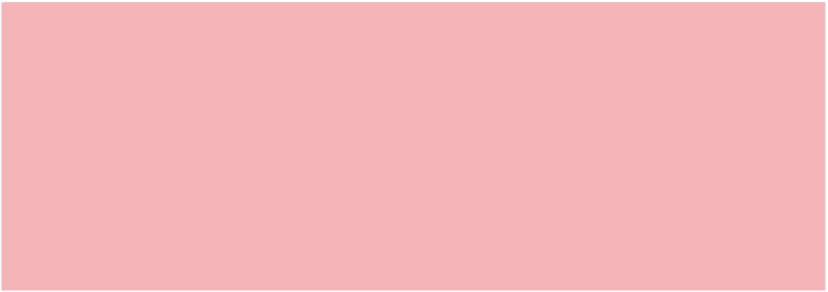}}} & \includegraphics[height = 0.13\linewidth]{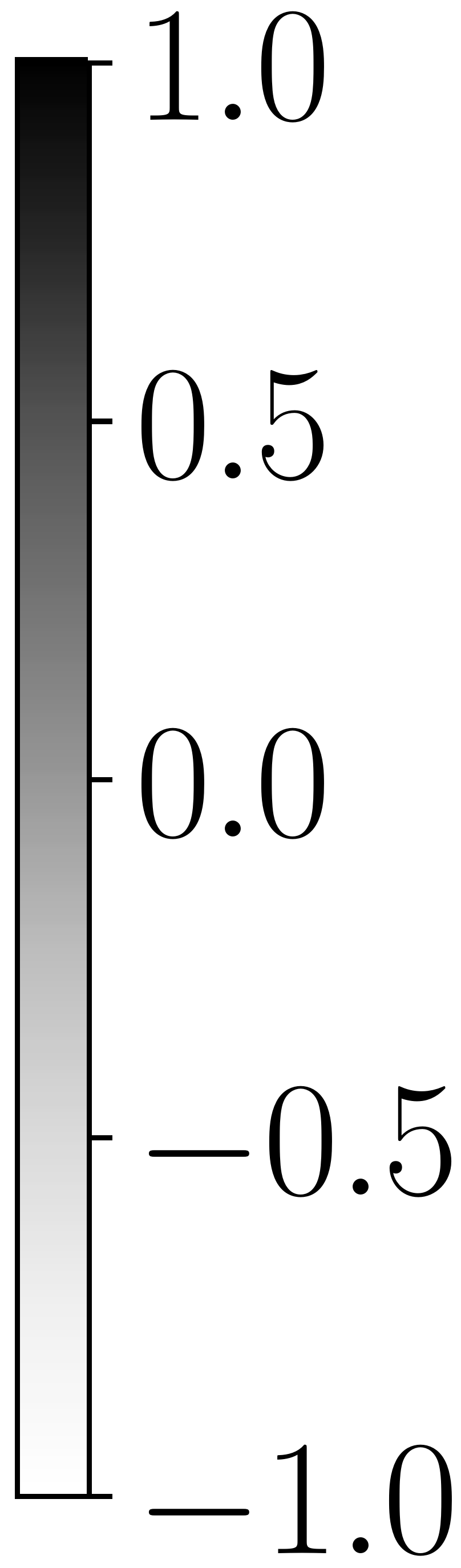} & \includegraphics[width = 0.13\linewidth]{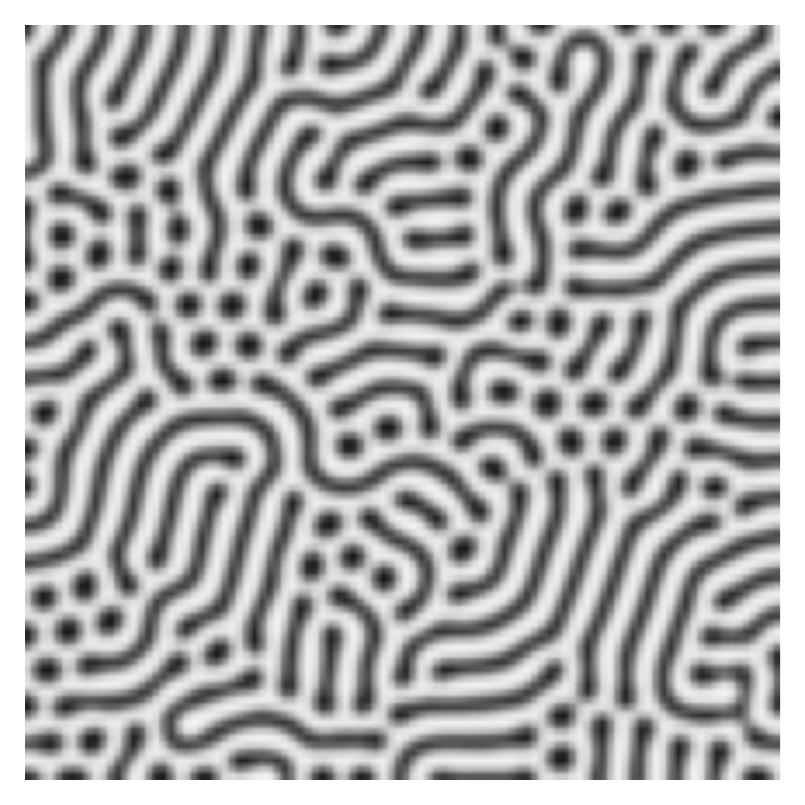}\llap{\raisebox{0.12\linewidth}{\includegraphics[width = 0.03\linewidth]{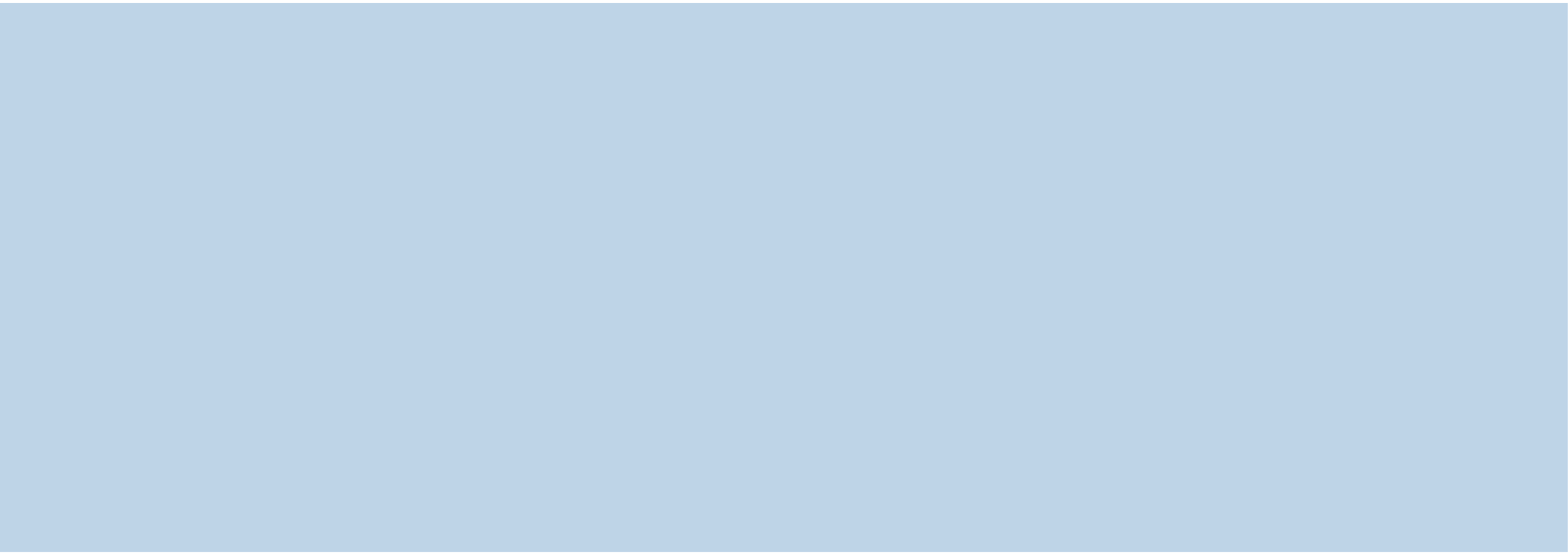}}} & \includegraphics[width = 0.13\linewidth]{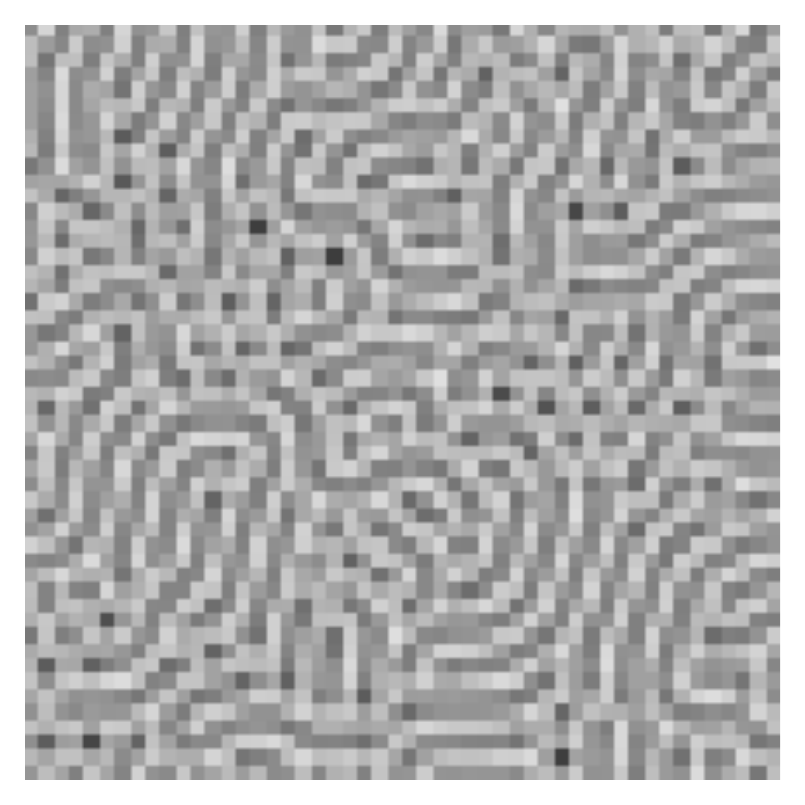}\llap{\raisebox{0.12\linewidth}{\includegraphics[width = 0.03\linewidth]{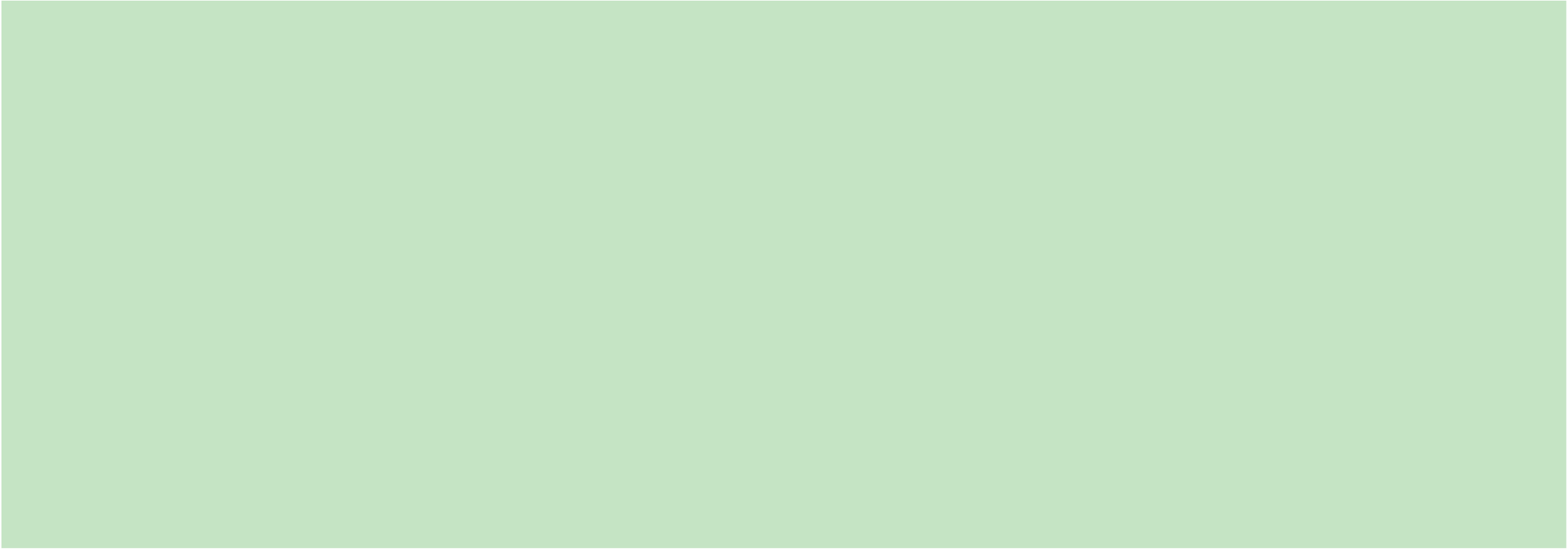}}} & \includegraphics[height = 0.13\linewidth]{colorbar_colorbar_op_25.pdf} & & \\\hline 
    
    \multicolumn{2}{|c|}{$\displaystyle \boldsymbol{u}_1$} & \multicolumn{3}{c|}{Contrast scaling: $\bdmc{C}(\boldsymbol{u}_1;c_1, c_2)$} & \multirow{3}{*}{\includegraphics[width = 0.26\linewidth]{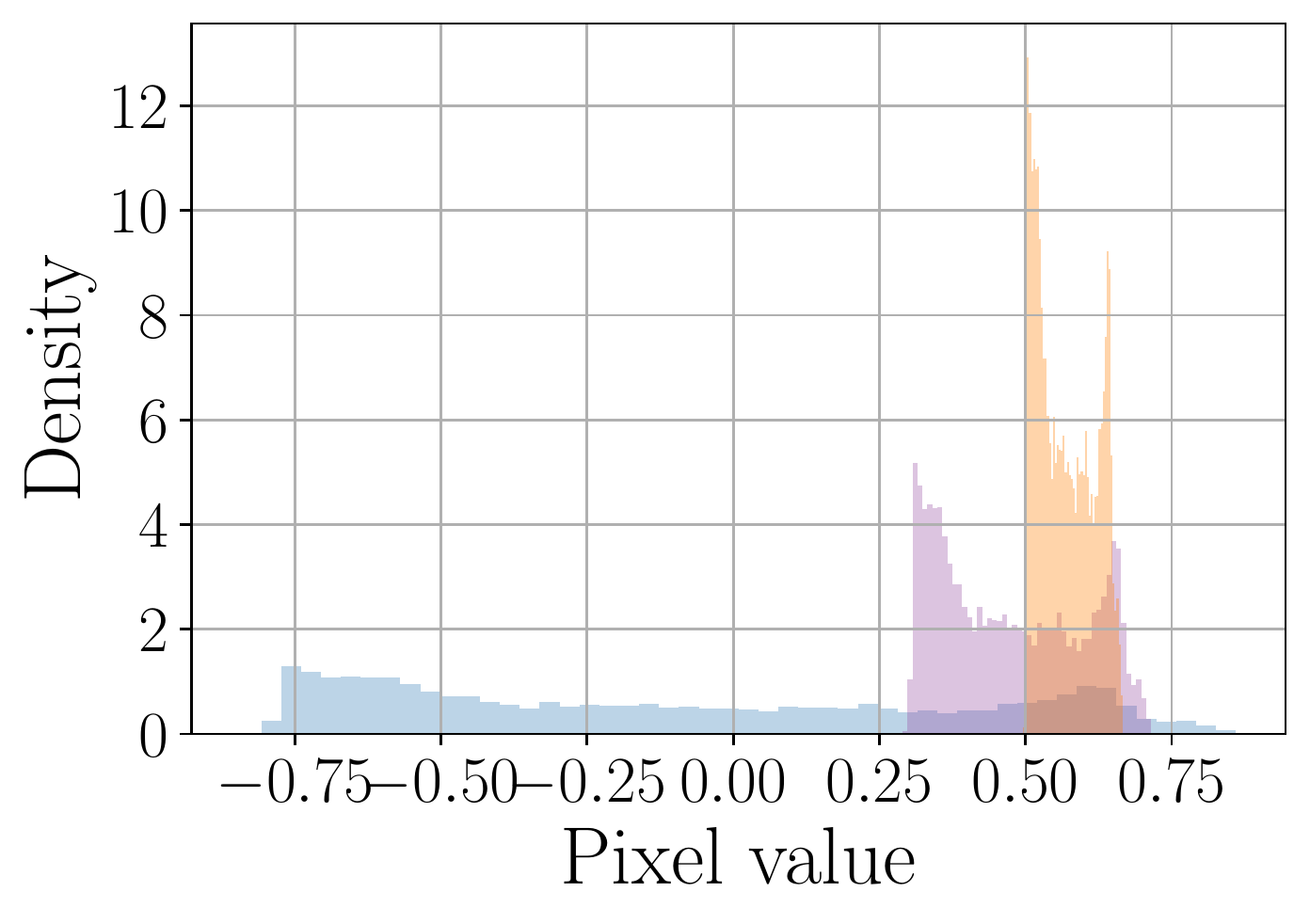}} & \multirow{3}{*}{\includegraphics[width = 0.26\linewidth]{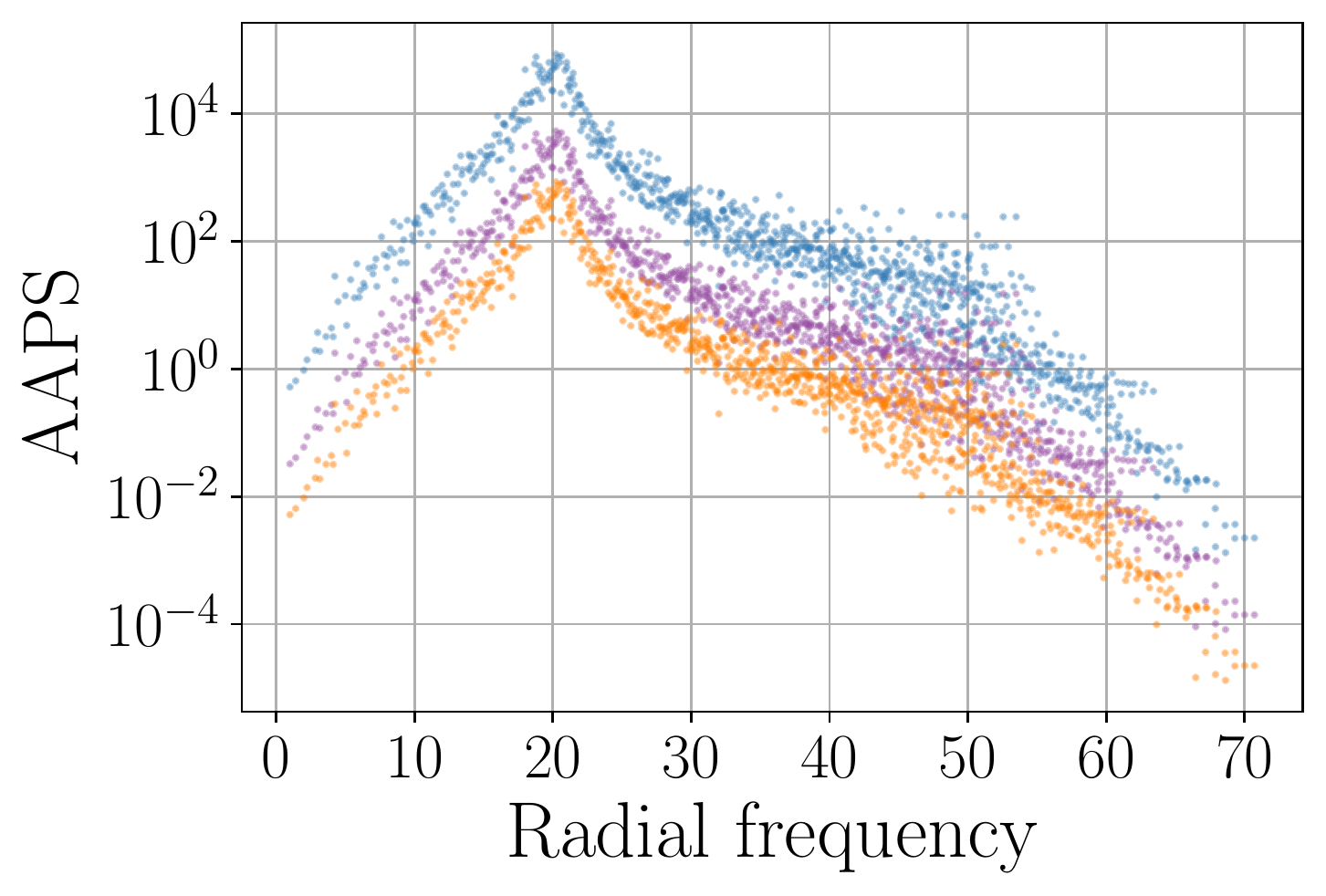}}\\\cline{1-5}
    & & $\displaystyle (c_1, c_2) = (0.25, 0)$ & $\displaystyle (c_1, c_2) = (0.1, 0.1)$ & & & \\
    \includegraphics[width = 0.13\linewidth]{corruptions_local_averaging_100.pdf}\llap{\raisebox{0.12\linewidth}{\includegraphics[width = 0.03\linewidth]{corruptions_legend_2.png}}} & \includegraphics[height = 0.13\linewidth]{colorbar_colorbar_op_25.pdf}& \includegraphics[width = 0.13\linewidth]{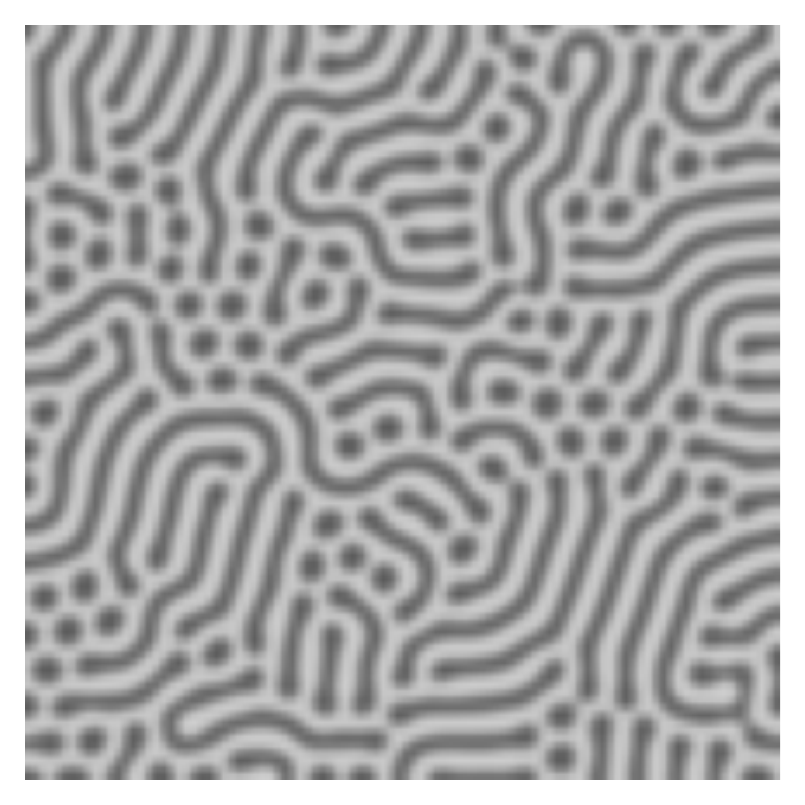}\llap{\raisebox{0.12\linewidth}{\includegraphics[width = 0.03\linewidth]{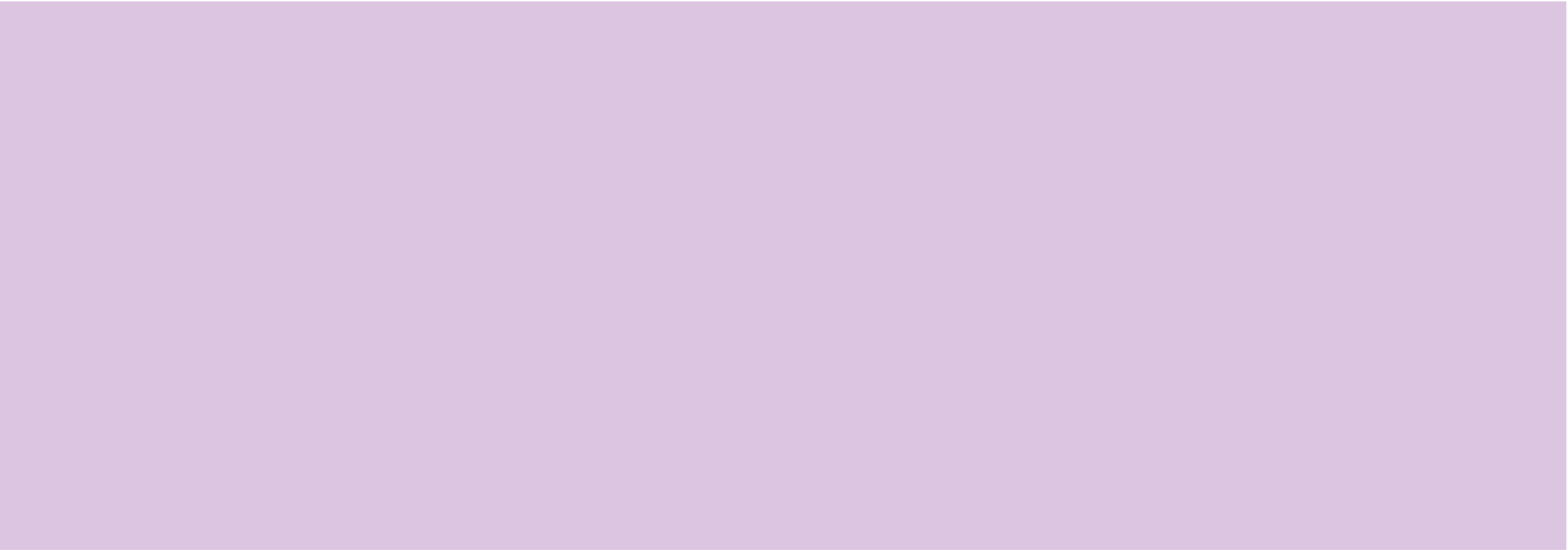}}} & \includegraphics[width = 0.13\linewidth]{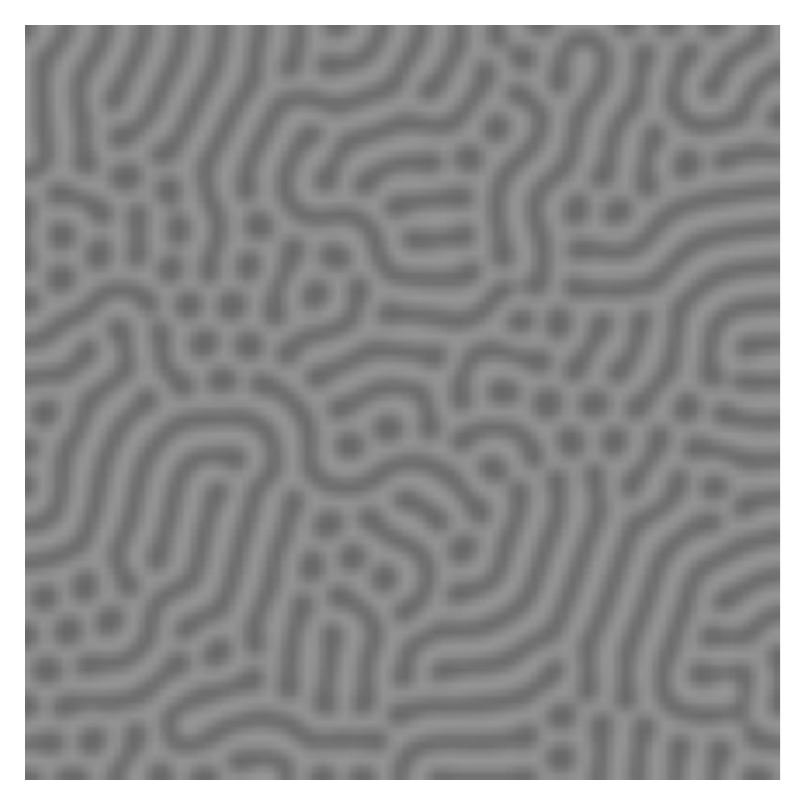}\llap{\raisebox{0.12\linewidth}{\includegraphics[width = 0.03\linewidth]{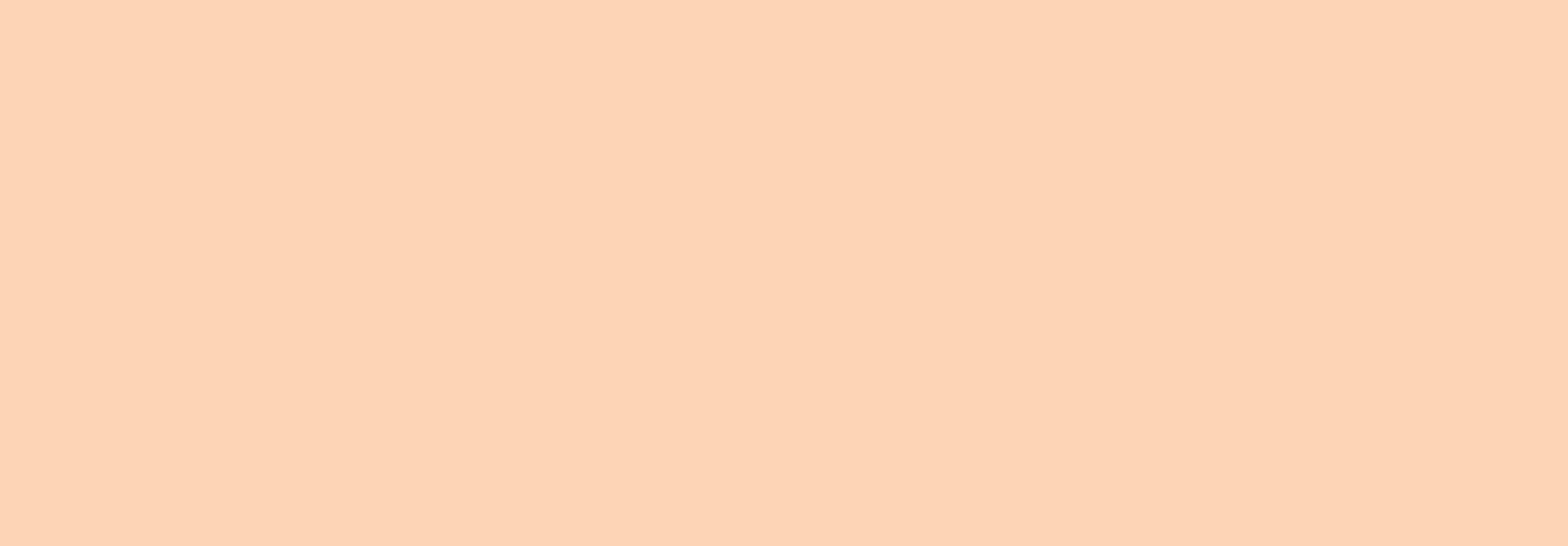}}} & \includegraphics[height = 0.13\linewidth]{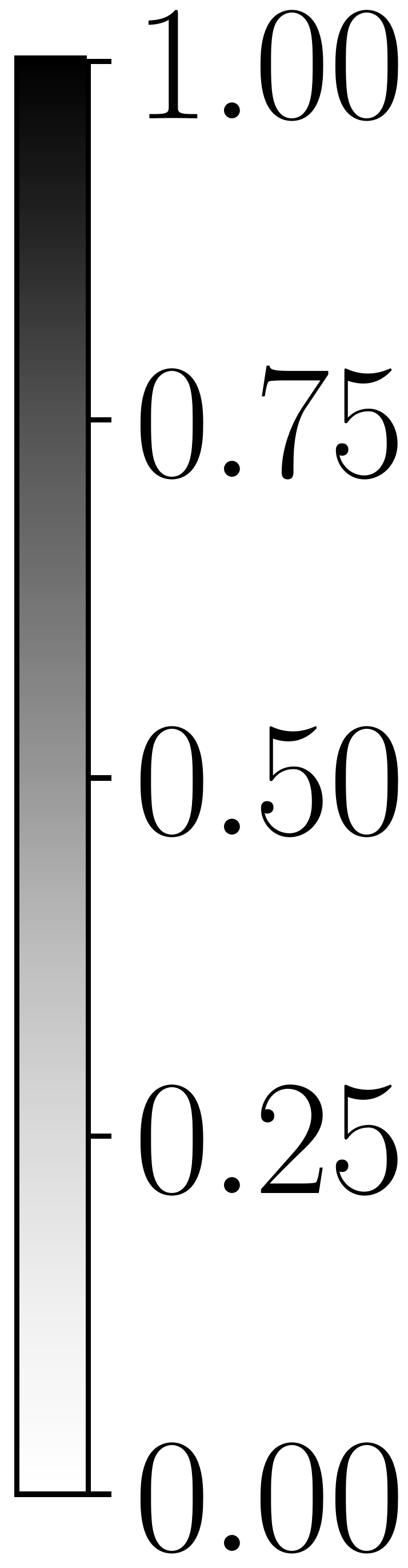}& & \\\hline
    
    \multicolumn{2}{|c|}{$\displaystyle \boldsymbol{u}_2$} & \multicolumn{3}{c|}{Blurring: $\bdmc{B}(\boldsymbol{u}_2;\sigma_b)$} & \multirow{3}{*}{\includegraphics[width = 0.25\linewidth]{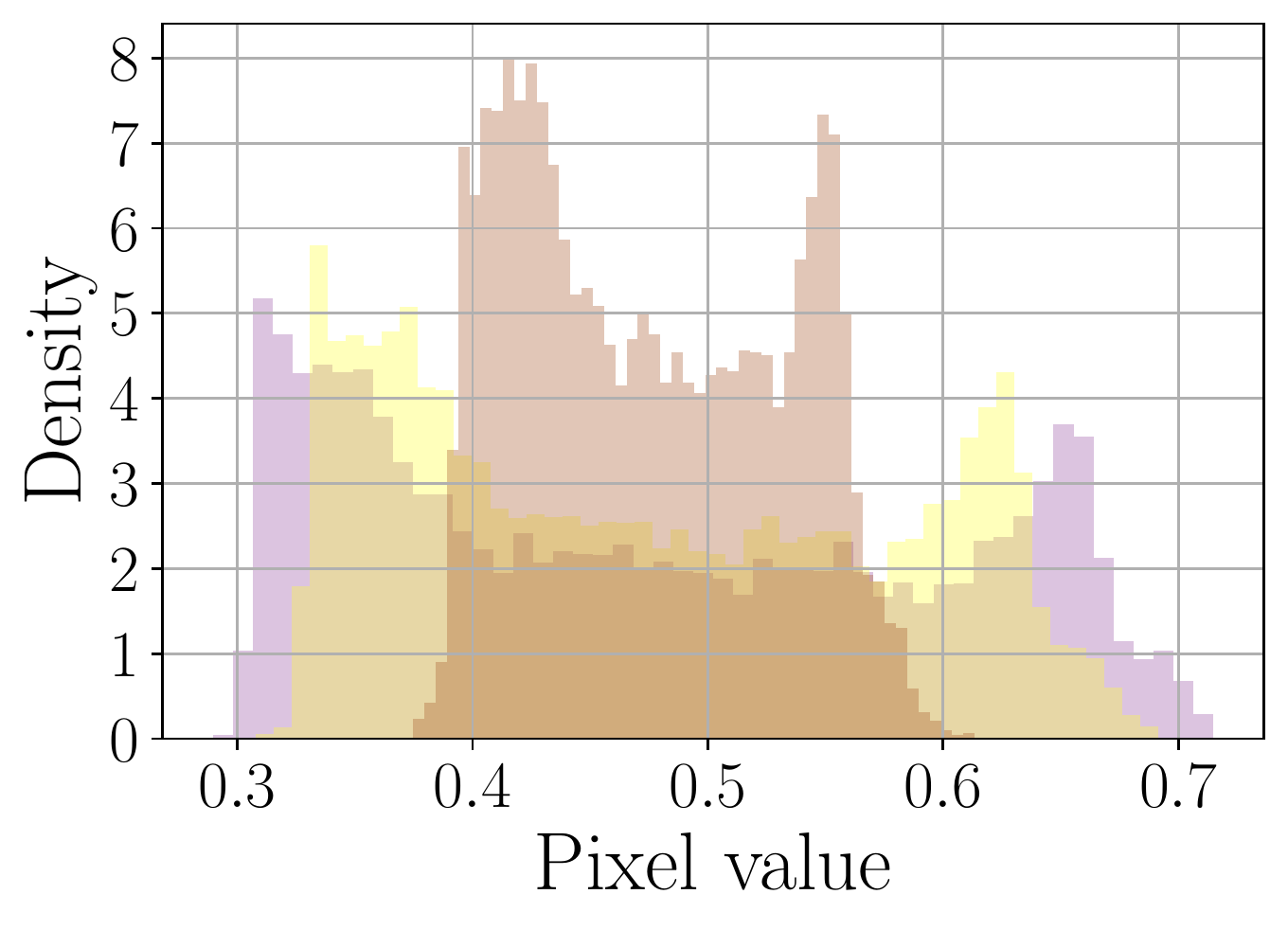}} & \multirow{3}{*}{\includegraphics[width = 0.26\linewidth]{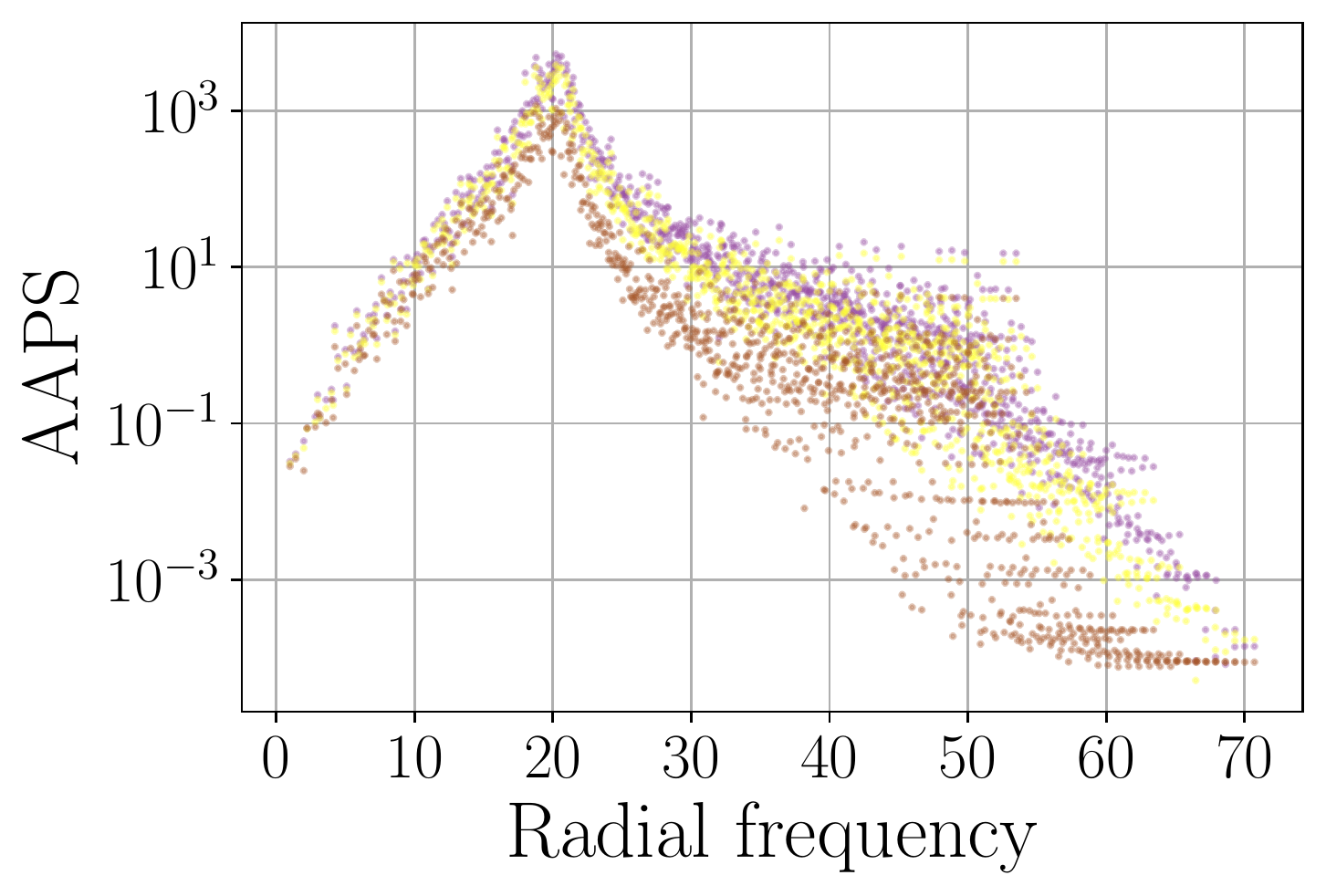}}\\\cline{1-5}
     & & $\displaystyle \sigma_b = 0.005$ & $\displaystyle \sigma_b = 0.01$ & & & \\
    \includegraphics[width = 0.13\linewidth]{corruptions_contrast_d25.pdf}\llap{\raisebox{0.12\linewidth}{\includegraphics[width = 0.03\linewidth]{corruptions_legend_4.png}}} & \includegraphics[height = 0.13\linewidth]{colorbar_colorbar_image_25.pdf} & \includegraphics[width = 0.13\linewidth]{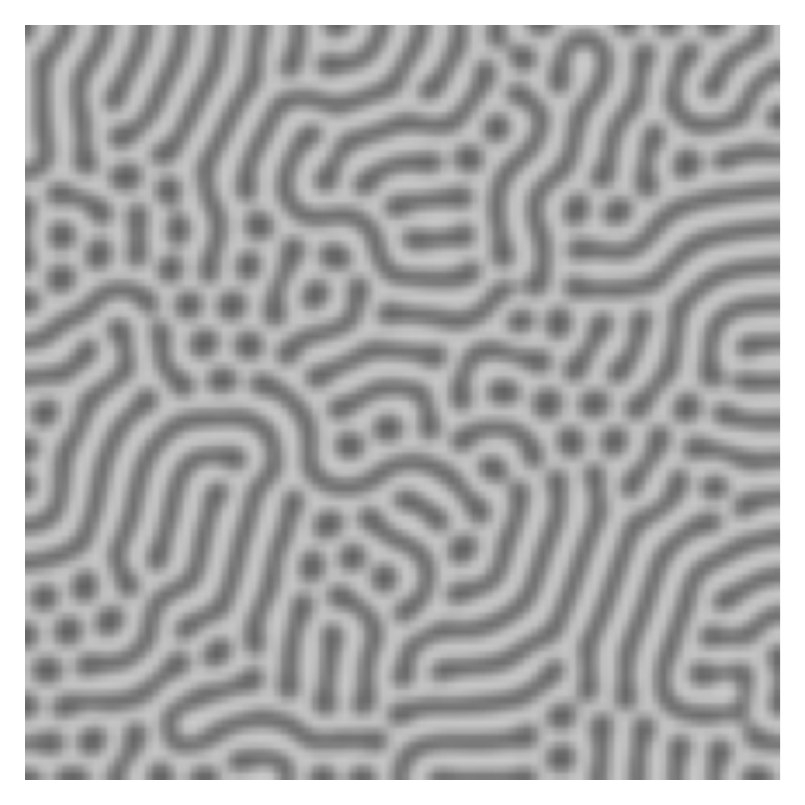}\llap{\raisebox{0.12\linewidth}{\includegraphics[width = 0.03\linewidth]{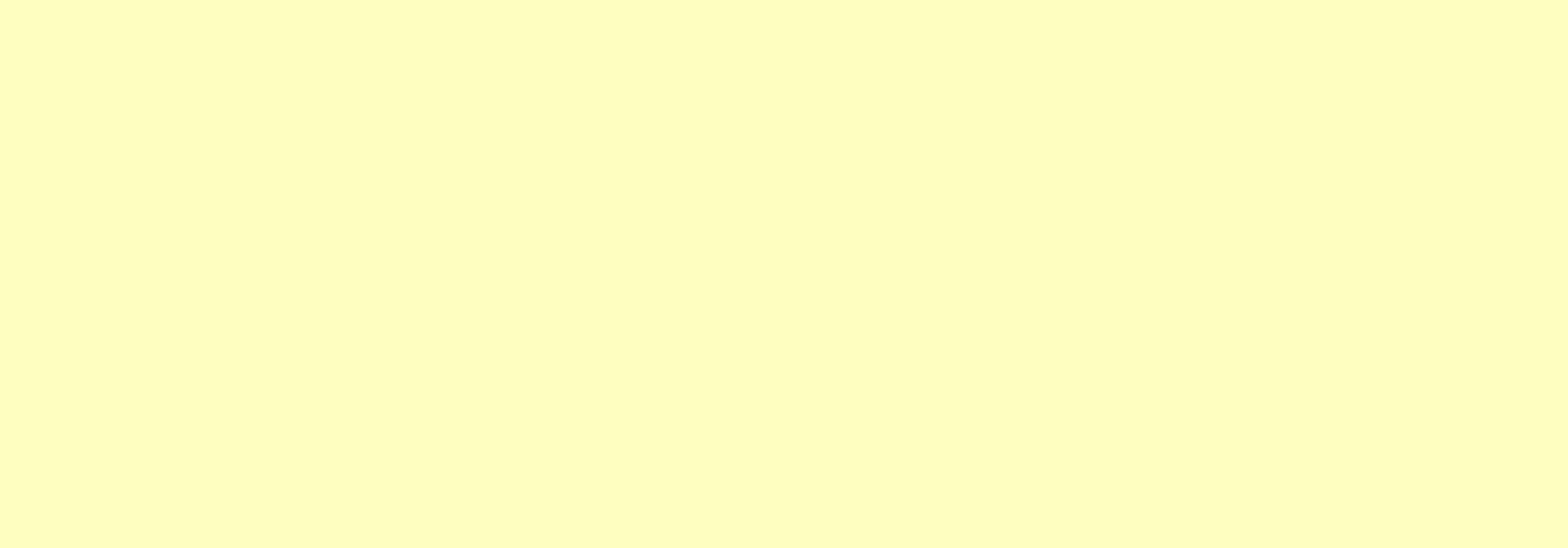}}} & \includegraphics[width = 0.13\linewidth]{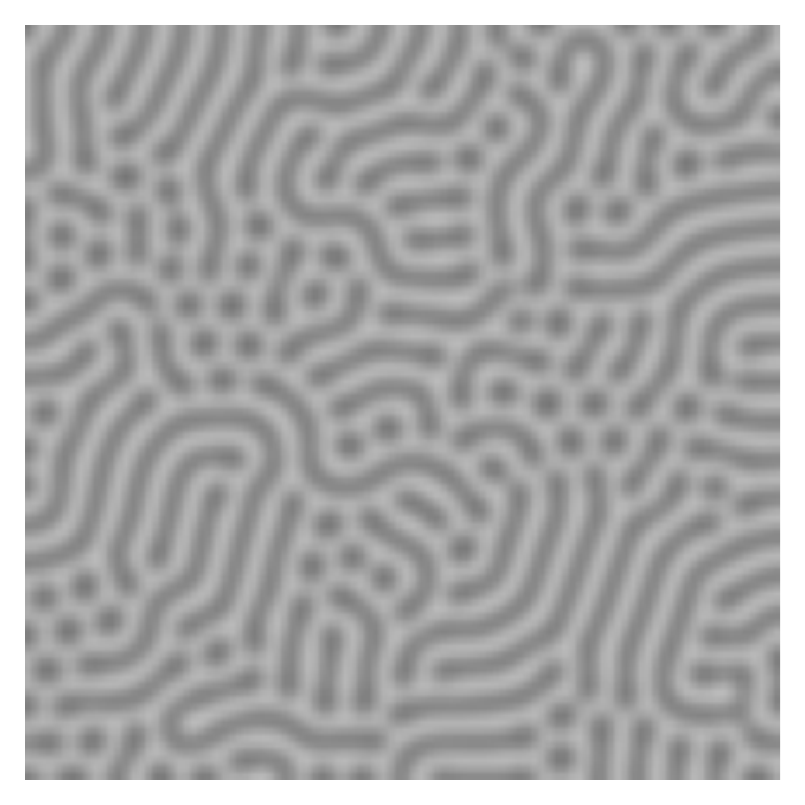}\llap{\raisebox{0.12\linewidth}{\includegraphics[width = 0.03\linewidth]{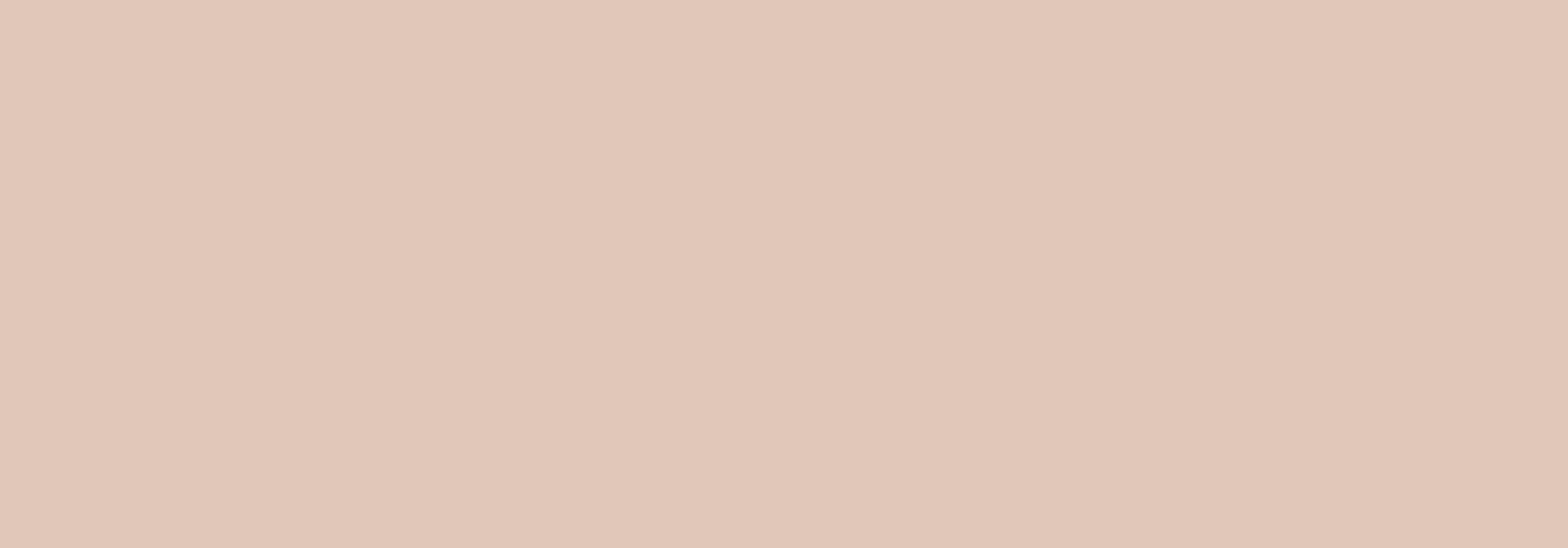}}} & \includegraphics[height = 0.13\linewidth]{colorbar_colorbar_image_25.pdf} & &\\\hline
    
    \multicolumn{2}{|c|}{$\displaystyle \boldsymbol{u}_3$} & \multicolumn{3}{c|}{Additive noise: $\boldsymbol{u}_3 + \boldsymbol{n}^j(\sigma_n)$} & \multirow{3}{*}{\includegraphics[width = 0.25\linewidth]{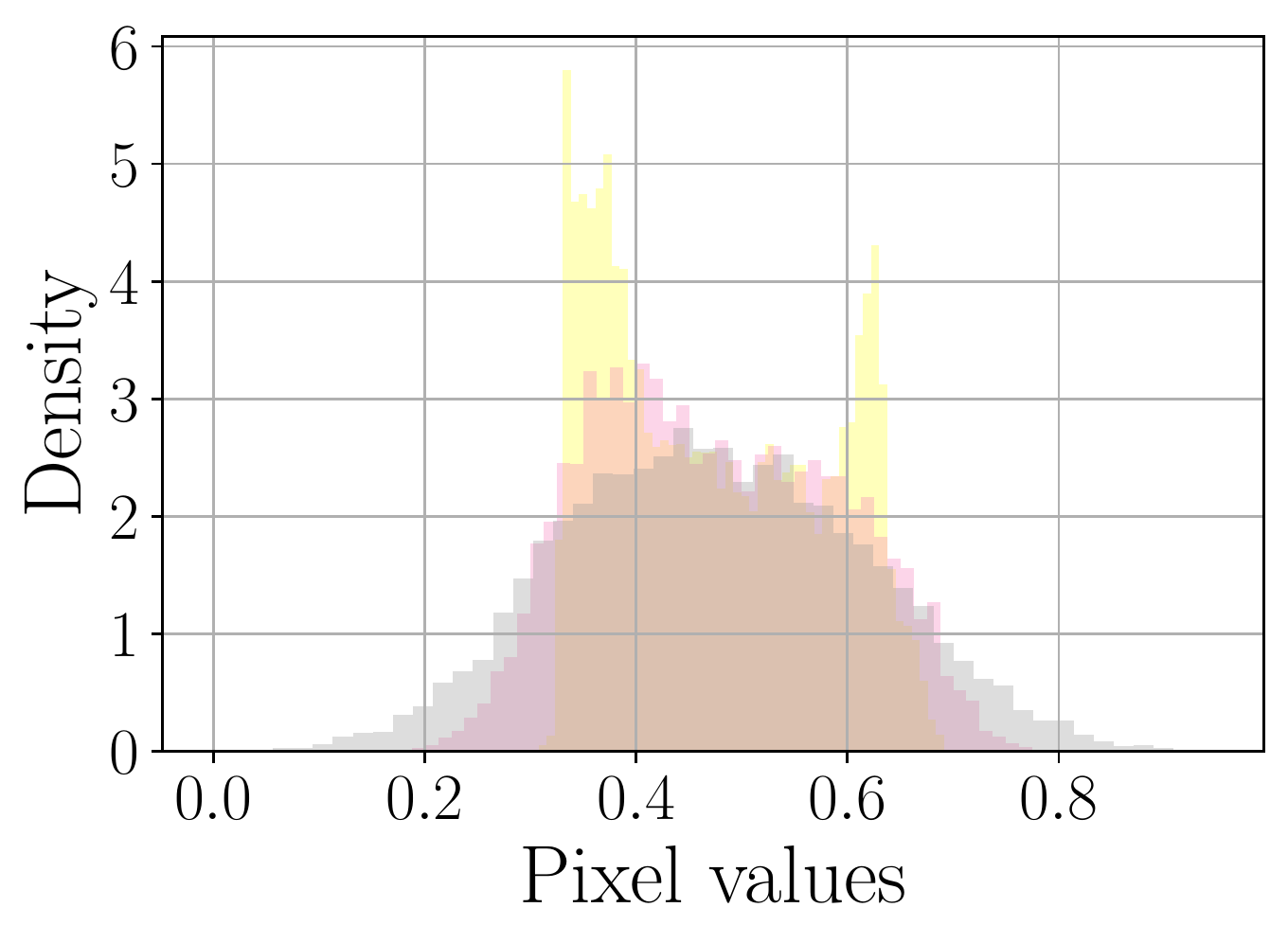}} & \multirow{3}{*}{\includegraphics[width = 0.26\linewidth]{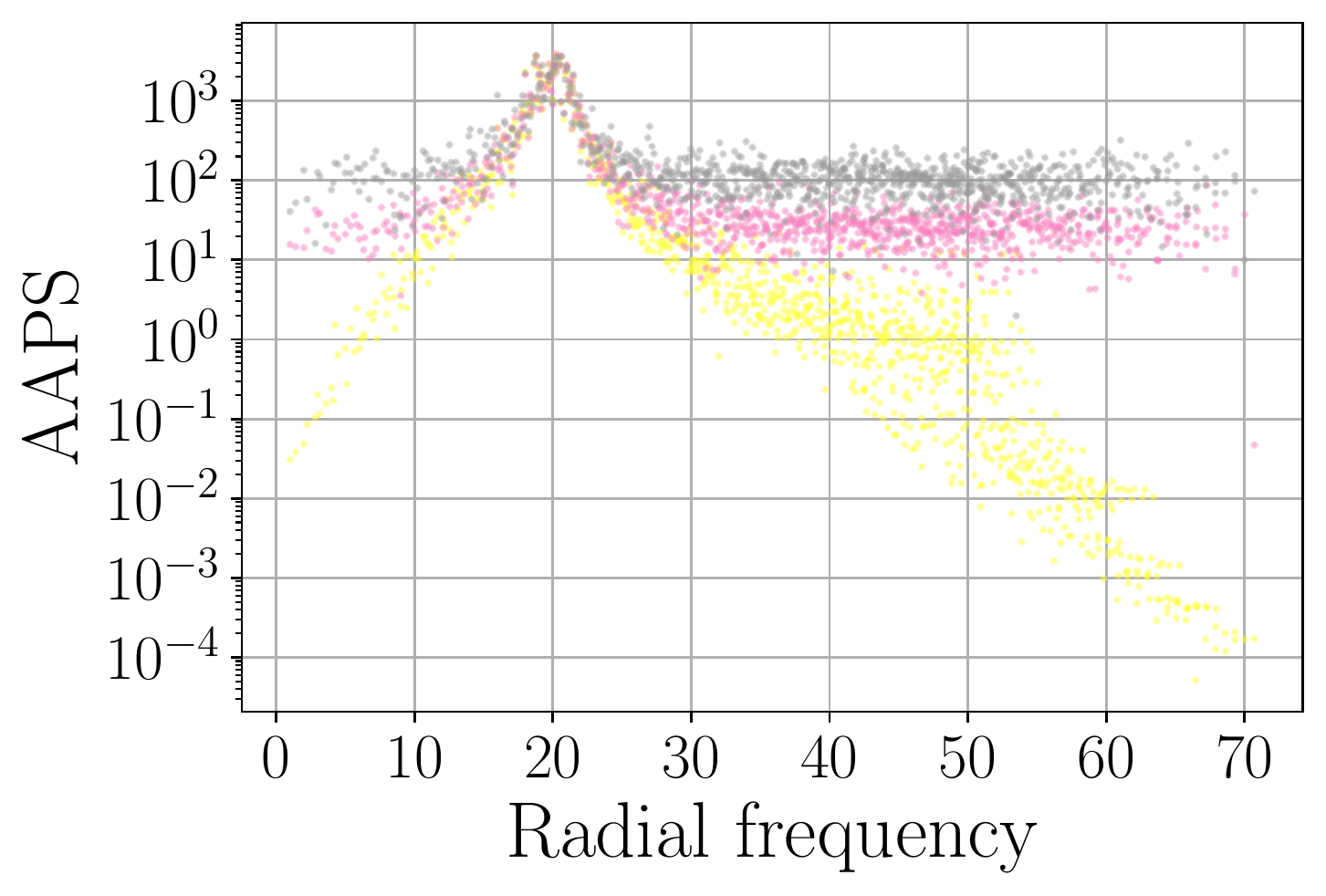}}\\\cline{1-5}
     & & $\displaystyle \sigma_n = 0.05$ & $\displaystyle \sigma_n = 0.1$ &  & & \\
    \includegraphics[width = 0.13\linewidth]{corruptions_blur_d5.pdf}\llap{\raisebox{0.12\linewidth}{\includegraphics[width = 0.03\linewidth]{corruptions_legend_6.png}}} & \includegraphics[height = 0.13\linewidth]{colorbar_colorbar_image_25.pdf} & \includegraphics[width = 0.13\linewidth]{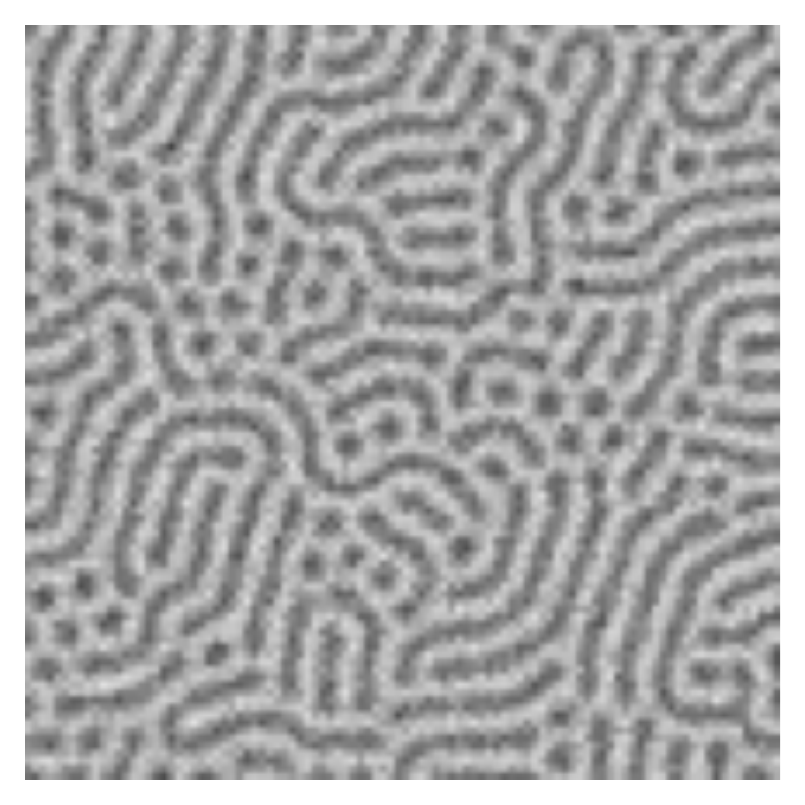}\llap{\raisebox{0.12\linewidth}{\includegraphics[width = 0.03\linewidth]{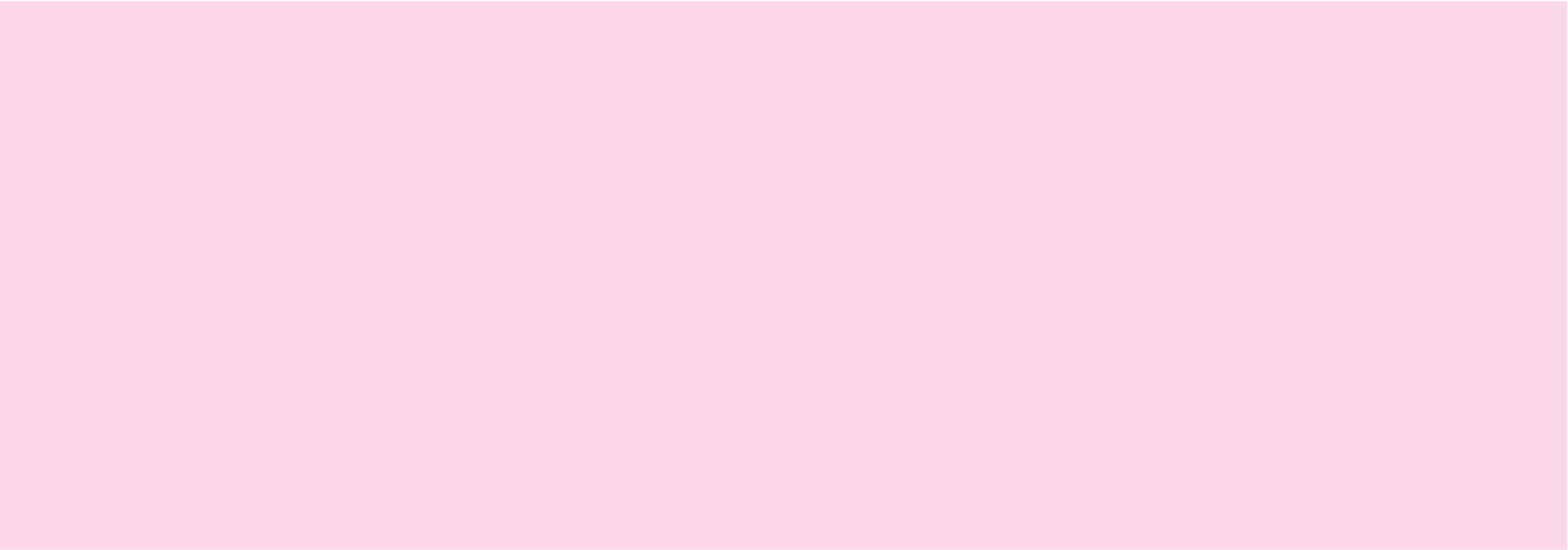}}} & \includegraphics[width = 0.13\linewidth]{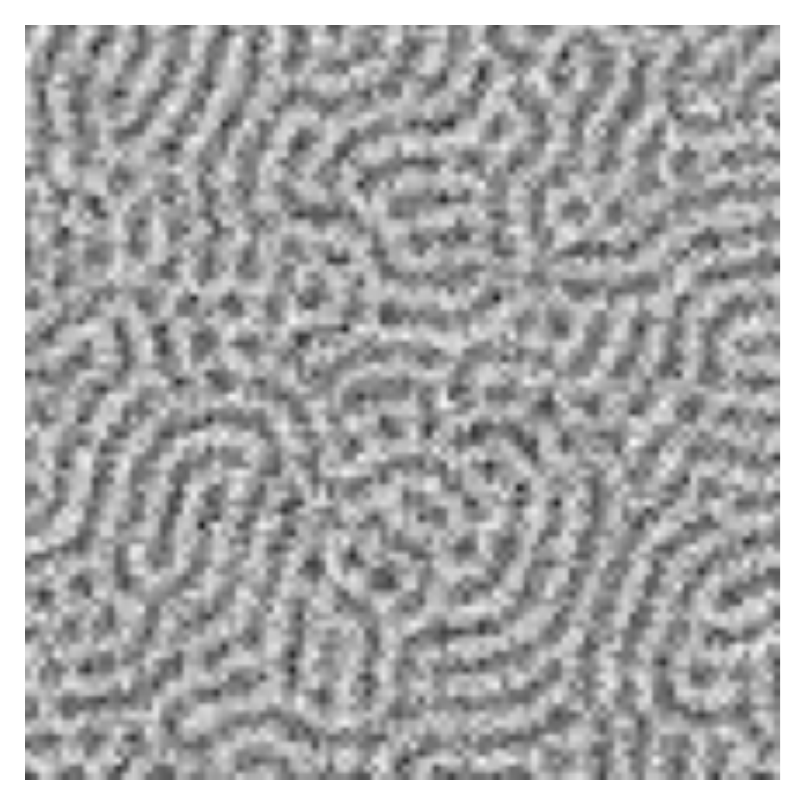}\llap{\raisebox{0.12\linewidth}{\includegraphics[width = 0.03\linewidth]{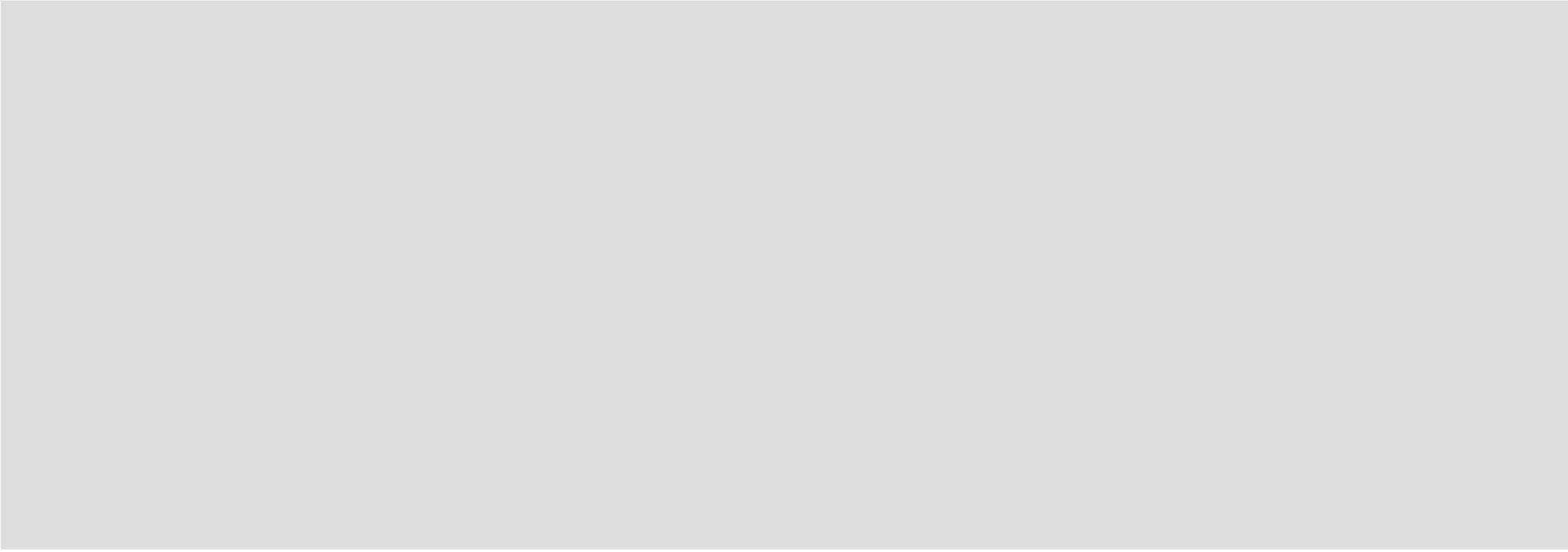}}} & \includegraphics[height = 0.13\linewidth]{colorbar_colorbar_image_25.pdf}& &\\\hline

\end{tabular}
}
\addtolength{\tabcolsep}{6pt} 
\caption{The effects of the four layers (local averaging $\bdmc{D}$, image contrast transformation $\bdmc{C}$, blurring $\bdmc{B}$, and additive noise) in the characterization model, as defined in Section~\ref{subsec:characterization_model}, are demonstrated through examples. From top to bottom, each row corresponds to a layer of the characterization model. In each row, we visualize (i) the input functions or images, (ii) and output images of the layer with varying parameters, (iii) the distribution of the Monte Carlo sample or pixel values of the input and output functions or images, (iv) the AAPS of the input and output images, as defined in Section \ref{subsec:aaps_computation}.}
\label{fig:image_corruptions}
\end{figure}

We adopt a simplified characterization model that maps the order parameter $u\in V^u(\Omega)$ describing the Di-BCP film top-surface morphologies to observed greyscale microscopy characterization images $\boldsymbol{d}\in[0, 1]^{M_1\times M_2}$:
\begin{equation}\label{eq:charac_model}
    \boldsymbol{d} = \boldsymbol{\mathcal{J}}(u) + \boldsymbol{n}\,,\quad \ \boldsymbol{\mathcal{J}}\coloneqq\boldsymbol{\mathcal{B}}\circ\bdmc{C}\circ\boldsymbol{\mathcal{D}}\,,
\end{equation}
where $\boldsymbol{\mathcal{J}}: V^u(\Omega)\to[0, 1]^{M_1\times M_2}$ is the \textit{state-to-image map} and other components are defined as follows:

\begin{enumerate}[label=(\roman*)]
\item The operator $\boldsymbol{\mathcal{D}}: V^u(\Omega)\to[-1,1]^{M_1\times M_2}$ applies local averaging and discretization introduced by a uniform grid of image-forming sensors labeled by indices $j =1,\dots, M_1$ and $k = 1,\dots, M_2$:
\begin{equation*}
\left(\boldsymbol{\mathcal{D}}\left(u\right)\right)_{jk} = \frac{M_1M_2}{L_1L_2}\int_{\Omega_{jk}} u(\boldsymbol{s})\,d\boldsymbol{s}\,,\quad \Omega_{jk} = \left[\left(j-1\right)L_1/M_1, jL_1/M_1\right]\times\left[\left(k-1\right)L_2/M_2, kL_2/M_2\right]\,.
\end{equation*}
\item The operator $\bdmc{C}:[-1, 1]^{M_1\times M_2}\to[0,1]^{M_1\times M_2}$ transforms the signals into grey-scale image signals in the range $[0,1]$. The operator should scale the range of order parameters to match the sample- and instrument-specific image brightness and contrast of the two monomer species in the Di-BCP film~\cite{Michler2008, Sawyer2008}. In this study, we consider a linear transformation as follows
\begin{equation*}
    \bdmc{C}(\boldsymbol{u}) = c_1\boldsymbol{u} -  2|c_1|c_2 + c_2 + \frac{1}{2}\,,\quad c_1\in\left[-\frac{1}{2}, \frac{1}{2}\right]\,,\quad c_2\in\left[-\frac{1}{2}, \frac{1}{2}\right]\,.
\end{equation*}
The parameters $c_1$ and $c_2$ account for contrast scaling and brightness shift, respectively.

\item The operator $\boldsymbol{\mathcal{B}}:[0,1]^{M_1\times M_2}\to[0,1]^{M_1\times M_2}$ applies blurring corruptions to the latent image. It is accomplished by convoluting\footnotemark\thinspace the latent image with a blur kernel specified by a point spread function (PSF). In particular, for electron microscopy \cite{Michler2008}, a technique often used for polymer characterization, the PSF is defined as the spatial distribution of electrons in a focused beam \cite{Orloff2009}. In this study, we model the PSF as a centered Gaussian function $G:\Omega\to\R_+$ with a constant diagonal covariance matrix:
\footnotetext{We use zero-padding for pixels near the boundary to compute the convolution.}
\begin{equation*}
    G(\boldsymbol{s}) = \frac{\sqrt{L_1L_2}}{2\pi\sigma_b}\exp\left(-\frac{L_1L_2\norm{\boldsymbol{s}}^2_2}{2\sigma_b^2}\right)\,.
\end{equation*}
The parameter $\sigma_b\in\R_+$ represents the level of the blurring corruption, normalized with respect to the field of view size, $L_1L_2$.

\item The term $\boldsymbol{n}\in\R^{M_1\times M_2}$ is the realization of a random matrix $\boldsymbol{N}$ that represents the stochastic sensor noise in the observed image. In this study, we assume that the entries of the stochastic noise are independent and identically distributed (i.i.d.). Furthermore, each entry is assumed to be normally distributed:
\begin{equation*}
    \boldsymbol{N}_{jk} \sim \mathcal{N}(0, \sigma_n^2)\;.
\end{equation*}
where the parameter $\sigma_n\in\R_+$ represents the level of the noise corruption.
\end{enumerate}
The effects of these components on the image data are demonstrated in Figure~\ref{fig:image_corruptions}.

Realistic characterization models are often much more complicated than the one presented above. For example, a realistic model of the PSF typically has the form of an Airy disk~\cite{Egerton2005}, and a realistic model of stochastic sensor noise is typically signal intensity--dependent, spatially correlated, and/or non-Gaussian~\cite{Roels2014,Haider2016}. We note that alternative linear contrast functions and PSFs are compatible with the pseudo-marginal approach for model calibration considered in this work. However, more complex noise models rarely admit analytical forms for the probability density of the observed image or its summary statistics. Alternative likelihood-free inference approaches \cite{cranmer2020frontier, Baptista2022} can be employed for model calibration in this setting.

In practice, the parameters in the state-to-image map $\bdmc{J}$, denoted $\w = (c_1, c_2, \sigma_b, \sigma_n)$, are epistemically uncertain, and their values can be determined with high precision before model calibration; see, e.g., \cite{Thong2001, Timischl2015, Zotta2018}. While $\w$ are \textit{nuisance parameters}, i.e., they are not the parameters of interest in our calibration task, the uncertainty of $\w$ greatly affects the outcome of model calibration. The treatment of these nuisance parameters is discussed in Section \ref{sec:bayesian_model_calibration}.

\subsection{Azimuthally-averaged power spectrum of image data} \label{subsec:aaps_computation}
To compute the AAPS of an observed image $\boldsymbol{d}\in\R^{M_1\times M_2}$, we first perform a discrete Fourier transform to obtain a 
$\hat{\boldsymbol{d}}\in\mathbb{C}^{M_1\times M_2}$:
\begin{equation*}
    \hat{\boldsymbol{d}}_{jk}\coloneqq\sum_{m_1=0}^{M_1-1}\sum_{m_2=0}^{M_2-1} \boldsymbol{{d}}_{jk}\exp{\left(-i \frac{2\pi jm_1}{M_1}\right)}\exp{\left(-i \frac{2\pi km_2}{M_2}\right)}\;.
\end{equation*}
Then the entries in $\hat{\boldsymbol{d}}$ are shifted to $\hat{\boldsymbol{d}}_s\in\mathbb{C}^{M_1\times M_2}$, where the zero frequency entry in $\hat{\boldsymbol{d}}$ is centered at $(\hat{\boldsymbol{d}}_s)_{\floor {M_1/2} \floor{M_2/2}}$ with $\floor{\cdot}$ being the flooring operator. We define $\boldsymbol{\mathcal{T}}_s:[0, 1]^{M_1\times M_2}\ni \boldsymbol{d}\mapsto \hat{\boldsymbol{d}}_s\in \mathbb{C}^{M_1\times M_2}$ as the map from an image to its zero-frequency-centered Fourier coefficient matrix. The zero-frequency-centered power spectrum of the image data is given by
\begin{equation}\label{eq:power_spectrum}
    \left(\boldsymbol{\mathcal{P}}\left(\boldsymbol{d}\right)\right)_{jk} \coloneqq \left\lvert\left(\boldsymbol{\mathcal{T}}_s\left(\boldsymbol{d}\right)\right)_{jk}\right\rvert^2\,,
\end{equation}
where $\boldsymbol{\mathcal{P}}:[0,1]^{M_1\times M_2}\to\R_+^{M_1\times M_2}$.

\begin{figure}[!hbt]
    \centering
    \includegraphics[width = 0.28\linewidth]{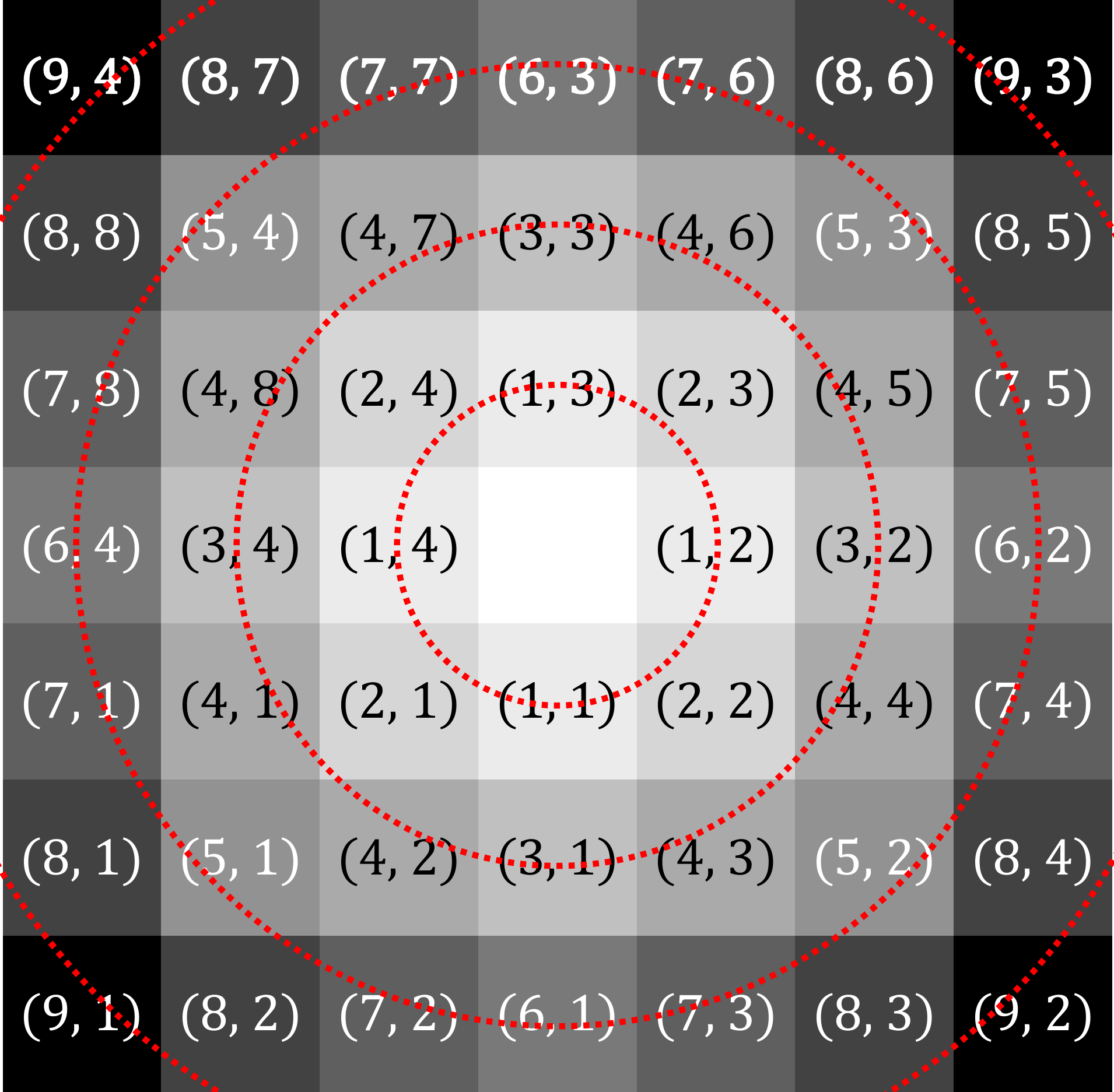}\includegraphics[width = 0.32\linewidth]{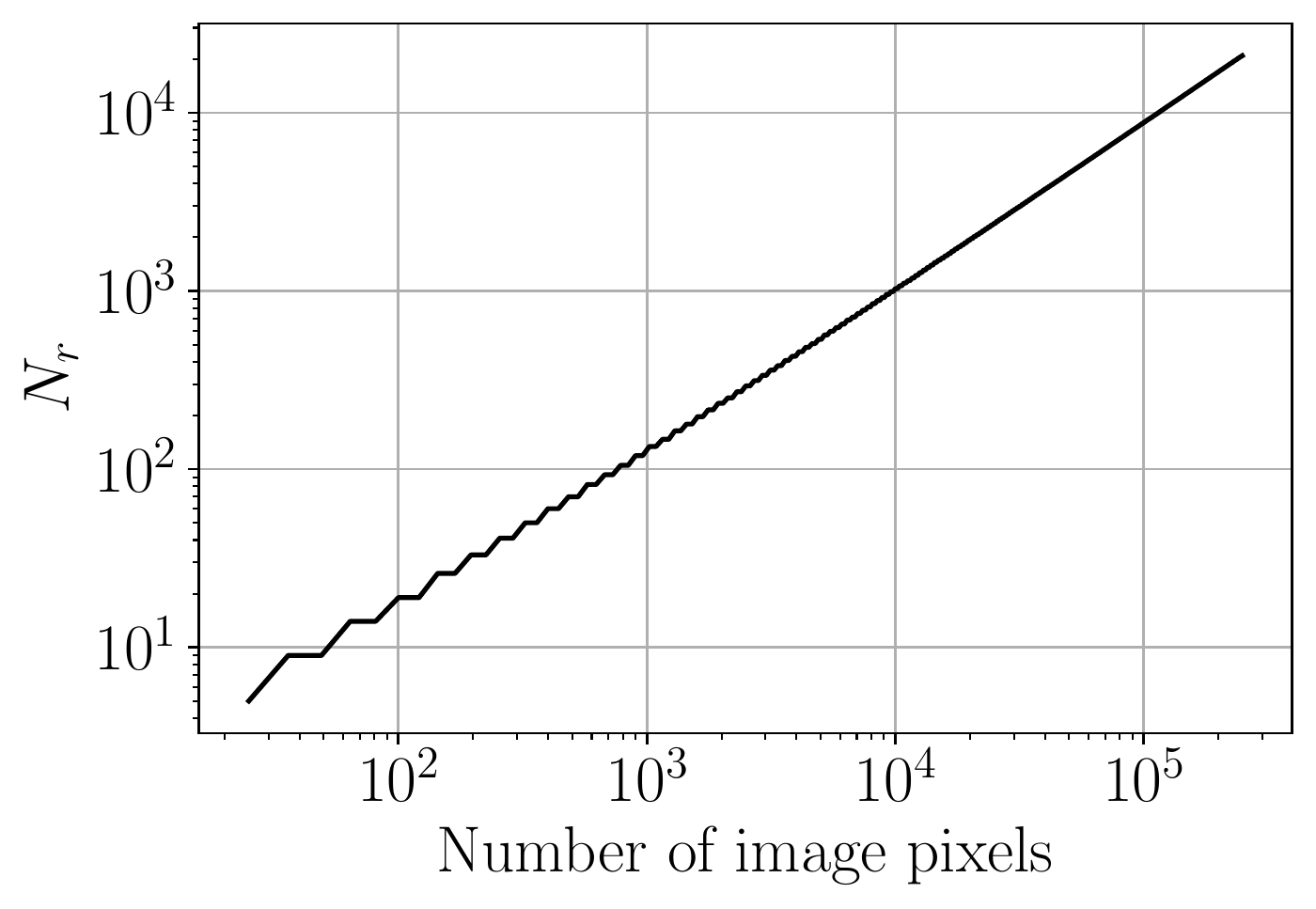}\includegraphics[width = 0.32\linewidth]{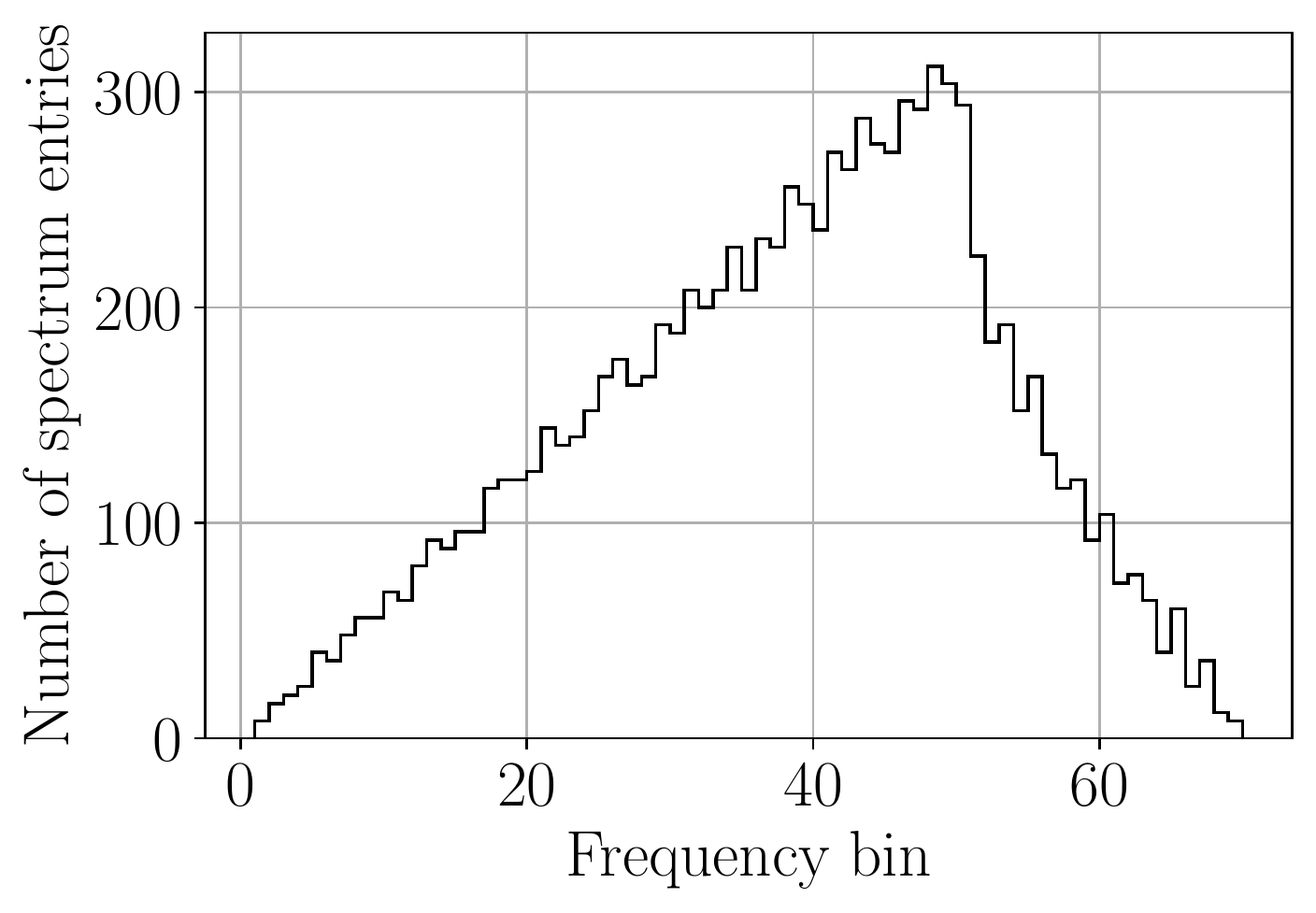}
    \caption{(\emph{left}) A $7\times 7$ matrix indexed according to the procedure described in Section~\ref{subsec:aaps_computation}. The entries with the same radial frequency are filled with the same shade. The dotted line depicts unit intervals of radial frequencies. (\emph{middle}) The number of unique radial frequencies $N_r$, i.e., the dimension of the full AAPS, versus the number of pixels in square images of varying sizes. (\emph{right}) The histogram of power spectrum entries of size $100\times 100$ sorted by their radial frequencies in unit size bins.}
    \label{fig:re-indexing}
\end{figure}

We re-index the shifted power spectrum to obtain the AAPS according to the radial coordinate. In particular, for each non-centered entry indexed $(j,k)$, its distance to the zero-frequency, or its \textit{radial frequency}, and orientation are computed:
\begin{equation*}
    r(j,k) = \sqrt{\left(j-\floor{M_1/2}\right)^2 + \left(k-\floor{M_2/2}\right)^2}\,,\quad \theta(j,k) = \atantwo\left(\left(j-\floor{M_1/2}\right),\left(k-\floor{M_2/2}\right)\right)\;.
\end{equation*}
The non-centered entries in the shifted power spectrum are then categorized according to their radial frequency values. Let $\boldsymbol{r}\in\R_+^{N_r}$ denote the vector of total $N_r$ unique radial frequencies, sorted by increasing radial frequency values. For a fixed radial frequency $\boldsymbol{r}_j$, let $\boldsymbol{\Theta}_{jk}\in \R^{(\boldsymbol{N}_{\theta})_j}$ denote the vector of total $(\boldsymbol{N}_{\theta})_j$ unique orientations, sorted by increasing angles. Finally, we assign new indexes $(j,k)$ for non-centered entries of the shifted power spectrum:
\begin{equation*}
    \boldsymbol{\mathcal{P}}(\boldsymbol{d})(\boldsymbol{r}_j, \boldsymbol{\Theta}_{jk})\,,\quad j = 1,\dots, N_r\,,\quad k = 1, \dots, (\boldsymbol{N}_\theta)_{j}\,.
\end{equation*}
The indexing of a $7\times 7$ matrix following the procedure above is demonstrated in Figure~\ref{fig:re-indexing}.

\begin{figure}[!hbt]
    \centering
    \begin{tabular}{cccc}
        Image & Power spectrum & AAPS & Smoothed AAPS \\
        $\boldsymbol{d}^*\in[0,1]^{200\times 200}$ & $\bdmc{P}(\boldsymbol{d}^*)$ & $\bdmc{P}_r(\boldsymbol{d}^*)$ & $\overline{\bdmc{P}_r}(\boldsymbol{d}^*)$ \\
        \includegraphics[width = 0.15\linewidth]{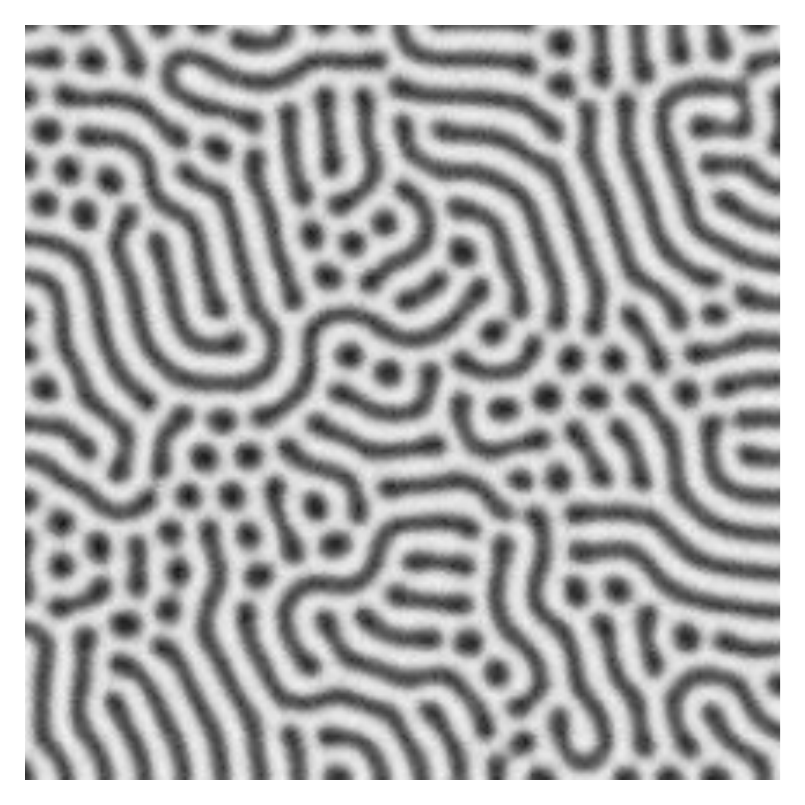}\includegraphics[height = 0.15\linewidth]{colorbar_colorbar_image_25.pdf} & \includegraphics[width = 0.20\linewidth]{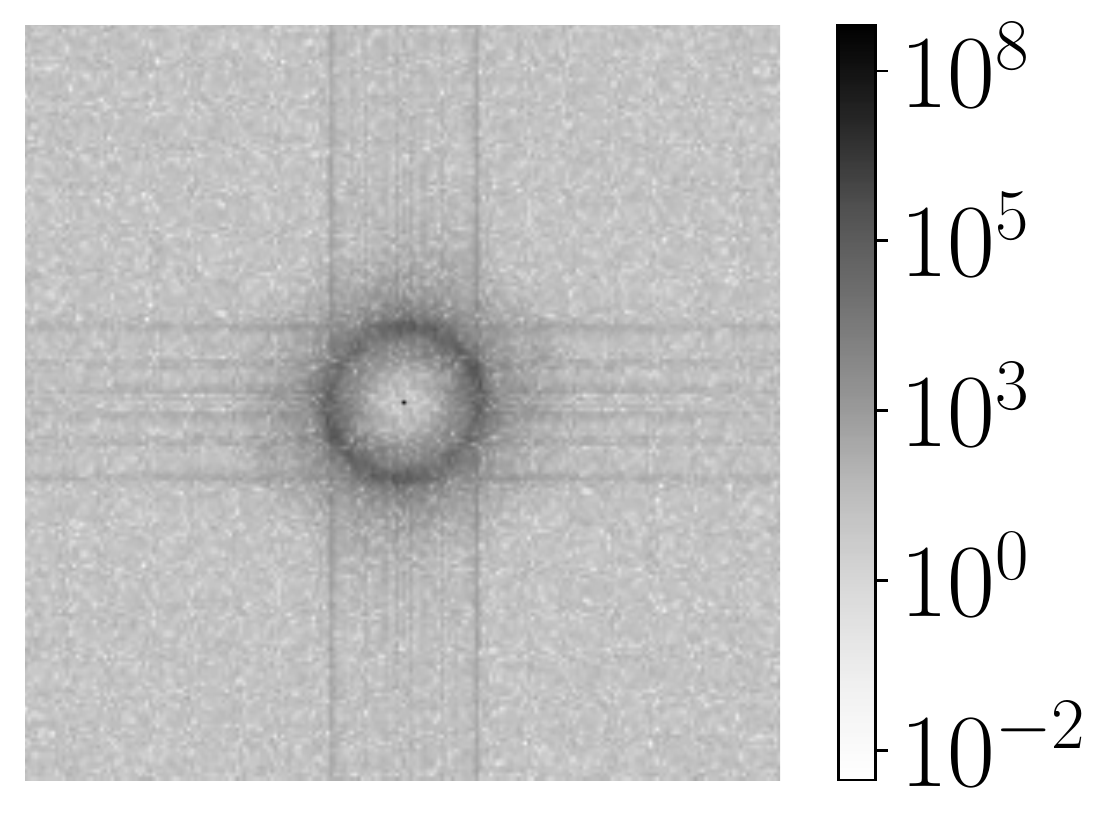}& \includegraphics[width = 0.22\linewidth]{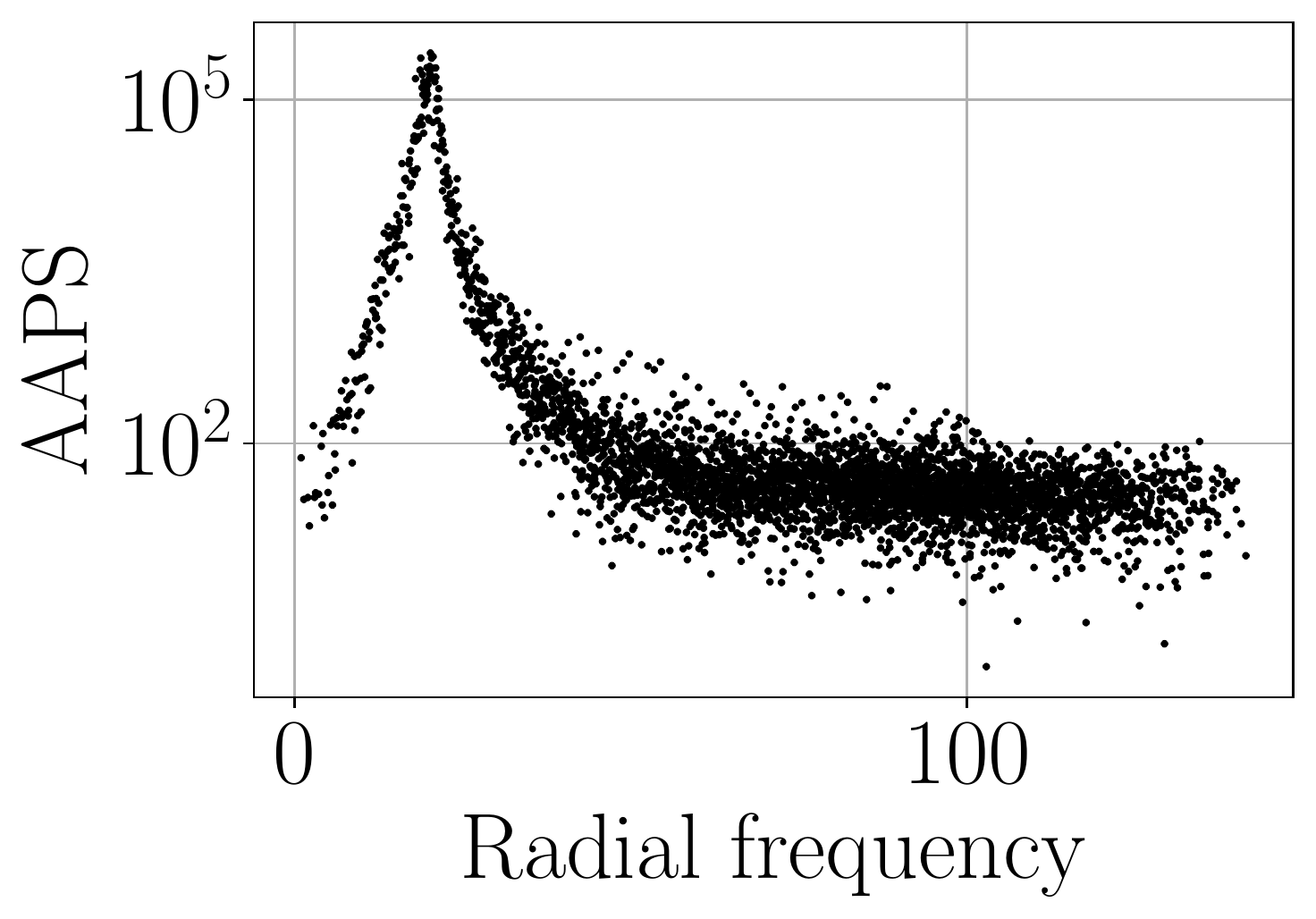} & \includegraphics[width = 0.22\linewidth]{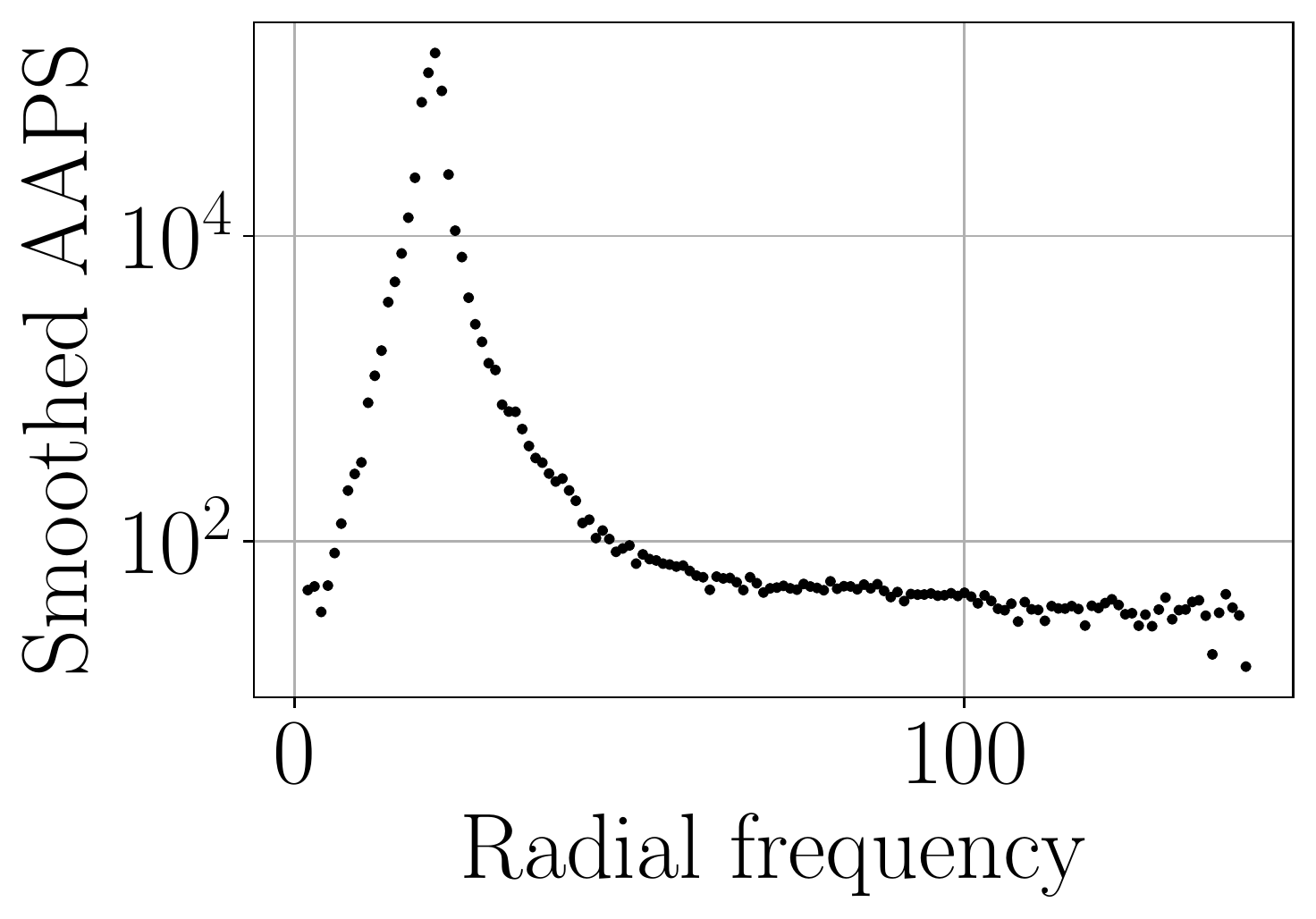}
    \end{tabular}
    \caption{A simulated top-down microscope image (with a high signal-to-noise ratio) of the top surface pattern of a Di-BCP thin film, its power spectrum, its AAPS, and its smoothed AAPS. These are used for the numerical experiments presented in Section~\ref{subsec:likelihood}, Figure~\ref{fig:likelihood_sensitivity_wz}, and Figure~\ref{fig:likelihood_sensitivity_x}.}
    \label{fig:likelihood_sensitivity}
\end{figure}

To marginalize the orientational dependence of the power spectrum, we average the entries in the shift power spectrum with the same radial frequency to obtain the AAPS of the image $\boldsymbol{d}$:
\begin{equation}\label{eq:aaps}
    \left(\boldsymbol{\mathcal{P}}_r(\boldsymbol{d})\right)_j = \frac{1}{(\boldsymbol{N}_\theta)_j} \sum_{k = 1}^{(\boldsymbol{N}_\theta)_j} \boldsymbol{\mathcal{P}}(\boldsymbol{d})(\boldsymbol{r}_j, \boldsymbol{\Theta}_{jk})\,,\quad j = 1,\dots, N_r\,,
\end{equation}
where $\boldsymbol{\mathcal{P}}_r:\R_+^{M_1\times M_2}\to \R_+^{N_r}$. We visualize the AAPS of simulated images in Figure~\ref{fig:image_corruptions} and~\ref{fig:likelihood_sensitivity}. The AAPS possesses a distinct shape with two peaks corresponding to the periodic and interfacial lengths of the Di-BCP film top surface morphology. Typically, only the former is visible in the presence of image corruption.

While the azimuthal averaging reduces the data dimension by one order of magnitude, as shown in Figure \ref{fig:re-indexing}, the dimension of the full AAPS is still large. If the overall shape of the AAPS is of primary interest, it is reasonable to apply data smoothing to the `fine-grained' and fluctuating AAPS spectrum to reduce the data dimension further. In this work, we consider local averaging over unit intervals of the radial frequencies in addition to the azimuthal averaging:
\begin{equation*}
    \left(\overline{\bdmc{P}_r}(\boldsymbol{d})\right)_j = \frac{1}{(\overline{\boldsymbol{N}_\theta})_j} \sum_{\boldsymbol{r}_k\in(j-1, j]}\sum_{l = 1}^{(\boldsymbol{N}_\theta)_k} \boldsymbol{\mathcal{P}}(\boldsymbol{d})(\boldsymbol{r}_k, \boldsymbol{\Theta}_{kl})\,,\quad j = 1,\dots, \ceil{\boldsymbol{r}_{N_r}}\,,
\end{equation*}
where $\overline{\bdmc{P}_r}:\R_+^{M_1\times M_2}\to\R_+^{\ceil{\boldsymbol{r}_{N_r}}}$, $\ceil{\cdot}$ is the ceiling operator, and $(\overline{\boldsymbol{N}_\theta})_j\coloneqq\sum_{\boldsymbol{r}_k\in(j-1, j]} (\boldsymbol{N}_\theta)_k$ is the number of power spectrum entries used to compute the $j$-th entry of the smoothed AAPS. We plot $\overline{\boldsymbol{N}_\theta}$ versus the radial frequency unit intervals in Figure \ref{fig:re-indexing} when $M_1 = M_2 = 100$.

\subsection{The probability distribution of the azimuthally-averaged power spectrum of image data}
Under the influence of the additive stochastic sensor noise random matrix $\boldsymbol{N}$, we further define the image data as a random matrix $\boldsymbol{D}$ for a given order parameter $u\in V^u(\Omega)$ following the top-down microscopy characterization model in \eqref{eq:charac_model}:
\begin{equation}\label{eq:distribution_image}
    \boldsymbol{D} = \boldsymbol{\mathcal{J}}(u) + \boldsymbol{N}\,,\quad
    \boldsymbol{D}_{jk}\sim \mathcal{N}\left(\left(\boldsymbol{\mathcal{J}}\left(u\right)\right)_{jk}, \sigma_n^2\right)\,.
\end{equation}
The additive white noise in \eqref{eq:charac_model} implies that the Fourier coefficients of the image data are also i.i.d.\ and normal after the discrete Fourier transforms,
\begin{equation*}
    \left(\boldsymbol{\mathcal{T}_s}(\boldsymbol{D})\right)_{jk}\sim \mathcal{CN}\left(\left(\left(\boldsymbol{\mathcal{T}}_s\circ\boldsymbol{\mathcal{J}}\right)\left(u\right)\right)_{jk}, M_1M_2\sigma_n^2\right)\,,
\end{equation*}
where $\mathcal{CN}(\cdot,\cdot)$ denotes the complex normal distribution. Consequently, the entries of the power spectrum are i.i.d and follow a non-centered chi-square distribution,
\begin{equation*}
    \frac{2}{M_1M_2\sigma_n^2}\left(\boldsymbol{\mathcal{P}}\left(\boldsymbol{D}\right)\right)_{jk}\sim \chi^2_2\left(\frac{2}{M_1M_2\sigma_n^2}\left(\left(\boldsymbol{\mathcal{P}}\circ\boldsymbol{\mathcal{J}}\right)\left(u\right)\right)_{jk}\right)\,,
\end{equation*}
where $\chi^2_k(\lambda)$ denotes the non-centered chi-square distribution with the non-centrality parameter $\lambda$ and the degrees-of-freedom parameter $k$. The analytical form of the probability density function for $\chi^2_k(\lambda)$ is
\begin{equation*}
    \pi_{\textrm{ncs}}(s;k,\lambda )={\frac {1}{2}}e^{-(s+\lambda )/2}\left({\frac {s}{\lambda }}\right)^{k/4-1/2}I_{k/2-1}({\sqrt {\lambda s}})\,, \quad 
    I_{\nu }(s)=(s/2)^{\nu }\sum _{j=0}^{\infty }{\frac {(s^{2}/4)^{j}}{j!\Gamma (\nu +j+1)}}\,,
\end{equation*}
where $I_{\nu}$ is the Bessel function of the first kind. The addition and scaling of noncentral chi-square random variables are still noncentral chi-square distributed. The AAPS of image data for a given order parameter $u$ thus has the following distribution:
\begin{equation}\label{eq:distribution_aaps}
    \frac{2(\boldsymbol{N}_{\theta})_j}{M_1M_2\sigma_{n}^2}\left(\boldsymbol{\mathcal{P}}_r\left(\boldsymbol{D}\right)\right)_j\sim \chi^2_{2(\boldsymbol{N}_{\theta})_j}\left(\frac{2(\boldsymbol{N}_{\theta})_j}{M_1M_2\sigma_{n}^2}\left(\left(\boldsymbol{\mathcal{P}}_r\circ\boldsymbol{\mathcal{J}}\right)\left(u\right)\right)_j\right)\,.
\end{equation}
Similarly, the smoothed AAPS of image data follows the noncentral chi-square distribution with a different number of degrees of freedom:
\begin{equation}\label{eq:distribution_smoothed_aaps}
    \frac{2(\overline{\boldsymbol{N}_\theta})_j}{M_1M_2\sigma_{n}^2}\left(\overline{\bdmc{P}_r}\left(\boldsymbol{D}\right)\right)_j\sim \chi^2_{2(\overline{\boldsymbol{N}_\theta})_j}\left(\frac{2(\overline{\boldsymbol{N}_\theta})_j}{M_1M_2\sigma_{n}^2}\left(\left(\overline{\bdmc{P}_r}\circ\boldsymbol{\mathcal{J}}\right)\left(u\right)\right)_j\right)\,.
\end{equation}

\begin{figure}[!hbt]
    \centering
    \addtolength{\tabcolsep}{-6pt}
    \begin{tabular}{c c c}
    \multicolumn{3}{c}{Information loss estimate for the normal approximation in \eqref{eq:normal_approx_smoothed_aaps}}\\
        $\scriptstyle \left(\left(\overline{\bdmc{P}_r}\circ\boldsymbol{\mathcal{J}}\right)\left(u\right)\right)_j = 1$ & $\scriptstyle \left(\left(\overline{\bdmc{P}_r}\circ\boldsymbol{\mathcal{J}}\right)\left(u\right)\right)_j = 10^{2}$& $ \scriptstyle\left(\left(\overline{\bdmc{P}_r}\circ\boldsymbol{\mathcal{J}}\right)\left(u\right)\right)_j = 10^{4}$\\
        \includegraphics[width = 0.34\linewidth]{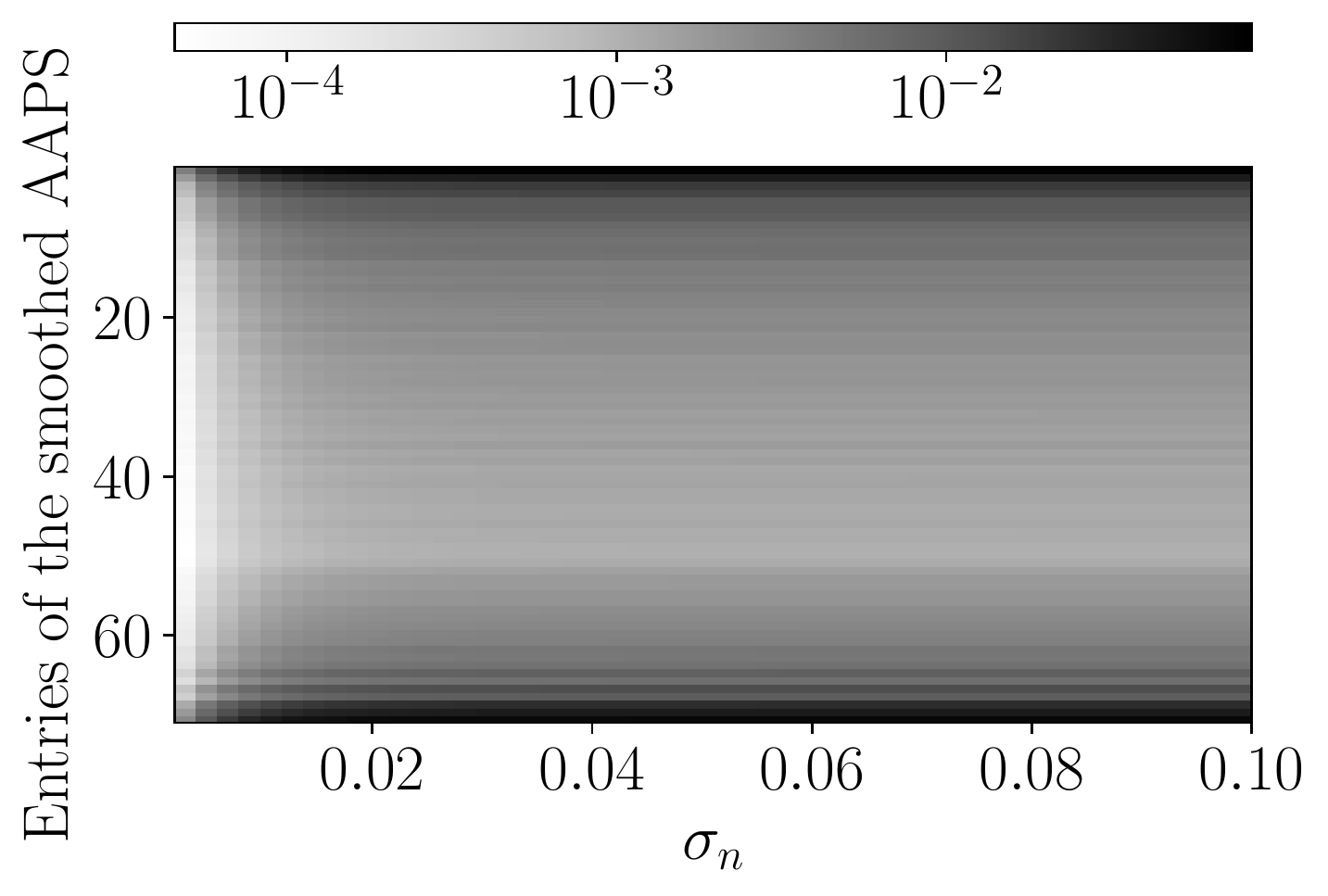} & \includegraphics[width = 0.3\linewidth]{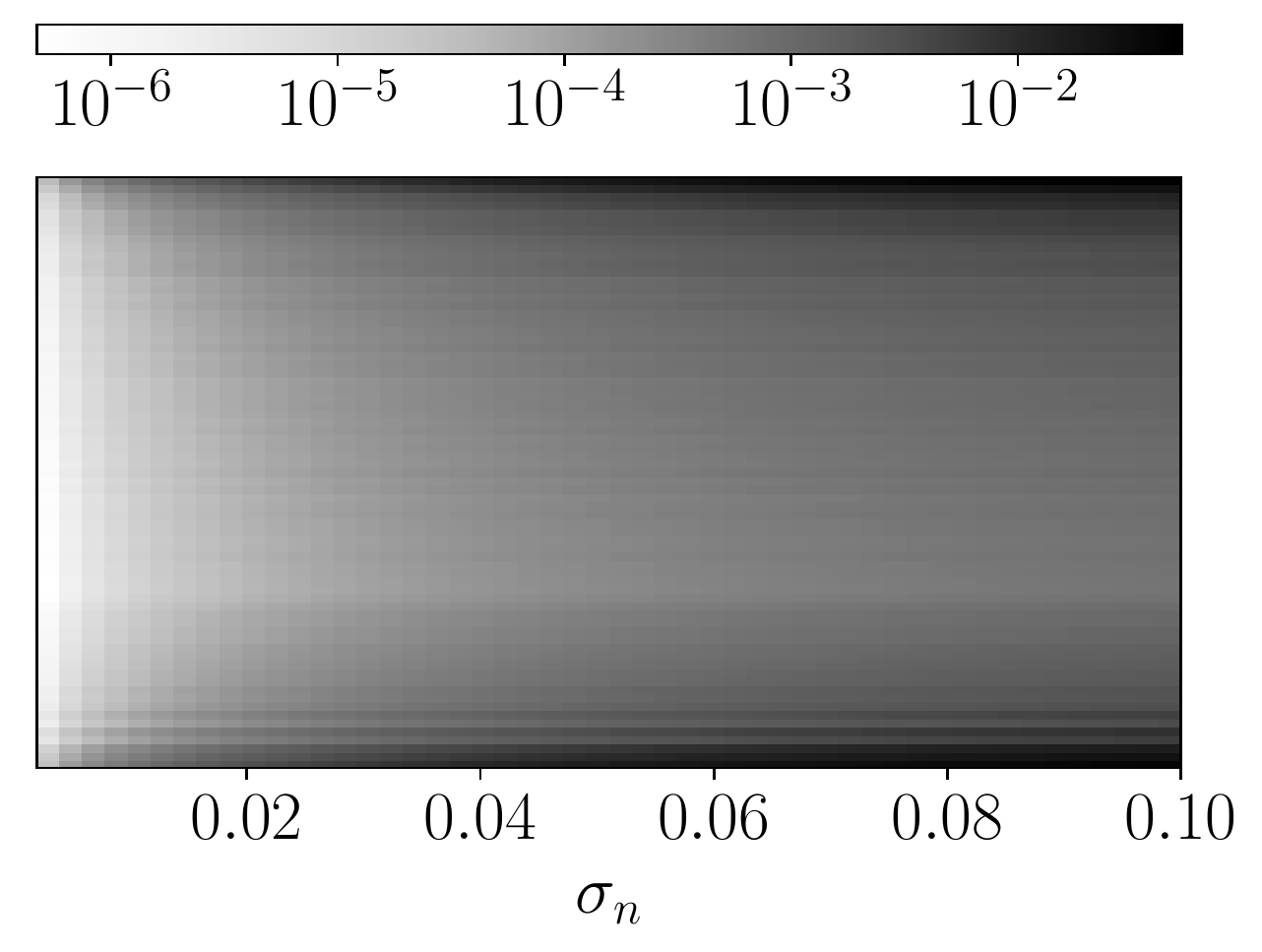} & \includegraphics[width = 0.3\linewidth]{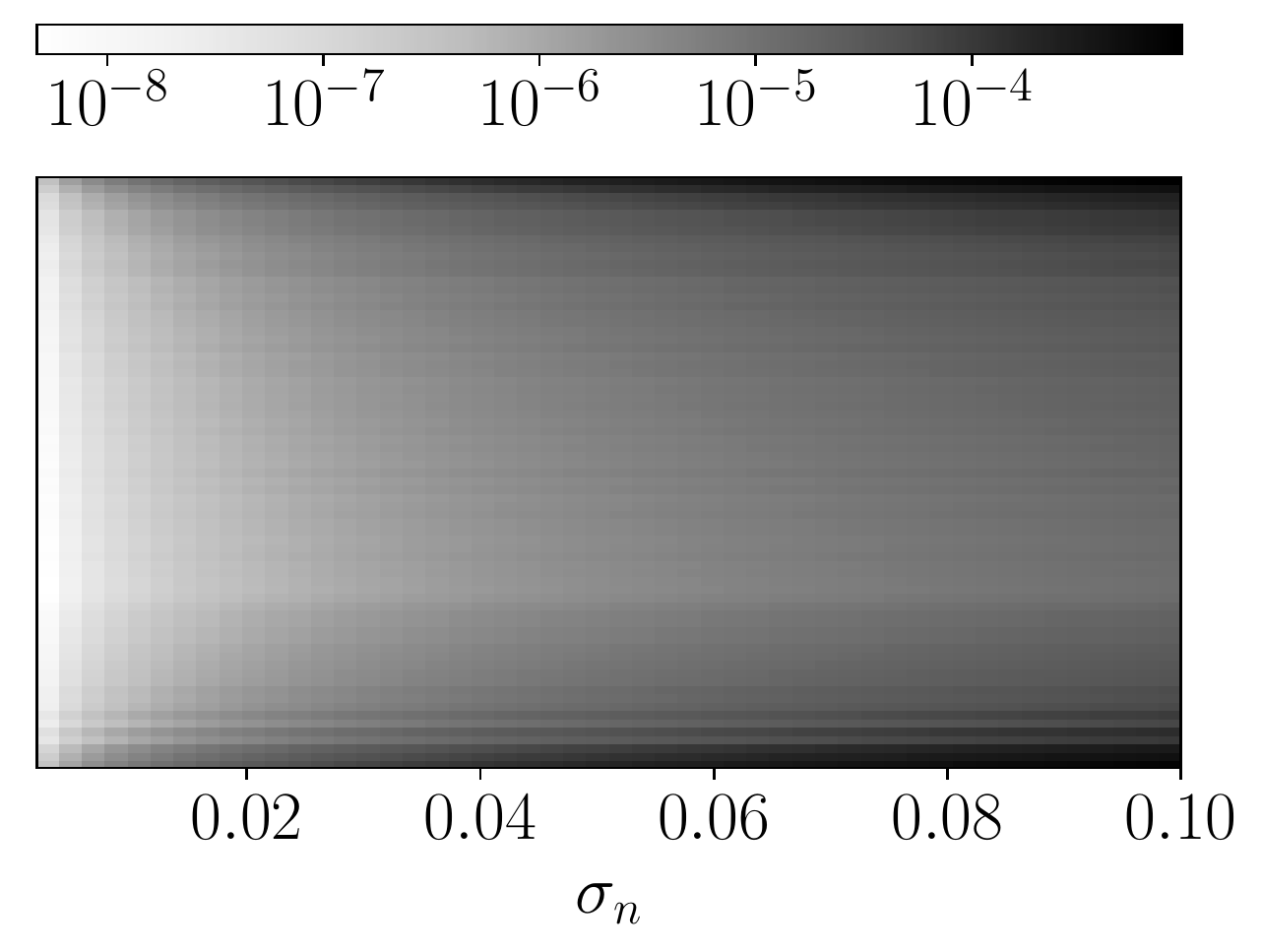}
    \end{tabular}
    \addtolength{\tabcolsep}{6pt}
    \caption{The estimated KL divergence values for using the approximate normal distribution instead of the noncentral chi-square distribution for the smoothed AAPS of image data. The estimated values are given at different image noise variances $\sigma_n$, different entries of the smoothed AAPS, and different magnitudes of the smoothed AAPS for the latent noise-free image $\left(\left(\overline{\bdmc{P}_r}\circ\boldsymbol{\mathcal{J}}\right)\left(u\right)\right)_j$.}
    \label{fig:normal_approximation}
\end{figure}

The noncentral chi-square distribution $\chi^2_k(\lambda)$ converges to the normal distribution $\mathcal{N}(k+\lambda, 2k+4\lambda)$ for as $k$ and/or $\lambda$ increases~\cite{Horgan2013,Seri2015}. As shown in Figure~\ref{fig:re-indexing}, the distributions of the smoothed AAPS entries mostly have large degrees-of-freedom parameter $2(\overline{\boldsymbol{N}_\theta})_j$, and they can be well-approximated by normal distributions:
\begin{equation}\label{eq:normal_approx_smoothed_aaps}
    \left(\overline{\bdmc{P}_r}\left(\boldsymbol{D}\right)\right)_j\mathrel{\dot\sim} \mathcal{N}\left(M_1 M_2\sigma_n^2 + 
    \left(\left(\overline{\bdmc{P}_r}\circ\boldsymbol{\mathcal{J}}\right)\left(u\right)\right)_j, \frac{M_1M_2\sigma_n^2}{2(\overline{\boldsymbol{N}_\theta})_j}\left(2M_1M_2\sigma_n^2 + \left(\left(\overline{\bdmc{P}_r}\circ\boldsymbol{\mathcal{J}}\right)\left(u\right)\right)_j\right)  \right)\,.
\end{equation}
To understand the accuracy of the normal approximation, we estimate and visualize in Figure~\ref{fig:normal_approximation} the Kullback–Leibler (KL) divergence estimates between the noncentral chi-square distributions of the smoothed AAPS entries in \eqref{eq:distribution_smoothed_aaps} and their normal approximations. We observe that estimated KL divergence values for the smooth AAPS's first and last few entries are comparatively large due to their relatively small degrees-of-freedom parameters. The information loss due to the approximation is more pronounced for large noise variances $\sigma_n$ and small magnitudes of the smoothed AAPS for the noise-free latent image $\left(\left(\overline{\bdmc{P}_r}\circ\boldsymbol{\mathcal{J}}\right)\left(u\right)\right)_j$.

\section{Diblock copolymer self-assembly model}\label{sec:forward_model}

This section briefly introduces a generic forward operator for statistical models describing the phenomenon of diblock copolymer self-assembly, mainly following the framework proposed and described in detail by Baptista et.\thinspace al~\cite{Baptista2022}. Furthermore, we introduced a particular continuum model of diblock copolymer self-assembly, the Ohta--Kawasaki (OK) model~\cite{Ohta1986, Uneyama2005, Muller2018, singh2023nonlocal}, used for our numerical studies. 

\subsection{The forward operator as a stochastic simulator}\label{subsec:stochastic_simulator}

Due to uncontrolled nucleation and domain growth during the phase transition of Di-BCP thin films, a rich variety of thin film patterns can be observed through microscopy for the same materials under the same experimental scenario. They are typically qualitatively consistent in phase and morphological characteristics, yet exhibit so-called random long-range disorder; see, e.g., Figure~\ref{fig:ok_model_samples}. To incorporate such aleatoric uncertainties into our definition of the forward operator, we assume that the latent top-surface Di-BCP thin film patterns $u$ of observed microscope images appear in~\eqref{eq:charac_model} are realizations of a $V^u(\Omega)$-valued random variable $U(S_E)$, where $S_E$ represents the relevant material, geometrical, or environmental factors of the experimental scenario.

A \textit{statistical model} of Di-BCP self-assembly can be generalized as a mathematical model derived from statistical polymer theory for specifying or characterizing the probability distribution, i.e., the \textit{statistical ensemble}, of Di-BCP structures for a range of experimental scenarios $S_E$ specified by the model parameter $\x\in \R^{d_x}$. It typically contains a model of polymer chains, models of the inter-polymers and polymer--environment interaction, and possibly a model of thermal fluctuation and/or dynamics of polymers during the self-assembly. A comprehensive review of these models can be found in \cite{Grosberg1994, Gompper2005, Fredrickson2006} and references therein.

Returning to the context of model calibration, the forward operator representing model-based predictions of the order parameter in~\eqref{eq:charac_model} can be treated as a \textit{stochastic simulator} in accordance with the definition of statistical models of Di-BCP self-assembly. Let us denote the forward operator as $\mathcal{F}:\R^{d_x}\times V^z\to V^u$, then the model assumes the following equality in distribution:
\begin{equation}\label{eq:state_variable}
    U(\x) = \mathcal{F}(\boldsymbol{x}, Z)\,,\quad Z\sim \nu_{Z}\,,
\end{equation}
where the $V^z$-valued random variable $Z$ with a probability distribution $\nu_Z$ represents the stochastic component of the forward operator that possibly depends on the model parameter. The random variable $Z$ may represent the sampling mechanism in Monte Carlo simulations~\cite{Binder2012}, or the infinite-dimensional stochastic noise due to thermal fluctuation in dynamical SCFT and DFT simulations~\cite{Fraaije1993, Schmid2011}. Though $Z$ is also a nuisance parameter, we refer to $Z$ specifically as an \textit{auxiliary variable} for distinction.

\subsection{The Ohta--Kawasaki model of diblock copolymer thin film self-assembly}\label{subsec:ok_model}

We again consider an incompressible Di-BCP thin film described by an order parameter. Under the thin film approximation, we may only consider the variation of the order parameter parallel to the film substrate. The OK model characterizes the order parameter $u$ of the Di-BCP film with a free energy $\mathcal{E}_{\text{OK}}$ parametrized by $\x = (\epsilon,\sigma, m)\in \R_+^2\times[-1, 1]$:
\begin{equation}\label{eq:ok_energy}
\mathcal{E}_{\text{OK}}(u; \x) \coloneqq \int_\Omega \left(\mathcal{W}\left(u(\boldsymbol{s})\right) + \frac{\epsilon^2}{2}|\grad u(\boldsymbol{s})|^2 + \frac{\sigma}{2}\left(u(\boldsymbol{s}\right)-m)(-\lap_{N})^{-1}\left(u(\boldsymbol{s})-m\right) \right)\,\textrm{d}\boldsymbol{s}\,,
\end{equation}
where $ \int_\Omega \left(u(\boldsymbol{s}) -m\,\right)\textrm{d}\boldsymbol{s} = 0$ is enforced as the melt incompressibility constraint, and $\mathcal{W}(u) =(1-u^2)^2/4$ is the double well potential. The action of the inverse Laplacian is defined by the solution of a Poisson equation with homogeneous Neumann boundary conditions\footnote{The Neumann boundary additionally models the effect of a bounded (confined) domain by assuming that polymer chains instantaneously reflect at the boundary~\cite{Hsu1985, cao2022thesis}.}:
\begin{equation*}\label{eq:lap_inv}
    w = (-\lap_N)^{-1}(u-m)\iff 
    \begin{cases}
        -\lap w(\boldsymbol{s}) = u(\boldsymbol{s})-m &\text{in }\Omega\;;\\
        \nabla w(\boldsymbol{s}) \cdot \boldsymbol{n} = 0 &\text{on }\partial\Omega\;,
    \end{cases}
\end{equation*}
where $\boldsymbol{n}$ is the unit vector normal to $\partial\Omega$.

The parameters in the OK model are connected to the commonly used material parameters of conformationally symmetric Di-BCPs. In particular, an appeal to the higher-fidelity SCFT model in \cite{Choksi2003} reveals the relationships:
\begin{equation} \label{eq:ok_parameter_maps}
    \epsilon^2 = \frac{l^2}{3f_A(1-f_A)\chi}\,,\quad\sigma = \frac{36}{f_A^2(1-f_A)^2l^2\chi N^2}\,, \quad m = 2f_A-1\,,
\end{equation}
where $N$ is the degree of polymerization of the polymer, $\chi$ is the Flory--Huggins parameter describing the strength of repulsion between the two monomers, $l$ is the statistical segment length of the polymer, and $f_A\in(0,1)$ represents the monomer segment number fraction of the monomer A block. The material parameters $f_A$ and $\chi N$ together determine the phase of Di-BCP equilibrium structures~\cite{Khandpur1995}.

The forward operator $\mathcal{F}$ of the OK model is defined as the map from parameters $\x = (\epsilon, \sigma, m)$ and an initial state $z$ to a local minimizer of the OK energy,
\begin{equation}\label{eq:ok_forward_operator}
u = \mathcal{F}(\x,z) \in \argmin_{v\in V^u(\Omega)} \mathcal{E}_{\text{OK}}(v;\x)\,,\quad     V^u(\Omega)\coloneqq\left\{v\in H^1(\Omega):\int_\Omega \left(v(\boldsymbol{s})-m\right)\textrm{d}\boldsymbol{s} = 0\right\}.
\end{equation}
As a result of the non-convex double well $\mathcal{W}$, the OK energy is non-convex and does not have a unique minimizer. The local minimizes consist of an ensemble of Di-BCP thin film metastable patterns. The forward operator $\mathcal{F}(\x,z)$ is evaluated by minimizing the OK free energy via a variational procedure starting from the initial state $z$. Conventionally, an $H^{-1}$ gradient flow approach is used, which involves solving the nonlocal Cahn--Hilliard equation \cite{Choksi2009, Choksi2011, Qin2013} until steady-state solutions are reached\footnotemark. While there are many sophisticated methods for solving the local or nonlocal Cahn--Hilliard equation \cite{gomez2008isogeometric, wu2014stabilized, shen2019new}, we adopt an alternative strategy for efficient numerical evaluations of the forward operator: a modified Newton method to minimize the OK energy proposed in \cite{Cao2022}.
\footnotetext{According to the dynamical DFT \cite{TeVrugt2020}, the gradient flow approach may faithfully represent the kinetic pathway of the self-assembly if an appropriate mobility operator \cite{Schmid2020} and/or a stochastic noise that satisfies the fluctuation-dissipation theorem \cite{Fraaije1993} are/is incorporated. However, the OK model is often treated as an equilibrium model, i.e., the thermal fluctuations and dynamics of Di-BCPs are not modeled.}

\begin{figure}[!hbt]
\addtolength{\tabcolsep}{-5pt} 
    \centering
    \begin{tabular}{c c c c}
        \multicolumn{4}{c}{$\displaystyle \x = (\epsilon,\sigma,m) = (2.5\times 10^{-3}, 4.8\times 10^{3}, 0)$}\\ \includegraphics[width = 0.14\linewidth]{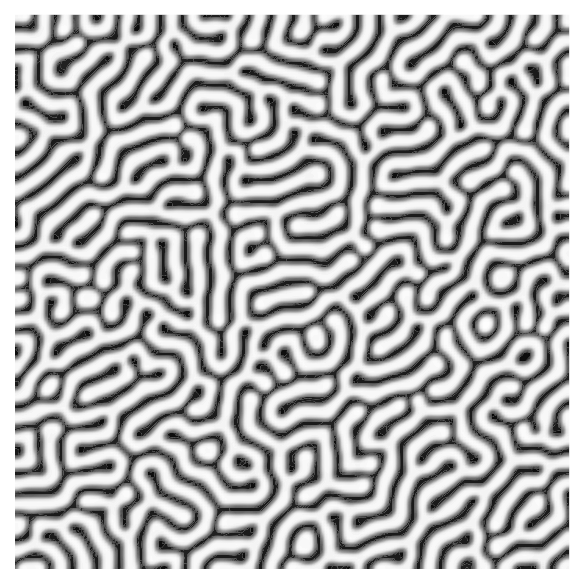} & \includegraphics[width = 0.14\linewidth]{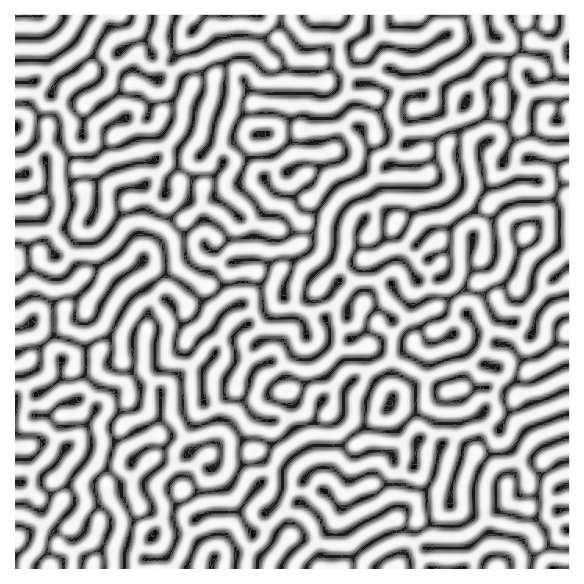} & \includegraphics[width = 0.14\linewidth]{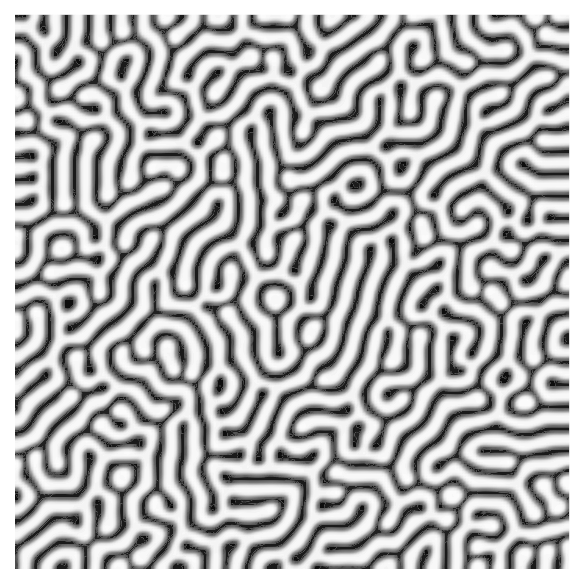} & \includegraphics[width = 0.14\linewidth]{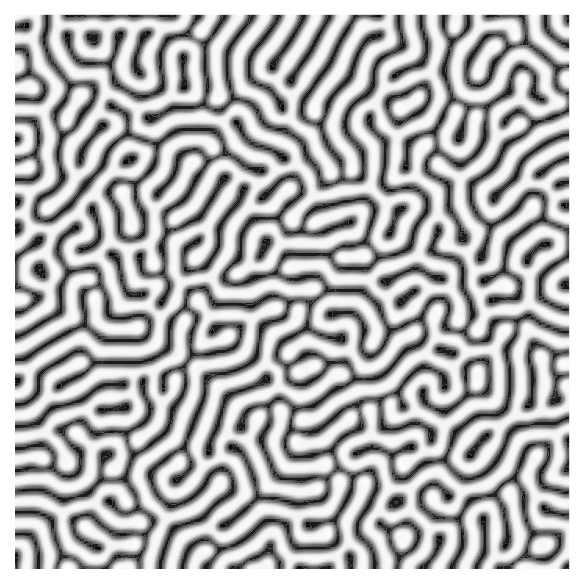}\includegraphics[height = 0.14\linewidth]{colorbar_colorbar_op_25.pdf}\\
        \multicolumn{4}{c}{$\displaystyle \x = (\epsilon,\sigma, m) = (3.75\times 10^{-3}, 4\times 10^{3}, -0.3)$}\\ \includegraphics[width = 0.14\linewidth]{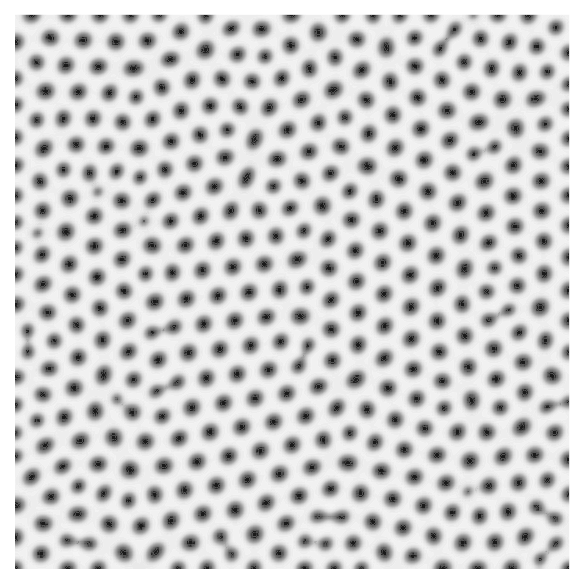} & \includegraphics[width = 0.14\linewidth]{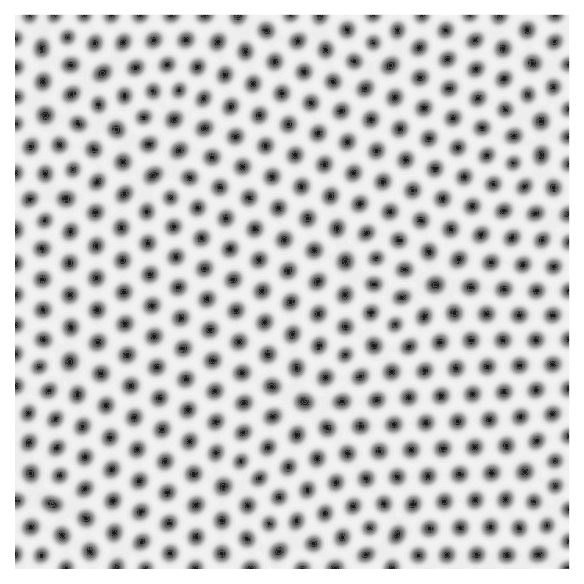} & \includegraphics[width = 0.14\linewidth]{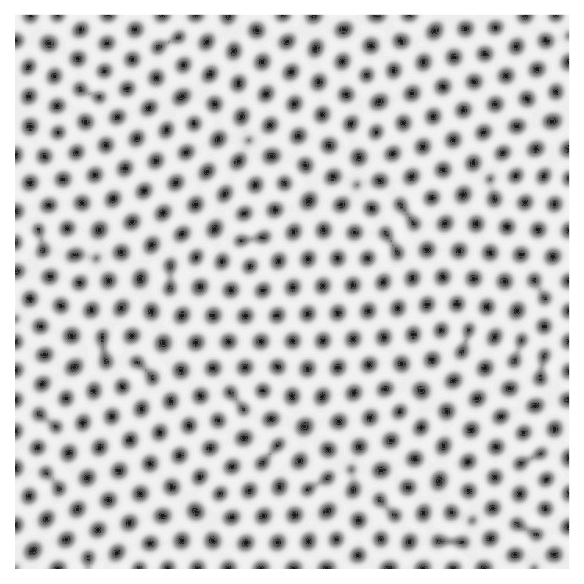} & \includegraphics[width = 0.14\linewidth]{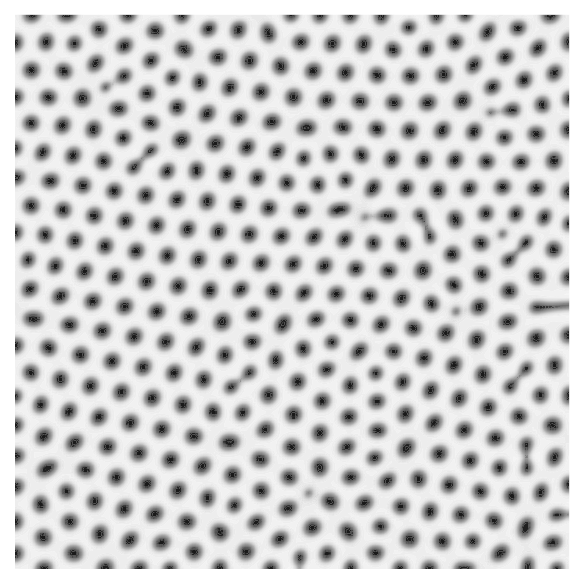}\includegraphics[height = 0.14\linewidth]{colorbar_colorbar_op_25.pdf}\\
        \multicolumn{4}{c}{$ \displaystyle \x = (\epsilon,\sigma, m) = (3.75\times 10^{-3}, 4\times 10^{3}, -0.15)$}\\ \includegraphics[width = 0.14\linewidth]{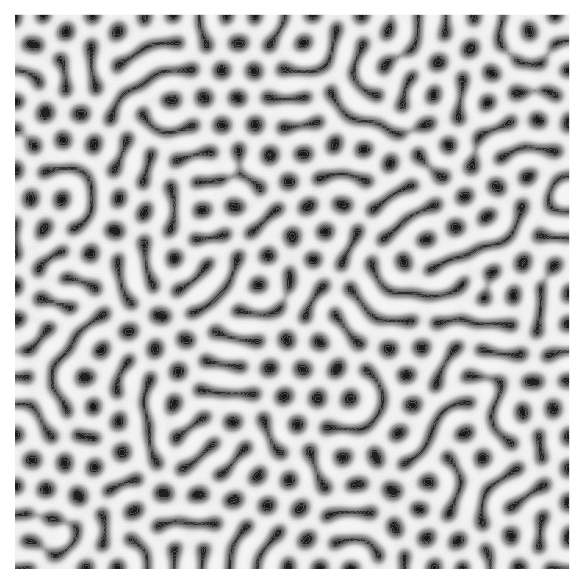} & \includegraphics[width = 0.14\linewidth]{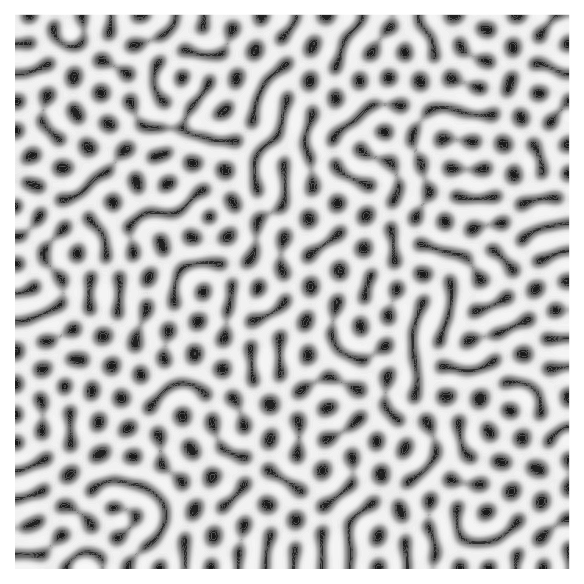} & \includegraphics[width = 0.14\linewidth]{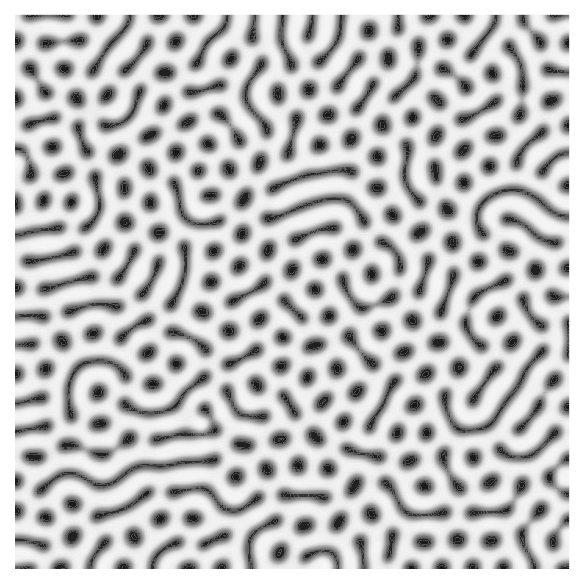} & \includegraphics[width = 0.14\linewidth]{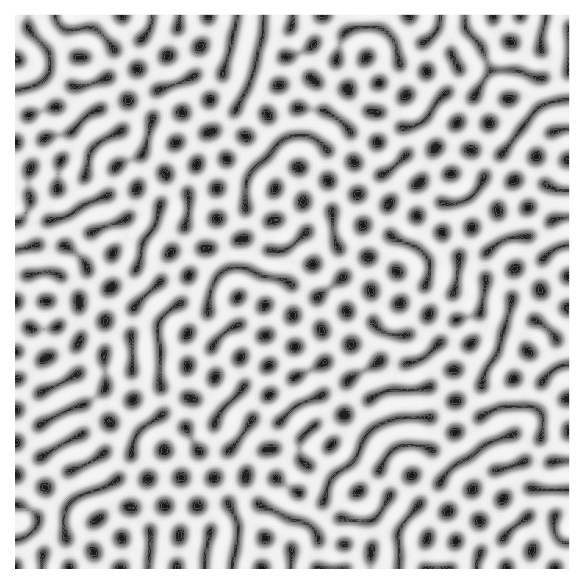}\includegraphics[height = 0.14\linewidth]{colorbar_colorbar_op_25.pdf}\\
    \end{tabular}
    \addtolength{\tabcolsep}{5pt} 
    \caption{Samples of morphologies (order parameters) generated by the forward operator of the OK model for Di-BCP thin film self-assembly \eqref{eq:ok_forward_operator} at three different sets of model parameters. From top to bottom, the patterns correspond to three different phases: stripes, spots, and mixtures. The patterns in each row are i.i.d.\ outcomes of the stochastic forward operator evaluation at a fixed set of model parameters, $\mathcal{F}(\x, Z)$. The variations of patterns in each row represent the aleatory uncertainty of Di-BCP thin film self-assembly.}
    \label{fig:ok_model_samples}
\end{figure}

Given many local minimizers for the OK energy, an approach is needed for modeling the initial state $z$. A commonly adopted initial state is the realization of a uniform white noise random field~\cite{Qin2013,Choksi2009,Choksi2011}. A more rigorous approach that uses transformed Gaussian random fields is proposed in \cite{Cao2022}, and it is adopted in our numerical examples:
\begin{equation*}
    Z=m+\frac{1}{2}\erf\left( U_G\right)\,,\quad U_G \sim \mathcal{N} (0,\mathcal{C}_0)\,,
\end{equation*}
where $\mathcal{C}_0$ is the covariance operator defined as the inverse of an elliptic differential operator. The parameters of the covariance operator can be tuned to mimic the random initial states produced by the uniform white noise random field for a given mesh while staying invariant to changes in spatial discretization. For additional mathematical details of the Gaussian random fields and their numerical implementation, see~\cite[Section 4.2]{Cao2022}. When coupled with the optimization procedure, the random initial states produce a variety of top-surface patterns of Di-BCP thin films. In Figure~\ref{fig:ok_model_samples}, we plot samples of Di-BCP patterns generated by the forward operator at different model parameter values. As described in Section~\ref{subsec:stochastic_simulator}, a physically-consistent model should satisfy the equality in distribution in~\eqref{eq:state_variable}. Such a modeling assumption can be validated once the models are calibrated; see, e.g., \cite{Wang2009, Hawkins2013, Farrel2015}. 

\begin{remark}[On parameter scaling and boundary effects]
To improve numerical conditioning, a rescaled computational domain $\Omega_c$ is often used instead of the physical domain $\Omega$, when implementing the forward operator. Since the model parameters $\epsilon$ and $\sigma$ are length-scale dependent, we present them after normalization with respect to the domain size:
\begin{equation}\label{eq:rescaling}
    \overline{\epsilon}= \epsilon/\sqrt{L_1 L_2} \,,\quad\overline{\sigma} = \sigma L_1L_2\,.
\end{equation}
The notation is omitted for the analysis and examples presented. Moreover, we minimize unwanted boundary effects in the model predictions by evaluating the forward operator on a larger domain and only keeping the centered subdomain matching the size of the field of view. Such an approach is adopted in the numerical results presented in Section~\ref{sec:numerical_results}.
\end{remark}

\section{Bayesian model calibration: Formulation and methodology}\label{sec:bayesian_model_calibration}
We now consider the formulation and inference methodologies for Bayesian model calibration. In particular, we now treat both the forward model parameters and the characterization model parameters as random variables denoted by $\bx$ and $\bw$. We assume that they have independent prior distributions denoted by $\nu_{\bx}$ and $\nu_{\bw}$ with probability density functions $\pi_{\bx}$ and $\pi_{\bw}$. The variability in the parameters represents their epistemic uncertainties. We denote the marginal variables of the OK model parameters as $\bx = (E, \Sigma, M)$, and the marginal variables of the characterization model parameters as $\bw = (C_1, C_2, \Sigma_b, \Sigma_n)$.

Firstly, we combine the characterization model and the forward operator to define an \textit{image model} under all inherent uncertainties,
\begin{equation}\label{eq:data_model}
    \boldsymbol{D} = \underbrace{(\bdmc{J}\circ\mathcal{F})}_{\bdmc{O}}(\bx, \bw ,Z) + \boldsymbol{N}\,,
\end{equation}
where $\bdmc{O}$ is referred to as the \textit{parameter-to-image map}. Let a particular observed image or its summary statistics, generally denoted as $\y^*\in\R^{d_y}$, as a realization of the data random variable $\by$. One may define the full image as data, $\by = \bd$, or summary statistics extracted from the image as data, $\by = \bdmc{Q}(\bd)$. Bayes' rule for updating the distribution of the parameters $\bx$ using $\y^*$, is given by
\begin{equation}\label{eq:bayes_rule}
    \pi_{\bx|\by}(\x|\y^*) = \frac{\mathcal{L}_I(\x;\y^*)\pi_{\bx}(\x)}{C_{\y^*}}\,,
\end{equation}
where $\pi_{\bx|\by}(\cdot|\y^*)$ is the posterior density of model parameters, $C_{\y^*} \coloneqq \expect{\bx}{\mathcal{L}_{I} (\bx;\y^*)}$ is a normalization constant, and $\mathcal{L}_I$ is the \textit{integrated likelihood function}.

The integrated likelihood function $\mathcal{L}_I(\x;\y^*)\coloneqq \pi_{\by|\bx}(\y^*|\x)$ describes the probability of observing $\y^*$ for all possible model predictions at the forward model parameter $\x$. It is defined by marginalizing the nuisance parameters $(\bw, Z)$ from the \textit{conditional likelihood function} $\mathcal{L}$:
\begin{equation}\label{eq:integrated_likelihood}
    \mathcal{L}_I(\x;\y^*)\coloneqq  \expect{\bw, Z}{\mathcal{L}(\x,\bw ,Z;\y^*)}\,,
\end{equation}
where $\mathcal{L}(\x,\w,z;\y^*)\coloneqq \pi_{\by|\bx,\bw,Z}(\y^*|\x,\w, z)$ describes the probability of observing $\y^*$ for a particular model predicted image $\bdmc{O}(\x,\w,z)$.

Equivalently, we can formulate Bayes' rule with the joint distribution of $(\bx,\bw)$,
\begin{equation}\label{eq:bayes_rule_joint}
    \pi_{\bx,\bw|\by}(\x, \w|\y^*) = \frac{ \mathcal{L}_I(\x, \w;\y^*)\pi_{\bx, \bw}(\x,\w)}{C_{\y^*}},\quad \mathcal{L}_I(\x,\w;\y^*)\coloneqq  \expect{Z}{\mathcal{L}(\x,\w ,Z;\y^*)}\,.
\end{equation}
To characterize the posterior distribution $\pi_{\bx|\by}(\cdot|\y^*)$, one can simply sample according to the joint density $\pi_{\bx,\bw|\by}(\cdot, \cdot|\y^*)$ and discard samples of the nuisance parameter. This approach is preferred if $\pi_{\bw}$ is not strongly informative.

In the following subsections, we first discuss the pseudo-marginal approach for sampling from the posterior distribution. Then we define the conditional likelihoods based on image data, $\by = \boldsymbol{D}$, and their AAPS, $\by =\bdmc{P}_r(\boldsymbol{D})$ or $\overline{\bdmc{P}_r}(\boldsymbol{D})$. The advantage of using entries of smoothed AAPS for the Bayesian model calibration task is demonstrated with numerical examples. Lastly, we propose a phase-informed prior distribution through mean pixel values of image data to account for the lost information in data compression.

\subsection{The pseudo-marginal method for posterior sampling with intractable likelihoods}\label{subsec:pseudo-marginal}

The integrated likelihood $\mathcal{L}_I(\x;\y^*)$ in~\eqref{eq:integrated_likelihood} has closed-form expressions only in very special situations. For example, when the map of $(\bx, \bw, Z)\mapsto \by$ is affine, and $\bw$, $Z$, $\boldsymbol{N}$ are normally distributed, we can analytically evaluate the integrated likelihood using Gaussian identities. However, the expectation with respect to the auxiliary variable is generally unavailable in closed form. Therefore, we must consider methods for sampling from or evaluating the posterior $\pi_{\bx|\by}(\cdot|\y^*)$ without directly evaluating the integrated likelihood. When the evaluation of conditional likelihoods is tractable, the \textit{pseudo-marginal method}~\cite{Andrieu2009} can be used to carry out such a task.

The pseudo-marginal method replace the integrated likelihoods $\mathcal{L}_I$ in Markov chain Monte Carlo (MCMC) methods~\cite{robert2004monte} for posterior sampling following Bayes' rule in~\eqref{eq:bayes_rule} with a Monte Carlo estimate $\widehat{\mathcal{L}_I}$ of the integrated likelihood,
\begin{equation}\label{eq:mc_estimator}
    \widehat{\mathcal{L}_I}(\x,\widehat{\bw},\widehat{\bz};\y^*) = \sum_{j=1}^{n_{\w}}\sum_{k=1}^{n_{z}} \frac{1}{n_{\w}n_z}\mathcal{L}(\x,\widehat{\bw}_j, \widehat{\bz}_k; \y^*)\,,\quad\widehat{\bw}\sim \otimes_{j=1}^{n_{\w}} \nu_{\bw}\,,\quad \widehat{\bz}\sim \otimes_{k=1}^{n_z} \nu_{Z}\,.
\end{equation}
For example, in the context of the Metropolis--Hastings algorithm, the integrated likelihood in the acceptance probability $\alpha(\x,\x')$ of the proposed chain position $\x'$ at the current chain position $\x$ are replaced with the estimator,
\begin{equation}\label{eq:pmmh_acce}
    \alpha(\x, \x') = \min\left(1, \frac{\mathcal{L}_I(\x';\y^*)\pi_{\bx}(\x')}{\mathcal{L}_I(\x;\y^*)\pi_{\bx}(\x)}\right)\longrightarrow
    \widehat{\alpha}(\x, \x') = \min\left(1, \frac{\widehat{\mathcal{L}_I}(\x', \widehat{\bw}, \widehat{Z};\y^*)\pi_{\bx}(\x')}{\widehat{\mathcal{L}_I}(\x, \widehat{\bw}, \widehat{Z};\y^*)\pi_{\bx}(\x)}\right)\,.
\end{equation}
The pseudo-marginal method is known to produce exact approximation, i.e., the stationary distribution of Markov chains generated by the estimate $\widehat{\mathcal{L}_I}$ is the exact posterior distribution; see~\cite{Andrieu2009, Andrieu2015, Warne2020}.

On the other hand, the acceptance probability $\widehat{\alpha}$ in the pseudo-marginal method becomes a random variable, and thus the effective sample efficiency and mixing time of generated Markov chains heavily depends on the quality of the Monte Carlo estimator. For example, if the integrated likelihood is excessively overestimated at a chain position, the chain will be reluctant to move from that position, leading to a long run of successive rejections. This phenomenon creates a trade-off between the number of Monte Carlo samples for constructing the estimator and the computational cost per effective sample of the generated Markov chains. Theoretical analysis and empirical studies imply that, under various idealistic assumptions on the form of the conditional likelihood and its distribution, the number of Monte Carlo samples in~\eqref{eq:mc_estimator} should be selected such that the variance of $\ln \widehat{\mathcal{L}}$ is lower than $2$ in regions with high posterior probability to maximize the computational efficiency~\cite{Pitt2012,Doucet2015, Sherlock2015}. This limitation suggests several desiderata for designing the data random variable $\by$ used in defining conditional likelihoods; three are listed below.
\begin{enumerate}
    \item \textbf{Low dimensional}: the dimension of $\by$ should be small, as the variance of the log--conditional likelihood defined by $\by$ increases as the dimension of $\by$ increases.
    \item \textbf{Insensitive to nuisance parameters}: the variability of the conditional likelihood defined by $\by$ in response to the variability of the nuisance parameters $(\bw, Z)$ should be low for effective construction of the estimator.
    \item \textbf{Sensitive to model parameters}: the Monte Carlo estimator based on the conditional likelihood defined by $\by$ should retain as much mutual information between the full image data $\boldsymbol{D}$ and the forward model parameters $\bx$ as possible.
\end{enumerate}

\subsection{The conditional likelihood functions: Full images versus their azimuthally-averaged power spectrum}\label{subsec:likelihood}
If the full image $\y^* = \boldsymbol{d}^*$ is used to define the conditional likelihood function $\mathcal{L}$, then the conditional likelihood has the form of a Gaussian function induced by the sensor noise according to~\eqref{eq:data_model},
   \begin{equation*}
   \begin{aligned}
    \mathcal{L}(\x,\w,z;\boldsymbol{d}^*) &\coloneqq \pi_{\boldsymbol{N}}(\boldsymbol{d} - \bdmc{O}(\x,\w,z)) \\
    &= \prod_{j = 1}^{M_1}\prod_{k=1}^{M_2} \pi_n\left(\boldsymbol{d}^*_{jk};\left(\bdmc{O}(\x,\w,z)\right)_{jk}, \sigma_n^2\right)\propto \exp\left(-\frac{1}{2\sigma_n^2}\norm{\boldsymbol{d}^* - \bdmc{O}(\x,\w,z)}_F^2\right)\,,
   \end{aligned}
\end{equation*}
where $\pi_n(s;\mu, \sigma^2)$ is the probability density function for a normal distribution with a mean of $\mu$ and a variance of $\sigma^2$, and $\norm{\cdot}_F$ is the Frobenius norm. 

If the conditional likelihood is defined based on the AAPS of the image data $\y^* = \aaps(\boldsymbol{d}^*)$, then we arrive at the following expression according to \eqref{eq:distribution_smoothed_aaps}:
\begin{equation*}
    \mathcal{L}\left(\x,\w,z;\boldsymbol{\mathcal{P}}_r\left(\boldsymbol{d}^*\right)\right) \coloneqq \prod_{j = 1}^{N_r}\pi_{\textrm{ncs}}\left(\frac{2\left(\boldsymbol{N}_{\theta}\right)_j}{M_1M_2\sigma_n^2}\left(\boldsymbol{\mathcal{P}}_r\left(\boldsymbol{d}^*\right)\right)_{j}; 2\left(\boldsymbol{N}_{\theta}\right)_j,\frac{2\left(\boldsymbol{N}_{\theta}\right)_j}{M_1M_2\sigma_n^2}\left(\boldsymbol{\mathcal{O}}_r\left(\x, \w, z \right)\right)_{j}\right)\,,
\end{equation*}
where $\boldsymbol{\mathcal{O}}_r\coloneqq\boldsymbol{\mathcal{P}}_r\circ\boldsymbol{\mathcal{J}}\circ\mathcal{F}$ is the \textit{parameter-to-spectrum map}. Similarly, we define the parameter-to-spectrum map for the smooth AAPS, $\overline{\boldsymbol{\mathcal{O}}_r}\coloneqq\overline{\boldsymbol{\mathcal{P}}_r}\circ\boldsymbol{\mathcal{J}}\circ\mathcal{F}$, and its conditional likelihood function $\mathcal{L}\left(\x,\w,z;\overline{\boldsymbol{\mathcal{P}}_r}\left(\boldsymbol{d}^*\right)\right)$ that has the same form as \eqref{eq:distribution_smoothed_aaps}, with $\boldsymbol{N}_{\theta}$ replaced with $\overline{\boldsymbol{N}_{\theta}}$.

The approximate conditional likelihood function can be used in the settings where the normal approximation in~\eqref{eq:normal_approx_smoothed_aaps} is preferred.
\begin{equation*}
\begin{aligned}
    \mathcal{L}_N(\x,\w, z;\overline{\bdmc{P}_r}({\boldsymbol{d}^*})) &\coloneqq \prod_{j = 1}^{\ceil{\boldsymbol{r}_{N_r}}}\pi_n\bigg(\left(\overline{\bdmc{P}_r}({\boldsymbol{d}^*})\right)_j; \left(\overline{\bdmc{O}_r}\left(\x, \w, z \right)\right)_j + M_1M_2\sigma_n^2, \\
    &\qquad\qquad \frac{M_1M_2\sigma_n^2}{2(\overline{\boldsymbol{N}_\theta})_j}\left(2M_1M_2\sigma_n^2 + \left(\overline{\bdmc{O}_r}\left(\x, \w, z \right)\right)_j\right)\bigg)\,.
\end{aligned}
\end{equation*}
We note that evaluating the probability density for noncentral chi-square distribution may be numerically ill-conditioned when the distribution's $\lambda$ and $k$ parameters are large. When the direct evaluation is inaccurate, the approximate conditional likelihood can be used instead without loss of accuracy, as shown in Figure~\ref{fig:normal_approximation}.

Comparing among the conditional likelihoods defined with $\by =\boldsymbol{D}$, $\by=\bdmc{P}_r(\boldsymbol{D})$, and $\by=\overline{\bdmc{P}_r}(\boldsymbol{D})$, we observe that
\begin{enumerate}
    \item The information loss due to image corruption is much more interpretable for the AAPS and the smoothed AAPS. As suggested by the normal approximation in \eqref{eq:normal_approx_smoothed_aaps}, when the simulated spectrum entries have magnitude values that are dominated by image noise,
    \begin{equation}\label{eq:noise_aaps_dimension_reduction}
        \left(\overline{\boldsymbol{\mathcal{O}}_r}(\bx,\bw, Z)\right)_j \ll M_1 M_2\sigma_n^2\,, \quad Z\sim \nu_{Z}\,,
    \end{equation}
    the information gain from these entries is negligible. This implies that a smaller pixel density, a smaller image contrast, a larger blurring intensity, and a larger noise intensity each lead to less uncertainty reduction in $\bx$. The hypothesis agrees with our observations in Figure~\ref{fig:image_corruptions}. We expect the information loss due to discarding the small and large radial frequency spectrum entries to be small due to their low magnitude values, allowing further dimension reduction of the AAPS.
    \item The conditional likelihood based on the AAPS of image data compares characteristic length scale information contained in data and model predictions. It is related to quantities such as the periodicity and interfacial length of the Di-BCP pattern. It is well-understood that these quantities are highly correlated to the values of Di-BCP material parameters~\cite{Leibler1980, choksi2001scaling}. Furthermore, the AAPS is relatively consistent in the presence of aleatoric uncertainties, as it discards the phase spectrum and marginalizes the orientational dependency of the power spectrum. On the other hand, the conditional likelihood based on full images compares the entry values of data and simulated images. In the presence of aleatory uncertainties, such comparisons do not characterize the fundamental relations between the image data and the model parameters. 
\end{enumerate}

\subsection{Sensitivity of conditional likelihood functions: Numerical experiments}

To quantitatively demonstrate the second observation above, we present two numerical experiments on the sensitivity of the conditional likelihoods with respect to the nuisance parameters and the parameters of the OK model, with data being a simulated microscope image $\boldsymbol{d}^*\in\R^{200\times200}$ generated at a set of reference parameters $(\x^*, \w^*, z^*)$ shown in Figure~\ref{fig:likelihood_sensitivity} with a high signal-to-noise ratio.

\begin{figure}[H]
\centering
    \begin{minipage}{0.63\linewidth}
        \includegraphics[width = \linewidth]{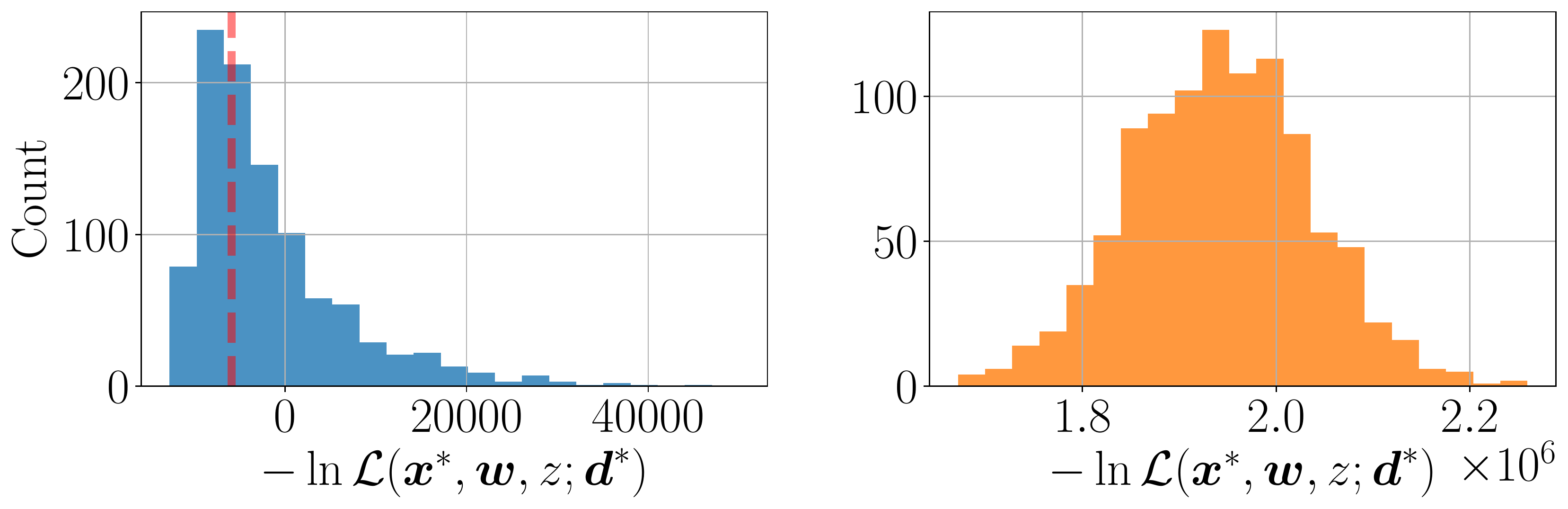}
    \end{minipage}
    \begin{tabular}{|c|c|}\hline
         \includegraphics[width = 0.03\linewidth]{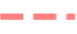} & $\w = \w^*$ and $z = z^*$ \\\hline 
         \includegraphics[width = 0.03\linewidth]{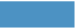} & $\w^{(j)}\sim \mathcal{N}(\w^*, \delta_w \boldsymbol{I}_w)$ and $z = z^*$ \\\hline 
         \includegraphics[width = 0.03\linewidth]{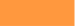} & $\w = \w^*$ and $z^{(j)}\sim \nu_{Z|\bx}(\cdot|\x^*)$\\\hline
    \end{tabular}
    \includegraphics[width=\linewidth]{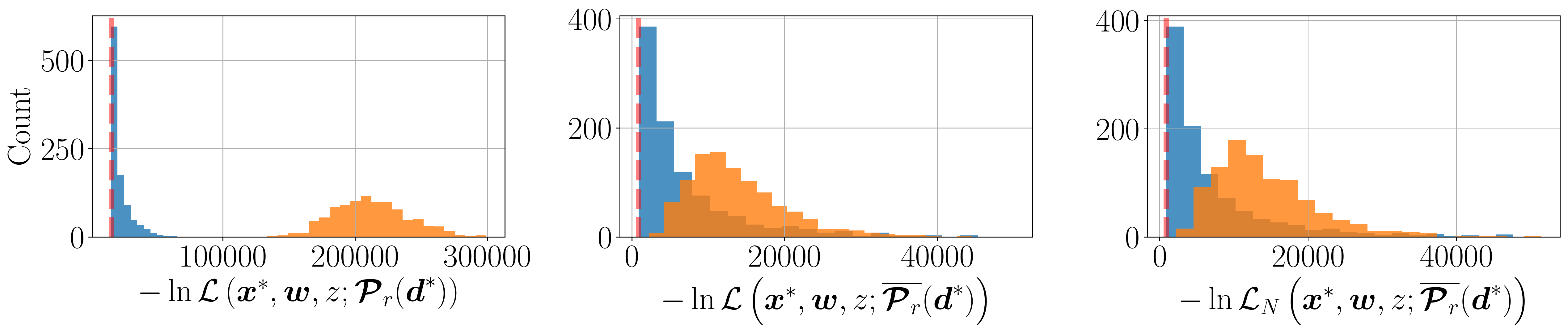}
    \caption{Conditional likelihood sensitivity to nuisance parameters. The histograms of different conditional likelihoods evaluated at the reference parameters $(\x^*, \w^*, z^*)$, with one of the nuisance parameters replaced by $10^3$ samples of $\w^{(j)}\sim \mathcal{N}(\w^*, \delta_w \boldsymbol{I}_w)$ or $z^{(j)}\sim \nu_{Z|\bx}(\x^*)$. The variance $\delta_w$ is set to a small value so that the resulting changes in the simulated images $\bdmc{O}(\x^*,\w^{(j)}, z^*)$ are barely discernible.}
    \label{fig:likelihood_sensitivity_wz}
\end{figure}

\begin{experiment}
    In Figure \ref{fig:likelihood_sensitivity_wz}, we visualize the response of different conditional likelihoods to variations in the nuisance parameters. For the log--conditional likelihood defined with full images (\emph{top row}), its variability in response to the auxiliary variable (\emph{orange}) is $3\sim 4$ orders of magnitude larger than that to a small variation in the characterization model parameters (\emph{blue}). Moreover, log--conditional likelihood at samples of the auxiliary variable is six orders of magnitude larger than the log--conditional likelihood value at the true parameter (\emph{red dashed line}). In comparison, the conditional likelihoods defined with the AAPS and the smoothed AAPS (\emph{bottom}) have much smaller variability in response to the auxiliary variable, and the conditional likelihood at samples of the auxiliary variable is much closer to the conditional likelihood value at the reference parameters.
\end{experiment}
\newline The results of the first numerical experiment imply that the AAPS is much more suitable for comparing model predictions and the image data in the presence of aleatoric uncertainties in Di-BCP self-assembly due to its insensitivity to the nuisance parameters. In particular, the enormous error in the sample-based estimation of the integrated likelihood defined with full images is unmanageable for the pseudo-marginal methods. 

\begin{figure}[H]
\centering
\begin{minipage}{0.32\linewidth}
    \includegraphics[width = \linewidth]{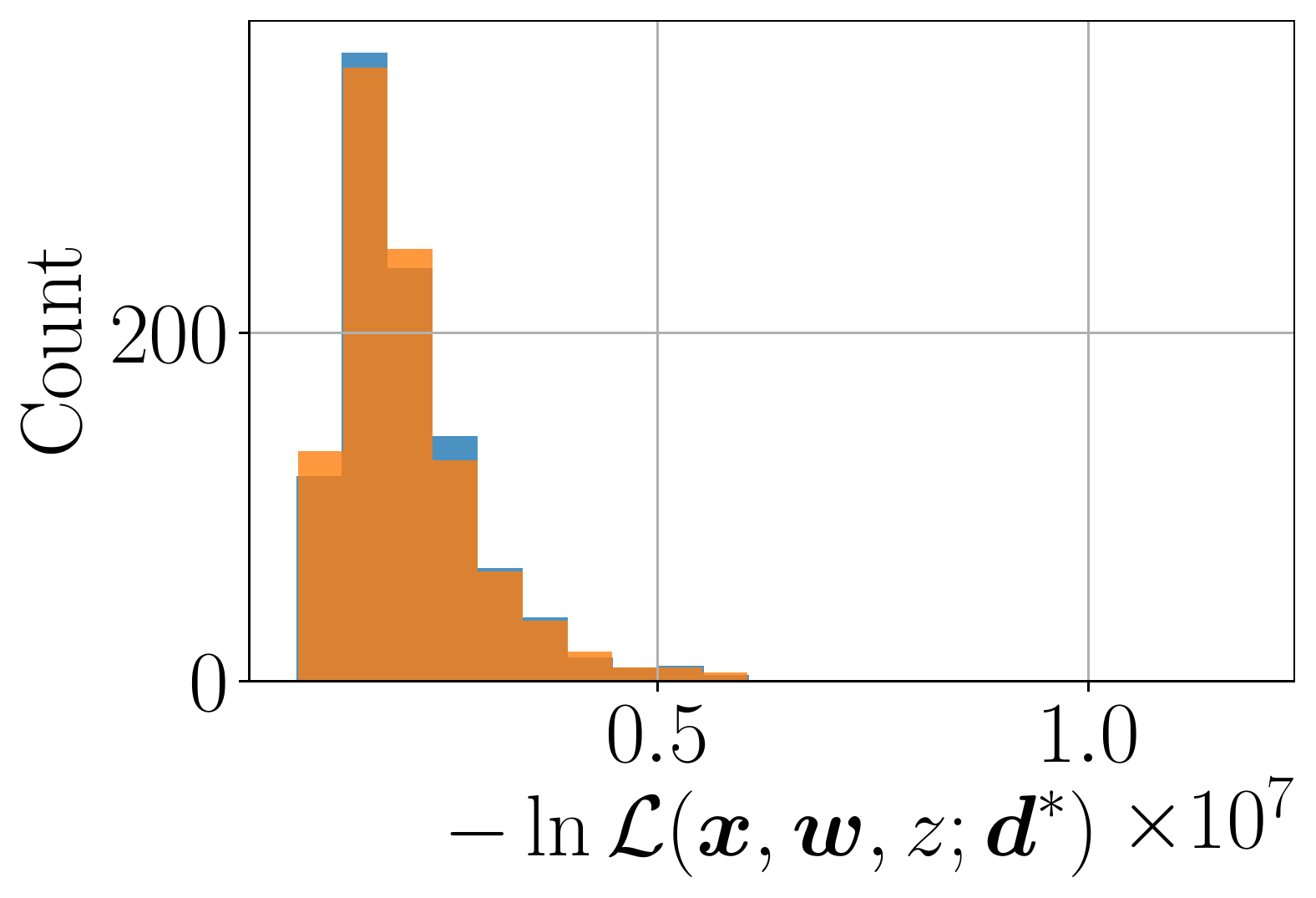}
\end{minipage}\hspace{10pt}
    \begin{tabular}{|c|c|}\hline
        \includegraphics[width = 0.03\linewidth]{likelihood_sensitivity_blue.png} & $\x =  \x^*, \w^{(j)}\sim \mathcal{N}(\w^*, \delta_w \boldsymbol{I}_w)$ and $z^{(j)}\sim \nu_{Z|\bx}(\x^*)$\\\hline
        \includegraphics[width = 0.03\linewidth]{likelihood_sensitivity_orange.png} & $\x^{(j)}\sim \mathcal{N}(\x^*, \delta_x \boldsymbol{I}_{x})$, $\w^{(j)}\sim \mathcal{N}(\w^*, \delta_w \boldsymbol{I}_{w})$ and $z^{(j)}\sim \nu_{Z|\bx}(\cdot|\x^{(j)})$\\\hline
    \end{tabular}
    \includegraphics[width=\linewidth]{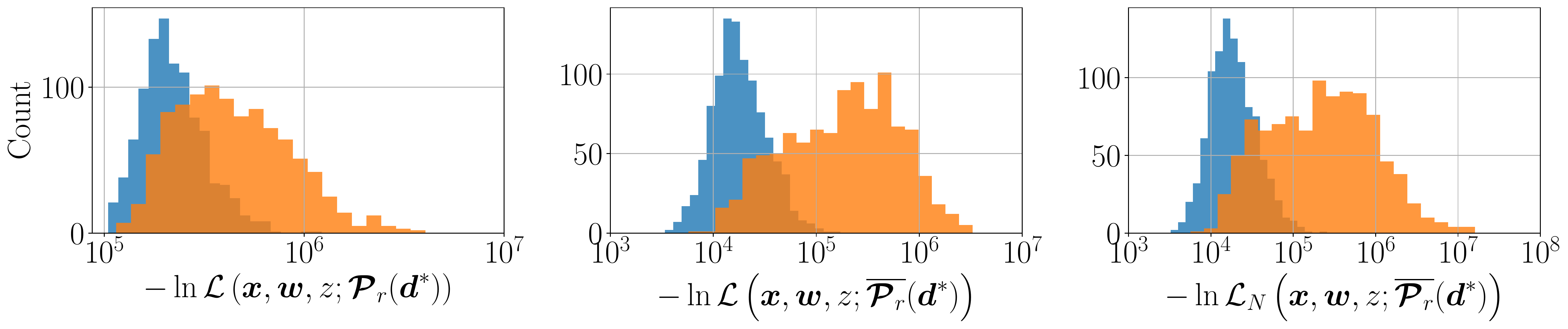}
    \caption{Conditional likelihood sensitivity to model parameters. The histograms of different conditional likelihoods evaluated at the reference model parameters $\x^*$ and $10^3$ samples of $\x^{(j)}\sim\mathcal{N}(\x^*, \delta_x\boldsymbol{I}_x)$, in the presence of variations in the nuisance parameters. The variance $\delta_x$ is set to a small value compared to the admissible parameter region~\cite{Choksi2011,Baptista2022} for the OK model.}
    \label{fig:likelihood_sensitivity_x}
\end{figure}

\begin{experiment}
In Figure~\ref{fig:likelihood_sensitivity_x}, we visualize the response of different conditional likelihoods to a small perturbation in the forward model parameters. For the log--conditional likelihoods defined with the full image (\emph{top row}), there is almost no visible difference between its response with or without the perturbation. On the contrary, the log--conditional likelihoods defined with the AAPS or the smoothed AAPS (\emph{bottom row}) show orders of magnitude separation in their response with or without the perturbation.
\end{experiment}
\newline The results of the second numerical experiment indicate that the AAPS is much more suitable for parameter inference in the presence of aleatoric uncertainties due to its sensitivity to the forward model parameters. In particular, the smoothed AAPS retains the sensitivity while vastly reducing the data dimension.

In conclusion, the numerical experiments demonstrate that the AAPS provides a good balance of insensitivity to nuisance parameters and sensitivity to model parameters. In addition, the smoothed AAPS is low dimensional and has the potential to be reduced further according to \eqref{eq:noise_aaps_dimension_reduction}, which makes it suitable for defining the conditional likelihood to be used in the pseudo-marginal method as discussed in Section~\ref{subsec:pseudo-marginal}. We use the entries of the smooth AAPS of image data with large magnitude values to carry out the model calibration studies in the rest of the paper.

\subsection{Accounting for the lost information: A phase-informed prior distribution}
The compression of data from full images to their smoothed AAPS results in reduced information gain from the model calibration procedure. In particular, the information associated with constant shifts in the image data is completely lost if only the smoothed AAPS of the image data is used to define the conditional likelihood for model calibration,
\begin{equation}\label{eq:constant_aaps}
    \overline{\bdmc{P}_r}(\boldsymbol{D} + \boldsymbol{c}) = \overline{\bdmc{P}_r}(\boldsymbol{D})\,,\quad \boldsymbol{c}_{jk} = c\in\R\,.
\end{equation}
We can partly recover the lost information by first inferring the parameters from the \textit{mean pixel values} of the image data.

Recall that the top-down microscopy characterization model in \eqref{eq:data_model} implies that the mean pixel value of the image random variable for a given Di-BCP material state follows a normal distribution:
\begin{equation*}
    \left\langle \boldsymbol{D}\right\rangle = \left\langle \bdmc{J}(u)\right\rangle + \langle \boldsymbol{N}\rangle \sim \mathcal{N}\left(\left\langle\bdmc{J}(u)\right\rangle, \frac{\sigma_n^2}{M_1 M_2}\right)\,,
\end{equation*}
where $\langle\cdot\rangle$ denotes the mean of the entry values in a matrix. By the commutative properties between the mean operator and each of the three operators in $\bdmc{J}$, we arrive at the following normal distribution conditioned on the characterization model parameters in~\eqref{eq:charac_model}:
\begin{equation*}
    \langle \boldsymbol{D}\rangle \sim \mathcal{N}\left(\frac{c_1}{L_1 L_2}\int_{\Omega} u(\boldsymbol{s}) \,\textrm{d}\boldsymbol{s}-2|c_1|c_2 + c_2 + \frac{1}{2}, \frac{\sigma_n^2}{M_1 M_2}\right)\,.
\end{equation*}
The incompressibility of Di-BCP films implies that the state space $V^u(\Omega)$, such as the one defined for the forward operator of the OK model in \eqref{eq:ok_forward_operator}, only contains order parameters with a constant spatial average that is typically specified by a model parameter. Therefore, evaluating the likelihood function for the mean pixel values of an observed image $\langle \boldsymbol{d}^*\rangle$ does not require simulating Di-BCP thin film self-assembly. For example, the likelihood function for $\langle \boldsymbol{d}^*\rangle$ based on the OK model is given by
\begin{equation} \label{eq:likelihood_mean}
    \mathcal{L}\left(\x, \w; \left\langle\boldsymbol{d}^*\right\rangle\right) = \pi_n\left(\langle\boldsymbol{d}^*\rangle; c_1m - 2|c_1|c_2+c_2+\frac{1}{2}, \frac{\sigma_n^2}{M_1 M_2}\right)\,.
\end{equation}

Recall that the OK model parameter $m$ is directly related to the monomer segment number ratio $f$ of Di-BCPs through \eqref{eq:ok_parameter_maps}, which is a material parameter commonly used in SCFT and other higher fidelity models of Di-BCP self-assembly. Moreover, the material parameter $f$ is highly correlated to the \textit{morphological phase} of the Di-BCP thin film~\cite{Khandpur1995}. In particular, if $|f-0.5|$ or $|m|$ is large, then the two monomer species occupy uneven portions of the top surfaces of Di-BCP thin films. This leads to the formation of spot patterns that enclose the monomer species with a smaller proportion. Therefore, inferring the material parameter $f$ through the mean pixel value of image data $\langle\boldsymbol{D}\rangle$ can be interpreted as assimilating the morphological phase information in image data.

We thus consider a sequential approach for Bayesian model calibration. First, we sample from or construct the conditional density $\pi_{\bx|\langle \boldsymbol{D} \rangle}(\cdot|\langle\data^*\rangle)$ with high precision and relatively negligible computational cost through Bayes' rule,
\begin{equation*}
    \pi_{\bx|\langle \boldsymbol{D}\rangle} (\x|\langle \boldsymbol{d}^*\rangle ) =  \frac{\mathcal{L}_I\left(\x;\left\langle \data^*\right\rangle\right)\pi_{\bx}(\x)}{C_{\langle \data^*\rangle}}\,,\quad \mathcal{L}_I\left(\x;\left\langle \data^*\right\rangle\right) = \expect{\bw}{ \mathcal{L}\left(\x, \w; \left\langle\boldsymbol{d}^*\right\rangle\right)}\,.
\end{equation*}
We refer to $\pi_{\bx|\langle \boldsymbol{D}\rangle} (\cdot|\langle \boldsymbol{d}^*\rangle )$ as the \textit{phase-informed prior distribution}. Then the posterior from the first stage inference is used as the prior for the second stage inference using an integrated likelihood defined with the AAPS of the image data:
\begin{equation*}
    \pi_{\bx|\overline{\aaps}(\boldsymbol{D}),\langle\boldsymbol{D}\rangle}\left(\x|\overline{\aaps}(\boldsymbol{d}^*), \langle\boldsymbol{d}^*\rangle\right) = \frac{\mathcal{L}_I(\x;\overline{\aaps}(\data^*))\pi_{\bx|\langle\boldsymbol{D}\rangle}(\x|\langle\boldsymbol{d}^*\rangle)}{C_{\overline{\aaps}(\data^*),\langle\data^*\rangle}}\,.
\end{equation*}
We expect that the first stage inference leads to increased information gain in components of the model parameters associated with the volume ratio of the two blocks in the Di-BCP film, such as the marginal parameter variable $M$ in the OK model, whereas the second stage inference leads to significantly increased information gain in the length scale--related parameters, such as the pair of marginal parameter $(E,\Sigma)$ in the OK model.

\section{A neural network surrogate of the parameter-to-spectrum map}\label{sec:nn_surrogate}

To accelerate the proposed Bayesian model calibration procedure and amortize the computational burden of repeated model simulations for each calibration task, we consider building a surrogate for the response of the parameter-to-AAPS map to the auxiliary variable, $\overline{\bdmc{O}_r}(\x, \w, Z)$. In particular, we formulate a surrogate with independent truncated normal distributions for the smooth AAPS entries:
\begin{equation}\label{eq:surrogate}
    \overline{\bdmc{O}_r}(\x, \w, Z)\mathrel{\dot\sim}\mathcal{TN}_{+}\left(\bdmc{S}_m(\x, \w), \bdmc{S}^2_{\sigma}(\x, \w)\boldsymbol{I}_{\ceil{\boldsymbol{r}_{N_r}}}\right)\,,
\end{equation}
where $\mathcal{TN}_+$ is the truncated normal distribution supported on $\R_+$ and $\bdmc{S}_m,\bdmc{S}_{\sigma}:\R^{d_x}\times \R^{d_w}\to\R_+^{\ceil{\boldsymbol{r}_{N_r}}}$ are the mean and the standard deviation predictors of the smoothed AAPS at the given input $(\x, \w)$. Note that the independence assumption allows us to bring the likelihood integration into the product with respect to the component of the likelihood:
\begin{equation}\label{eq:likelihood_integration}
    \expect{Z}{\prod_{j}\mathcal{L}\left(\x,\w,Z;\left(\overline{\bdmc{P}_r}(\boldsymbol{d})\right)_j\right)} \xrightarrow{~\eqref{eq:surrogate}} \prod_{j}\expect{Z}{\mathcal{L}(\x,\w,Z;\left(\overline{\bdmc{P}_r}(\boldsymbol{d})\right)_j)}\,.
\end{equation}
 This approximation is equivalent to a sequential approach for Bayesian inference using the smoothed AAPS of image data, where only one of the entries is used at each stage. The expectation in \eqref{eq:likelihood_integration} can be computed efficiently for the surrogate in \eqref{eq:surrogate} using the Gauss–Hermite quadrature as follows:

\begin{gather}\label{eq:surrogate_likelihood}
    \expect{Z}{\mathcal{L}(\x,\w,Z;\left(\overline{\bdmc{P}_r}(\boldsymbol{d})\right)_j)} \approx \sum_{k}\frac{w_k^{\text{GH}}}{\sqrt{\pi}\Phi_j} \pi_{\textrm{ncs}}\bigg(\frac{2\left(\overline{\boldsymbol{N}_{\theta}}\right)_j}{M_1M_2\sigma_n^2}\left(\overline{\boldsymbol{\mathcal{P}}_r}\left(\boldsymbol{d}\right)\right)_{j}; 2\left(\overline{\boldsymbol{N}_{\theta}}\right)_j, \frac{2\left(\overline{\boldsymbol{N}_{\theta}}\right)_j}{M_1M_2\sigma_n^2}\widetilde{\xi}_{jk}\bigg)\boldsymbol{1}_{\widetilde{\xi}_{jk}\geq 0}\,,\\
    \widetilde{\xi}_{jk} = \sqrt{2}\xi_k\left(\bdmc{S}_{\sigma}(\x,\w)\right)_j + \left(\bdmc{S}_{m}(\x,\w)\right)_j\,,\quad \Phi_j = \frac {1}{2}\left(1-\text{erf} \left(-\frac{\left(\bdmc{S}_m(\x,\w)\right)_j}{\sqrt{2}\left(\bdmc{S}_{\sigma}(\x,\w)\right)_j}\right)\right)\,,\nonumber
\end{gather}
where $(w_k^{\text{GH}}, \xi_k^{^{\text{GH}}})$, $k\in\N$, are the weights and positions of the Gauss--Hermite quadrature, $\boldsymbol{1}$ is the indicator function, and $\Phi_j$ is the scaling constant in the truncated normal distribution. The computed integrated likelihood estimate can be used to efficiently perform parameter inference through the formulation described in Section~\ref{sec:bayesian_model_calibration}.

Taking the approximation in \eqref{eq:surrogate}, the information concerning the correlation of the spectrum magnitude across different frequencies in response to the auxiliary variable is discarded. Moreover, the impact of the normal approximation in \eqref{eq:surrogate} on the consistency of parameter inference needs to be examined. Consequently, we expect bias and a lower uncertainty reduction in Bayesian model calibration performed through the surrogate parameter-to-spectrum map, particularly when the aleatoric uncertainty induced by $Z$ dominates other uncertainties induced by $\bw$ and $\boldsymbol{N}$. These issues are numerically studied in Section~\ref{sec:numerical_results}. 

In this section, we present mean and standard deviation predictors $\bdmc{S}_m$ and $\bdmc{S}_{\sigma}$ obtained through training multilayer perceptrons (MLPs) using data generated from the OK model introduced in Section~\ref{subsec:ok_model}.

\begin{remark}
    To avoid the normal approximation in \eqref{eq:surrogate}, one should consider alternative techniques in conditional sampling or conditional density estimation, such as conditional generative adversarial network~\cite{mirza2014conditional}, normalizing flow~\cite{papamakarios2021normalizing}, etc.
\end{remark}

\subsection{The loss function and training data generation}\label{subsec:surrogate_formulation}
We consider the following approximation problems in the logarithmic scale of the smooth AAPS:
\begin{gather*}
    \nn(\x,\w;\boldsymbol{t}_m) \approx \ln(\bdmc{S}_m(\x,\w))\,,\quad \nn(\x,\w;\boldsymbol{t}_{\sigma}) \approx \ln(\bdmc{S}_{\sigma}(\x,\w))\,,\\
    \bdmc{S}_m(\x,\w) = \mathbb{E}_Z\left[\overline{\bdmc{O}_r}(\x,\w, Z)\right]\,,\quad \bdmc{S}_{\sigma}(\x,\w) = \sqrt{\text{diag}\left(\mathbb{V}_Z\left[\overline{\bdmc{O}_r}(\x,\w, Z)\right]\right)}\,,
\end{gather*}
where $\boldsymbol{t}_m,\boldsymbol{t}_{\sigma}\in\R^{d_{t}}$ parameterizes the surrogate $\nn$. We use loss functions $\mathcal{H}_{m}:\R^{d_{t}}\to\R_+$ and $\mathcal{H}_{\sigma}:\R^{d_{t}}\to\R_+$ for optimizing $\boldsymbol{t}$ via minimizing mean square errors:
\begin{gather*}
    \boldsymbol{t}_m = \argmin_{\boldsymbol{t}\in\R^{d_{t}}} \mathcal{H}_m(\boldsymbol{t})\,,\quad \boldsymbol{t}_{\sigma} = \argmin_{\theta\in\R^{d_{t}}} \mathcal{H}_{\sigma}(\boldsymbol{t})\,,\\
    \begin{cases}
        \mathcal{H}_m(\boldsymbol{t}) = \cfrac{1}{2}\mathbb{E}_{\bx_s, \bw_s}\left[\norm{\nn(\bx_s,\bw_s;\boldsymbol{t}) - \ln\left(\bdmc{S}_m(\bx_s,\bw_s)\right)}_2^2\right]\,,\\
    \mathcal{H}_{\sigma}(\boldsymbol{t}) = \cfrac{1}{2}\mathbb{E}_{\bx_s, \bw_s}\left[\norm{\nn(\bx_s,\bw_s;\boldsymbol{t}) - \ln\left(\bdmc{S}_{\sigma}(\bx_s,\bw_s)\right)}^2_2\right]\,,\\
    \end{cases}
\end{gather*}
where $(\bx_s,\bw_s)\sim \nu_{\bx_s}\otimes\nu_{\bw_s}$ is the input distribution for the surrogate.

To estimate the loss function and its derivative, we generate samples of $\bx_s$, $\bw_s$, and $Z$ in order to estimate $\bdmc{S}_m$ and $\bdmc{S}_{\sigma}$. We consider the following training data generation procedure.

\begin{enumerate}
    \item Generate samples $\{\x^{(j)}\}_{j=1}^{t_x}$ using a space-filling design, e.g., Latin hypercube sampling.
    \item Perform simulations to obtain $\{u^{(jk)}\}_{j,k=1}^{t_x,t_z}$, with $u^{(jk)} = \mathcal{F}(\x^{(j)},z^{(jk)})$ and $\{z^{(jk)}\}_{j,k=1}^{t_x,t_z}\sim \nu_{Z|\bx}(\cdot|\x^{(j)})$.
    \item For each $\x^{(j)}$, generate samples $\{\w^{(jl)}\}_{l=1}^{t_w}\sim\nu_{\bw_s}$ using a space-filing design.
    \item Compute samples of the smooth AAPS, $\overline{\bdmc{O}_r}(\x^{(j)},\w^{(jl)}, z^{(jk)}) = (\overline{\aaps}\circ\bdmc{J})(u^{(jk)};\w^{(jl)})$.
    \item Compute the mean and standard deviation estimates, $\widehat{\bdmc{S}}_m(\x^{(j)},\w^{(jl)})$ and $\widehat{\bdmc{S}}_{\sigma}(\x^{(j)},\w^{(jl)})$, via marginalizing the $k$ component associated with $z^{(j,k)}$ at each sample.
    \item Use the samples $\x^{(j)}$ and $\w^{(jl)}$ and the mean and standard deviation estimates at these samples to evaluate the empirical estimation of the loss functions and its derivative.
\end{enumerate}

The offline cost of training data generation is dominated by the $t_x\times t_z$ number of computer simulations using the Di-BCP thin film self-assembly model. Therefore, efficient schemes for placing samples or quadrature points for $(\bx_s, Z)$ are crucial for reducing the offline cost of the surrogate construct. We leave this topic for future work.

\subsection{A neural network surrogate based on the Ohta--Kawasaki model}\label{subsec:surrogate_results}

This subsection describes a surrogate parameter-to-spectrum map based on the Ohta--Kawasaki model in Section~\ref{subsec:ok_model} constructed using the formulation described in Section~\ref{subsec:surrogate_formulation}. 

We use a uniformly distributed $\bx_s \sim\mathcal{U}(\mathcal{A}_x)$, where $\mathcal{A}_x\subset\R^{3}$ is an admissible region of the parameters $(\epsilon,\sigma, m)$ in which model predictions exhibit phase separation, and lower bounds are enforced on the interface size and periodicity length of the simulated patterns; see \cite[Section 7.1]{Baptista2022} for a detailed description of such a parameter distribution. We consider a stratified sampling approach to generate $t_x = 800$ space-filling samples and solve the OK model at each sample using $t_z = 50$ samples of random initial states. This leads to a total of $40,000$ model simulations.

We set $M_1 = M_2 = 100$, i.e., simulated images of size $100\times 100$, and sample from a uniformly distributed $(C_1, \Sigma_b)\sim\mathcal{U}(\mathcal{A}_w)$ for evaluating the state-to-image map~\eqref{eq:charac_model} based on the simulated order parameters. The region $\mathcal{A}_w = [0,1/2]\times[0, 0.08]$ consists of a wide range of possible contrast scaling (no material contrast to sharp material contrast) and blurring level (no blurring to $\sigma_b\sim 8\%$ of the image size). Note that we exclude the nuisance parameters $c_2$ and $\sigma_n$ associated with brightness shift and the noise level. The former does not impact the smooth AAPS of simulated images due to~\eqref{eq:constant_aaps}, and the latter does not enter the parameter-to-spectrum map. We consider a stratified sampling approach to generate around $t_w = 10$ samples for each $\x^{(j)}$. This leads to a total of $8000$ samples for training and validation of the parameter-to-spectrum surrogate. 

\begin{figure}[!hbt]
    \centering
    \addtolength{\tabcolsep}{-5pt} 
    \begin{tabular}{F G}
        \includegraphics[width = \linewidth]{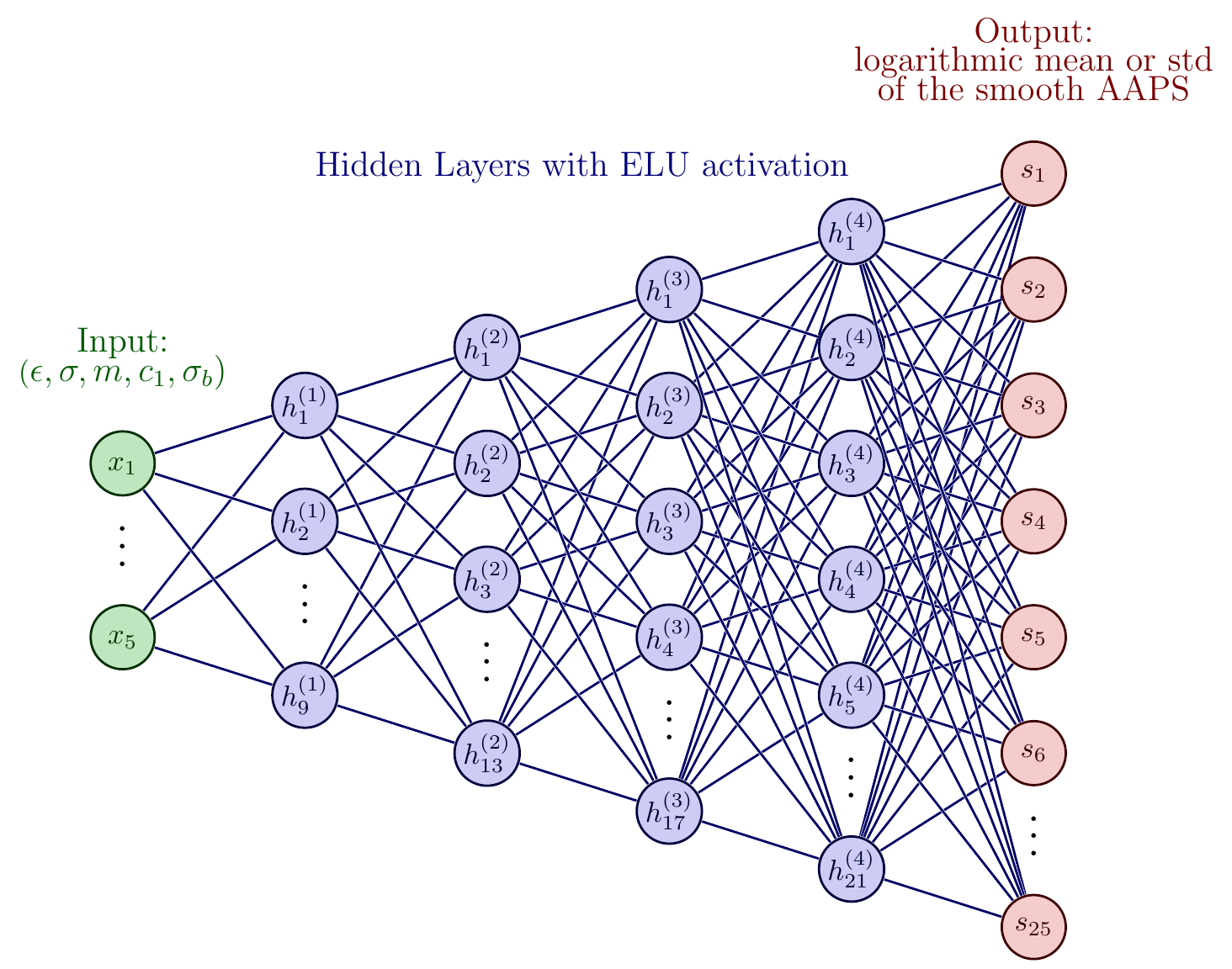} & \includegraphics[width =\linewidth]{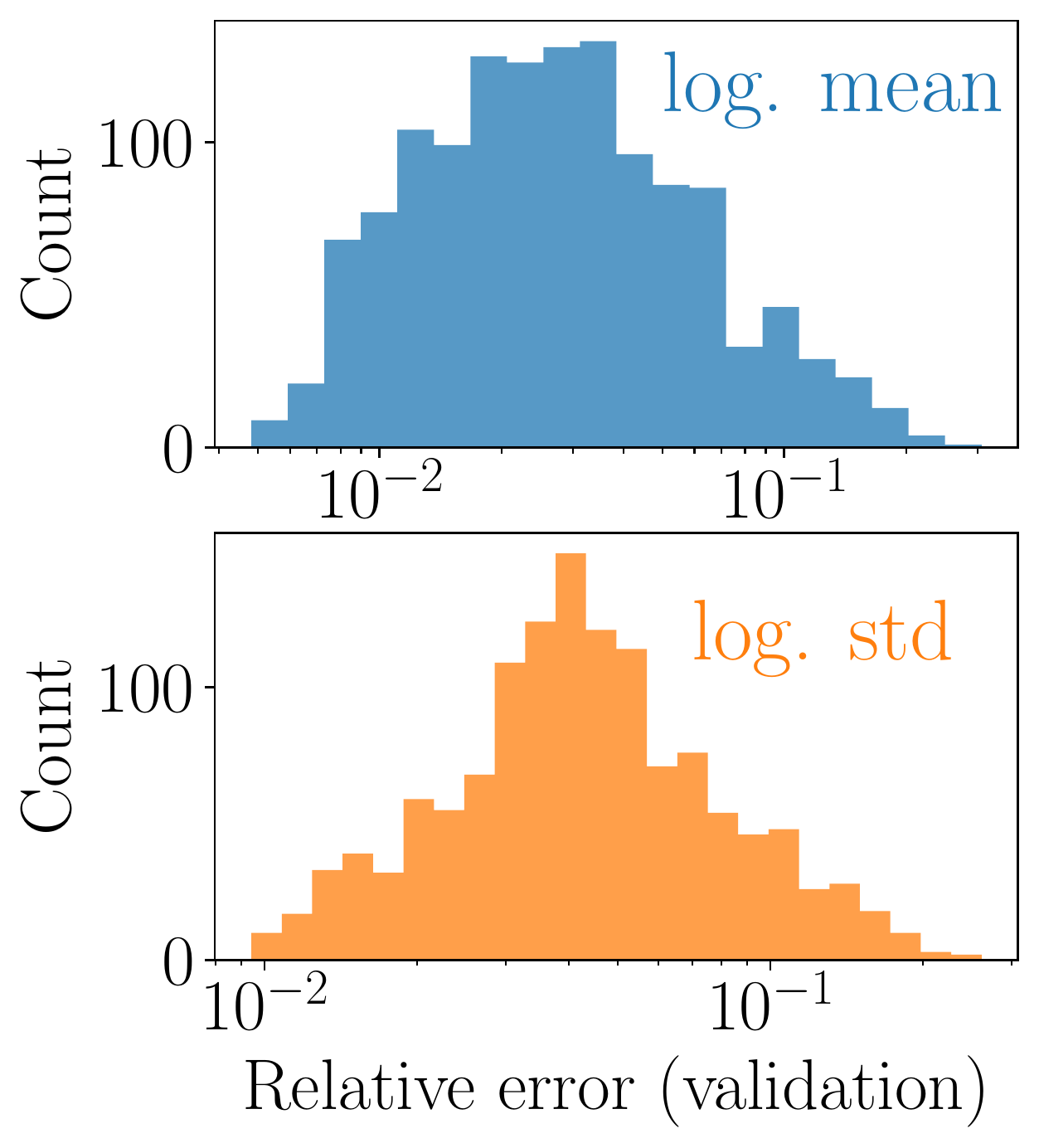}
    \end{tabular}
    \addtolength{\tabcolsep}{5pt} 
    \caption{(\emph{left}) The architecture of the MLP used for constructing the surrogates that predict the logarithmic mean and standard deviation of the smooth AAPS given model and nuisance parameters. (\emph{right}) The histogram of the averaged relative error measured over the validation set is presented in the logarithmic scale.}
    \label{fig:neural_network_and_valdiation}
\end{figure}

The surrogate $\widetilde{\bdmc{S}}$ is constructed using MLPs~\cite{Goodfellow2016} that map the input parameters to the logarithmic mean and standard deviation for the first $25$ entries of the smoothed AAPS:
\begin{align*}
    \widetilde{\bdmc{S}}_m(\cdot;\boldsymbol{t}):\R^5\ni(\epsilon,\sigma, m, c_1,\sigma_b)\mapsto \ln(\bdmc{S}_m(\x,\w))_{1:25}\in\R^{25}\,,\\
    \widetilde{\bdmc{S}}_{\sigma}(\cdot;\boldsymbol{t}):\R^5\ni(\epsilon,\sigma, m, c_1,\sigma_b)\mapsto \ln(\bdmc{S}_{\sigma}(\x,\w))_{1:25}\in\R^{25}\,.
\end{align*}
It consists of $4$ hidden layers with an increasing breadth of $9$, $13$, $17$, and $21$; see Figure \ref{fig:neural_network_and_valdiation}. The exponential linear unit (ELU) activation function~\cite{clevert2016fast} is used. The simulated data is partitioned according to the $\bx_s$ samples; we use $6400$ samples for training and $1600$ samples for validation. The training is performed with the Adam algorithm~\cite{kingma2017adam} in \texttt{PyTorch}~\cite{paszke2019pytorch}. The performance of the trained MLPs on the validation set is visualized in Figure \ref{fig:neural_network_and_valdiation}. The averaged relative error measured on the validation set is $3.4\%$ for the logarithmic mean predictor and $5.1\%$ for the logarithmic standard deviation predictor.

\section{Numerical results}\label{sec:numerical_results}

\begin{figure}[!hbt]
    \centering
    \addtolength{\tabcolsep}{-5pt} 
    \begin{tabular}{c c c c c c c c}
         $\displaystyle u_1^*\sim\mathcal{F}(\x_1^*, Z) $ & &$\displaystyle \data_1^* \in[0,1]^{400\times 400}$ & & $\displaystyle u_2^*\sim\mathcal{F}(\x_2^*, Z)$ & &$\displaystyle \data_2^* \in[0,1]^{400\times 400}$ & \\
         \includegraphics[width = 0.18\linewidth]{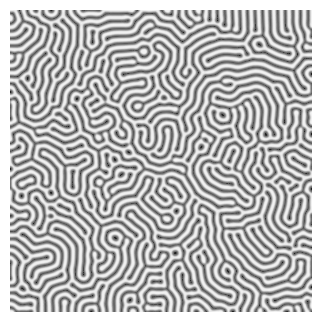} & \includegraphics[height = 0.18\linewidth]{colorbar_colorbar_op_25.pdf} & \includegraphics[width = 0.18\linewidth]{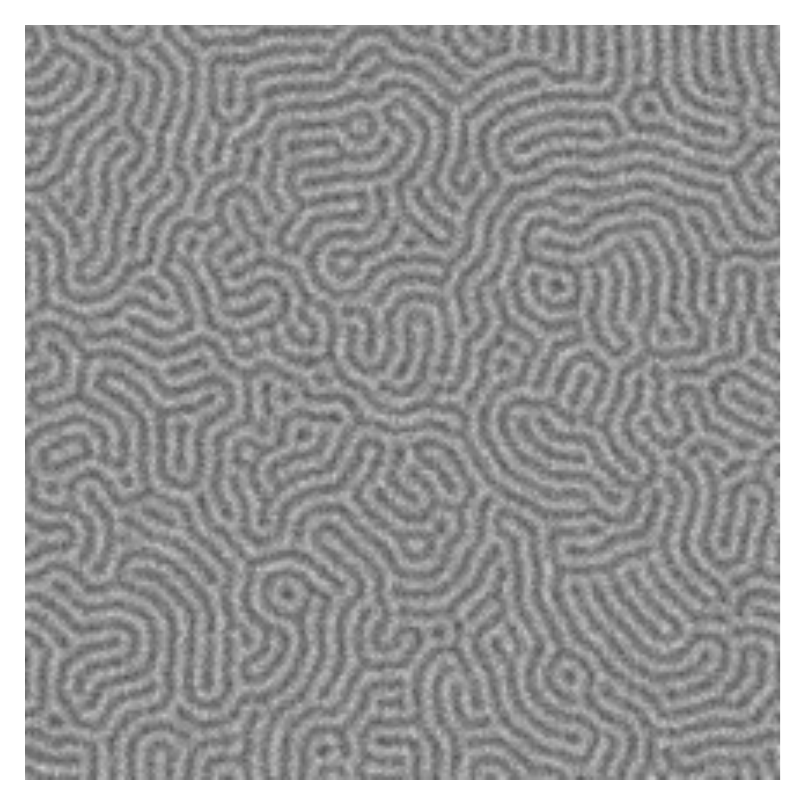} & \includegraphics[height = 0.18\linewidth]{colorbar_colorbar_image_25.pdf} & \includegraphics[width = 0.18\linewidth]{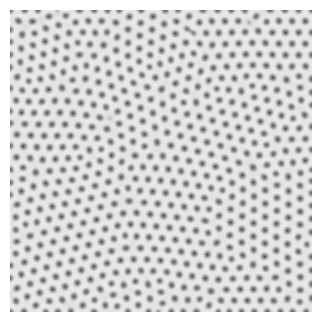} & \includegraphics[height = 0.18\linewidth]{colorbar_colorbar_op_25.pdf} & \includegraphics[width = 0.18\linewidth]{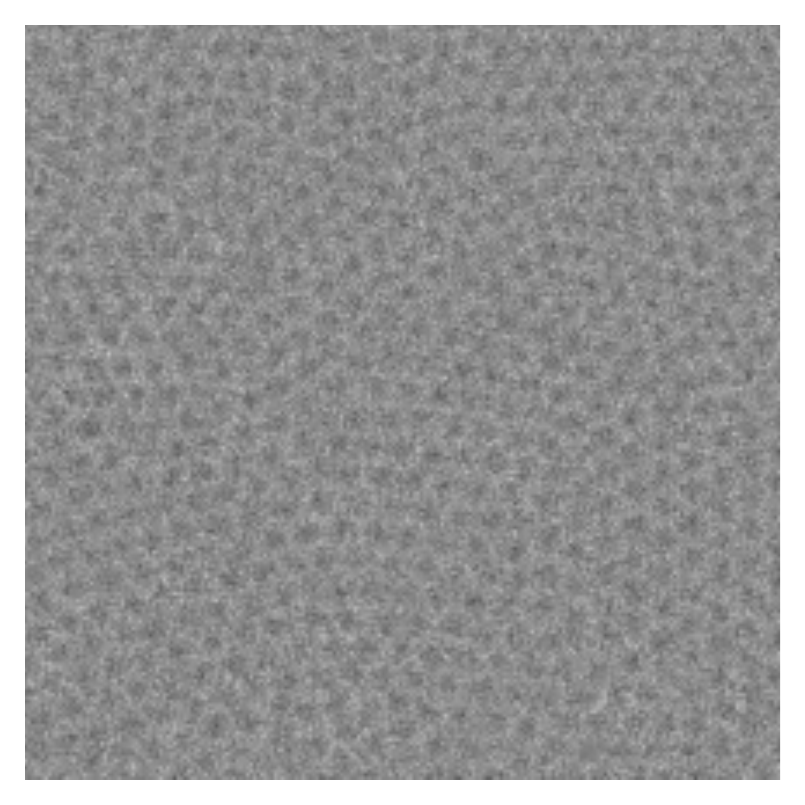} & \includegraphics[height = 0.18\linewidth]{colorbar_colorbar_image_25.pdf}
    \end{tabular}
    \addtolength{\tabcolsep}{5pt} 
    \caption{Two sets of simulated order parameter $u_j^*$ and synthetic microscopy images $\data_j^*$ of size $400\times 400$, $j= 1,2$. The resulting patterns are stripes (\emph{left}) and spots (\emph{right}). The order parameters and images are generated by the OK model (Section~\ref{subsec:ok_model}) and the state-to-image map (Section~\ref{subsec:characterization_model}) at $\x_1^* = (2\times 10^{-3}, 1.696\times 10^4, -0.03)$, $\x_2^* = (3.25\times10^{-3},9.6\times 10^{3}, -0.32)$, $\w_1^* = (0.21, 0.07, 0.39, 0.091)$, and $\w_2^* = (0.16, 0.13, 0.85, 0.1)$.}
    \label{fig:synthetic_data}
\end{figure}

In this section, we consider two numerical examples, with synthetic image data $\data_i^*\in\R^{400\times 400}$ sampled at two sets of parameters $(\x^*_j,\w^*_j, Z \sim \nu_Z)$, $j = 1, 2$ via the image model \eqref{eq:data_model} and the OK model~\eqref{eq:ok_forward_operator}. They are shown in Figure~\ref{fig:synthetic_data}. The resulting patterns consist primarily of stripes or spots. The full images are used for the first stage inference that constructs phase-informed prior distributions, where the marginal parameter $M$ related to the copolymer composition is inferred through the mean pixel value of the images. Sub-images of size $100\times 100$ are extracted\footnotemark\thinspace and used for the second stage inference for the full parameter $\bx$. The length scale--related parameters are rescaled to match the size of the full image according to \eqref{eq:rescaling}.
\footnotetext{By taking sub-images of a full image for parameter inference using the smooth AAPS entries, the inverse problem becomes ill-posed and challenging, yet easier from a computational perspective due to the reduced simulation cost on a smaller domain.}

We first visualize the posterior samples and posterior predictive samples generated by the pseudo-marginal method and conditional likelihood evaluations via simulating the OK model, demonstrating the viability of the model calibration procedure proposed in Section~\ref{sec:bayesian_model_calibration}. We then compare these samples to those generated by the parameter-to-spectrum surrogate. For the first part of this section, we assume that the characterization model parameter $\bw = \w_j^*$, $j=1,2$, is known. We consider Bayesian model calibration with the parameter-to-spectrum surrogate and an uncertain $\bw$ for the second part.

In the numerical examples, we assume an uninformative uniform prior distribution with independent components, i.e., $\pi_{\bx}(\x) = \pi_{E}(\epsilon)\pi_{\Sigma}(\sigma)\pi_{M}(m)$, and wide support. In particular, we set a upper bound of $(6.25\times 10^{-3}, 5.12\times 10^4)$ and lower bound of $(3.125\times 10^{-4}, 2.56\times 10^3)$ for $(E,\Sigma)$. They can be interpreted as prior knowledge of the field of view size relative to the interfacial length and periodicity of the observed Di-BCP morphology~\cite[Section 7.5]{Baptista2022}. We also set a uniform prior for $M$ supported in $[-0.2, 0]$ or $[-1/\sqrt{3}, -0.2]$ for the stripe or spot pattern. They are assigned based on the phase diagram for the OK model in 2D and assuming that phase separation is most likely observed in experiments \cite{Cao2022, Choksi2011, Baptista2022}. The posterior samples are produced with adaptive Metropolis proposals~\cite{harrio2001adaptive}. Model simulations via the finite element method are implemented through \texttt{FEniCS} \cite{AlnaesBlechta2015a} and the MCMC sampling is implemented partly through \texttt{hIPPYlib} \cite{Villa2018, villa2021hippylib}. The code for these numerical experiments is open-sourced\footnotemark.
\footnotetext{The BCPUQ project: \url{https://bitbucket.org/lcao11/workspace/projects/BCPUQ}.}

\subsection{Bayesian model calibration with known nuisance parameters}

Assuming $\nu_{\bw} = \delta_{\w^*}$ and a marginal uniform prior distribution for $M$, we can evaluate the marginal conditional density $\pi_{M|\langle \boldsymbol{D}\rangle}(\cdot|\langle\boldsymbol{d}\rangle)$ directly through the likelihood function in \eqref{eq:likelihood_mean}:
\begin{equation}\label{eq:phase_informed_prior_known}
    \pi_{M|\langle\boldsymbol{D}\rangle}(m|\langle\data\rangle) = \begin{cases}
    \mathcal{L}(\x,\w^*;\langle\data\rangle) + 1 -\int_{\text{supp}(M)} \mathcal{L}(\x,\w^*;\langle\data\rangle)\,\textrm{d}m\,,& m\in \text{supp}(M)\,;\\
    0\,, & m\not\in \text{supp}(M)\,.
    \end{cases}
\end{equation}
where $\text{supp}(M)$ is the support of the uniform marginal prior distribution for $M$. Note that \eqref{eq:phase_informed_prior_known} follows a truncated Gaussian distribution. Consequently, the phase-informed prior can be analytically expressed as
\begin{equation*}
    \pi_{\bx|\langle\boldsymbol{D}\rangle}(\x|\langle\data\rangle) = \pi_{E}(\epsilon)\pi_{\Sigma}(\sigma)\pi_{M|\langle\boldsymbol{D}\rangle}(m|\langle \data\rangle)\,.
\end{equation*}

\subsubsection{Posterior sampling via the pseudo-marginal method and simulating the Ohta--Kawasaki model}\label{subsubsec:sampling_model_known}
\begin{figure}[!hbt]
    \centering
\addtolength{\tabcolsep}{-5pt} 
    \begin{tabular}{c c c c}
        $\displaystyle \data_1^* \in[0,1]^{400\times 400}$ & $\displaystyle \widetilde{\data_1^*} \in [0,1]^{100\times 100}$& & \multirow{2}{*}{\includegraphics[width = 0.34\linewidth]{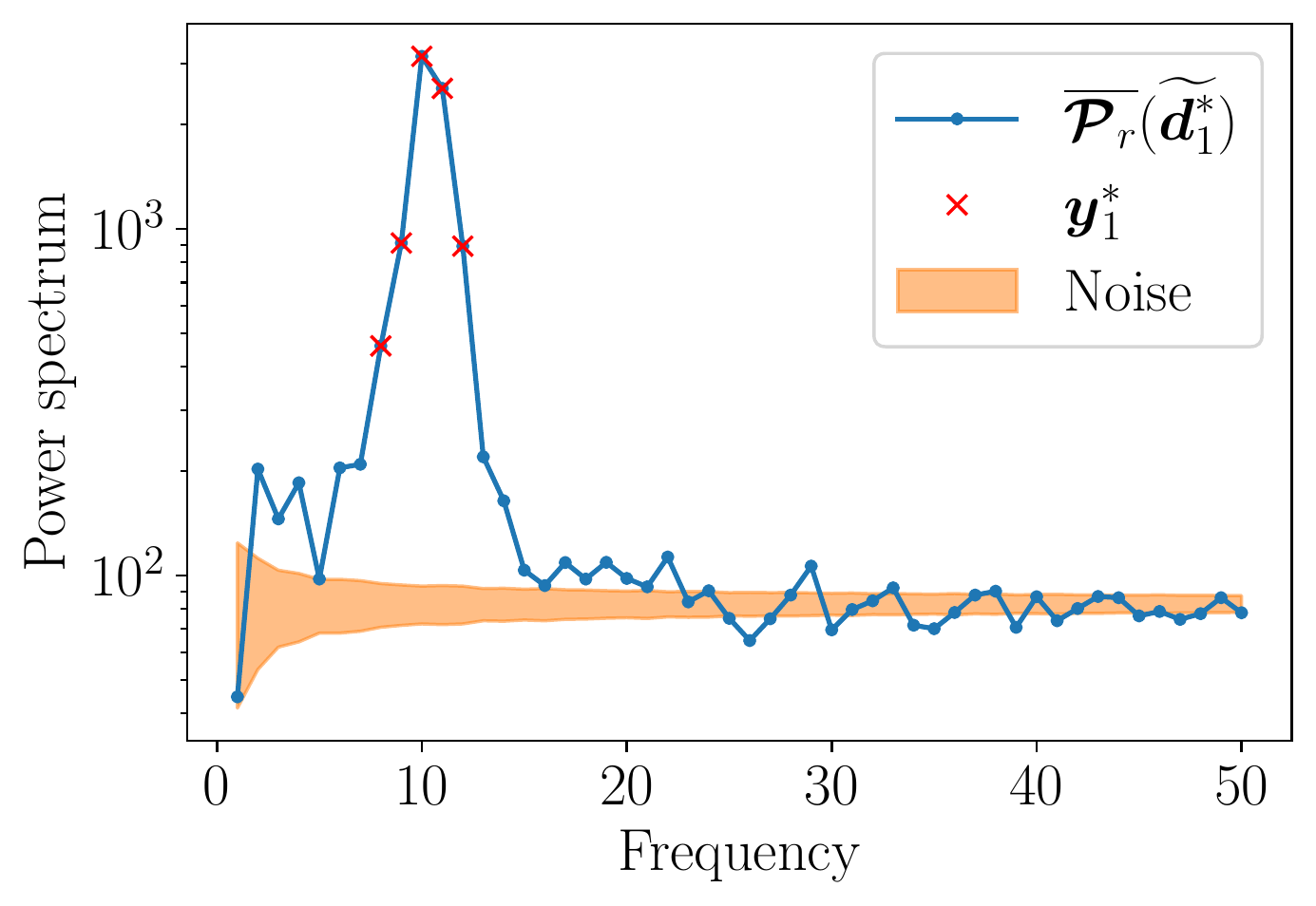}} \\
        \includegraphics[width = 0.18\linewidth]{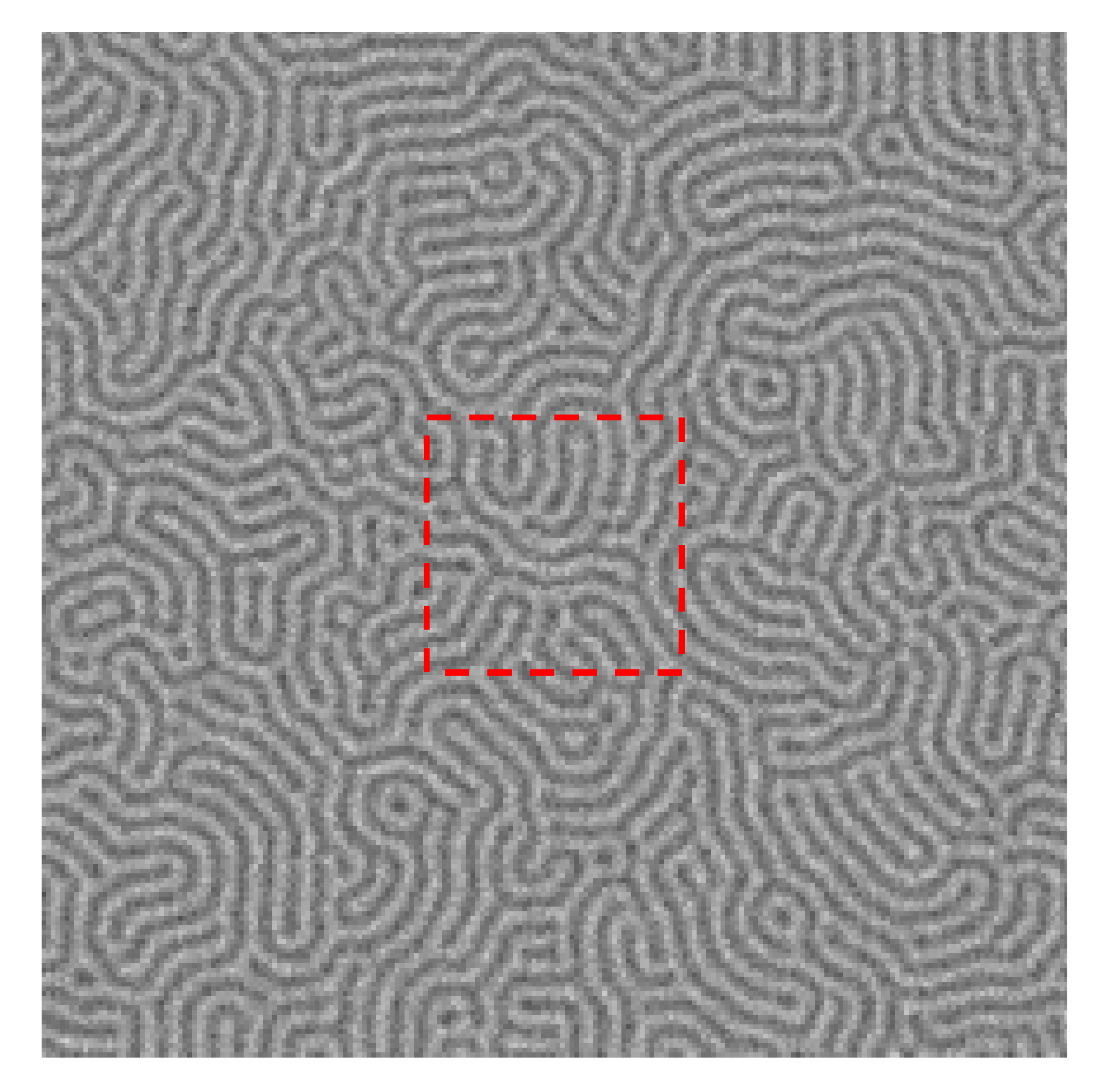} & \includegraphics[width = 0.18\linewidth]{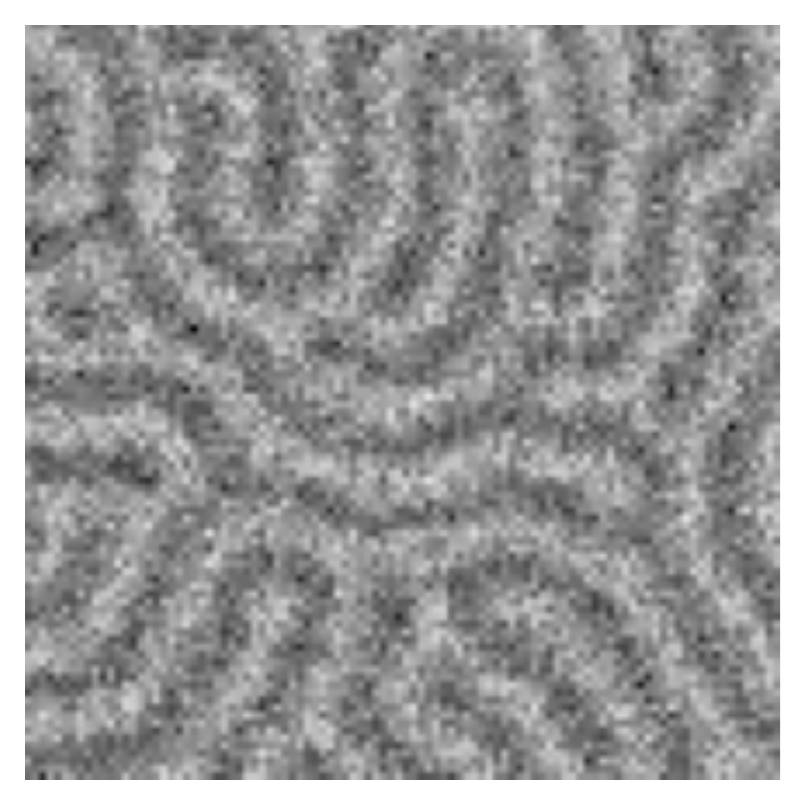} & \includegraphics[height = 0.18\linewidth]{colorbar_colorbar_image_25.pdf} & \\
         $\displaystyle \data_2^* \in[0,1]^{400\times 400}$ & $\displaystyle \widetilde{\data_2^*} \in [0,1]^{100\times 100}$ & & \multirow{2}{*}{\includegraphics[width = 0.34\linewidth]{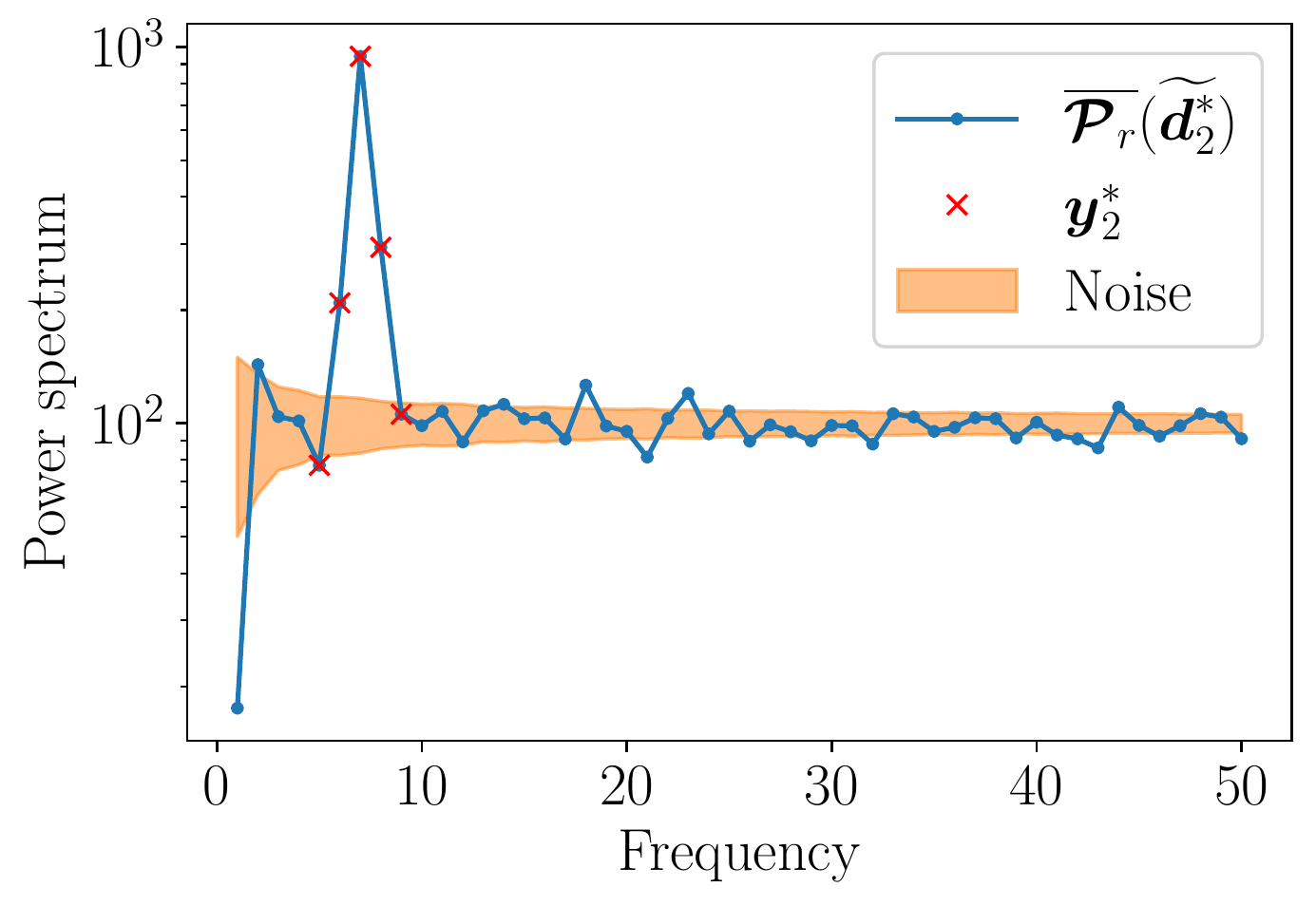}} \\
         \includegraphics[width = 0.18\linewidth]{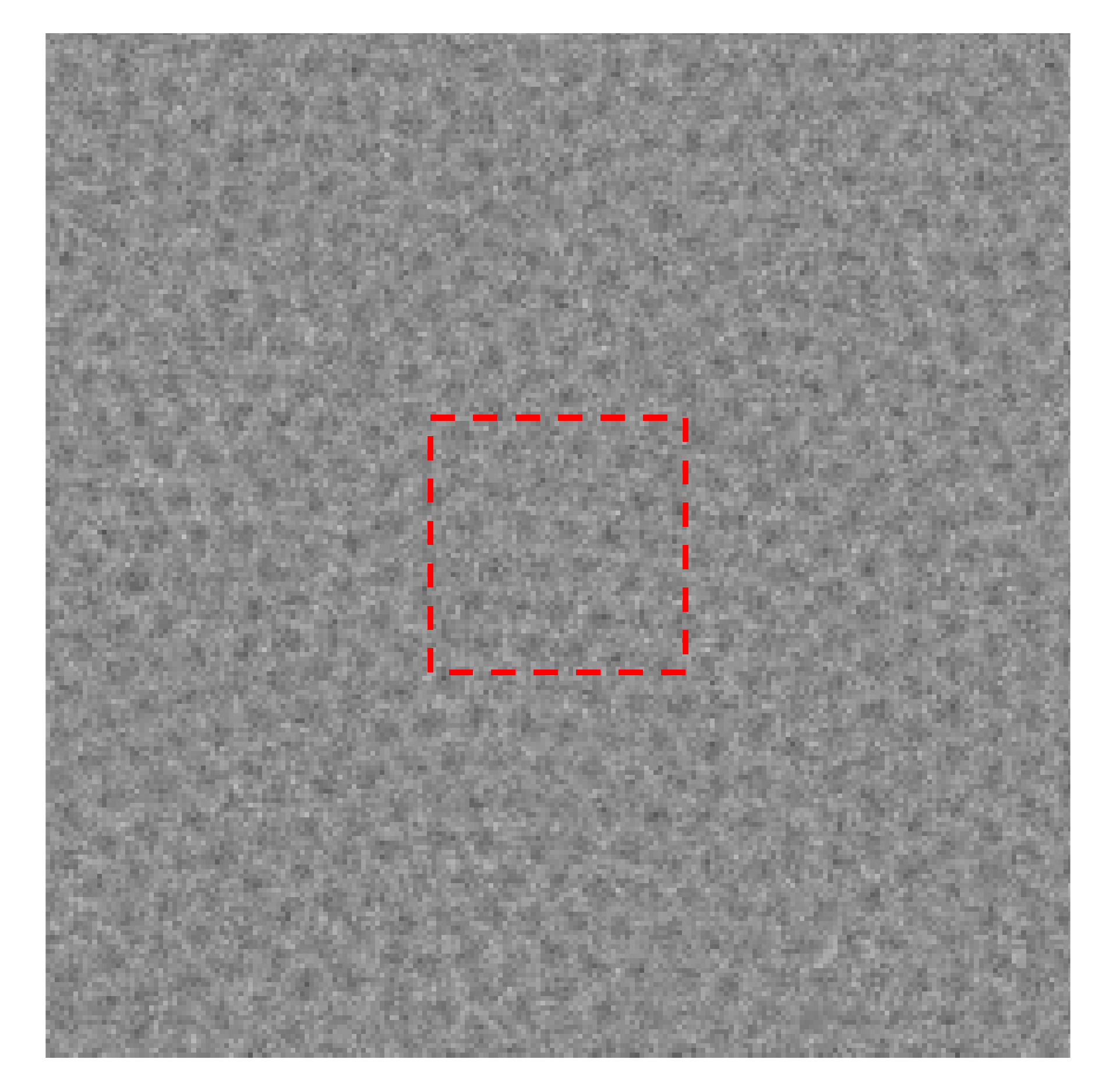} &\includegraphics[width = 0.18\linewidth]{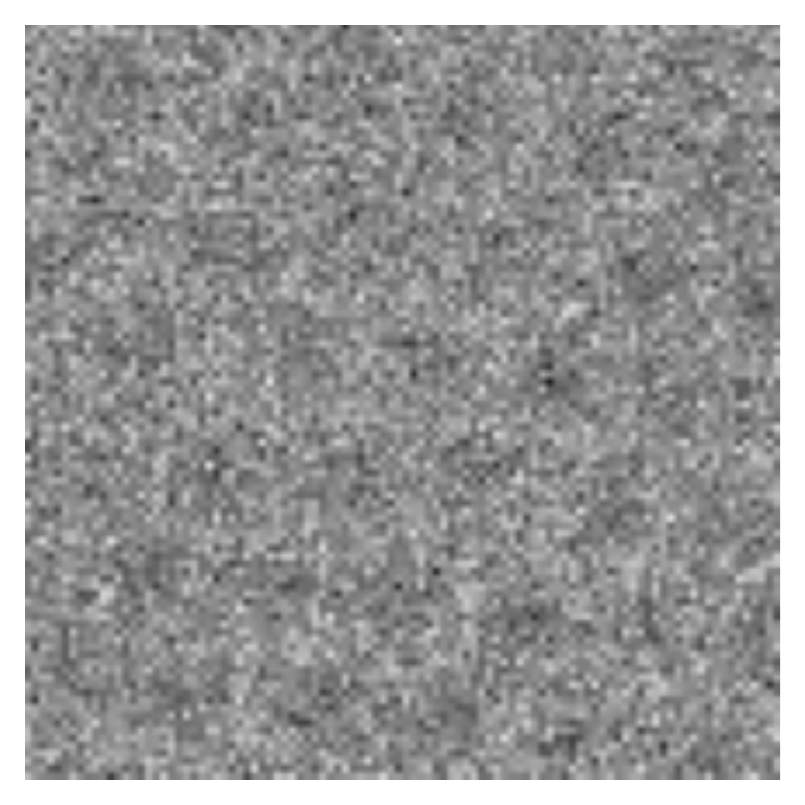} &\includegraphics[height = 0.18\linewidth]{colorbar_colorbar_image_25.pdf} & \\
    \end{tabular}
\addtolength{\tabcolsep}{5pt} 
    \caption{Sub-images $\widetilde{\data_j^*}$, $j= 1,2$, of size $100\times 100$ (\emph{red dashed line and middle column}) are extracted and used for the second stage inference. For the second stage inference, $5$ entries of the smooth AAPS near the entry with the largest magnitude value are extracted as data $\y_j^*\in\R_+^5$, $j=1,2$ (\emph{right column}).}
    \label{fig:synthetic_data_spectrum}
\end{figure}

This subsection presents numerical results for Bayesian model calibration via the pseudo-marginal method and the Ohta--Kawasaki model. Centered sub-images $\widetilde{\data_j^*}$, $j= 1,2$, of size $100\times 100$ are extracted, and $5$ entries of their smooth AAPS near the entry with the largest magnitude are used as the data $\y_j^*\in\R_+^5$, $j=1,2$ for Bayesian inference; see Figure~\ref{fig:synthetic_data_spectrum}. We construct the integrated likelihood estimator in \eqref{eq:mc_estimator} using $n_{z} = 20$ and simulate multiple MCMC chains to obtain $10,000$ MCMC samples (after burn-in) for each problem. The average acceptance rate for the adaptive metropolis is $3.9\%$ from the stripe pattern and $9.6\%$ from the spot pattern. The simulated Markov chains have total effective sample size estimates~\cite{vats2019multivariate} of around $405$ for the stripe pattern and $440$ for the spot pattern.

\begin{figure}[!hbt]
\centering
    \addtolength{\tabcolsep}{-4pt} 
    \begin{tabular}{c c}
    The stripe pattern in Figures \ref{fig:synthetic_data} and \ref{fig:synthetic_data_spectrum} & The spot pattern in Figures~\ref{fig:synthetic_data} and \ref{fig:synthetic_data_spectrum}\\
    $\langle\boldsymbol{D}\rangle = \langle\data_1^*\rangle$ and $\by = \y_1^*$ & $\langle\boldsymbol{D}\rangle = \langle\data_2^*\rangle$ and $\by = \y_2^*$ \\
        \includegraphics[width = 0.49\linewidth]{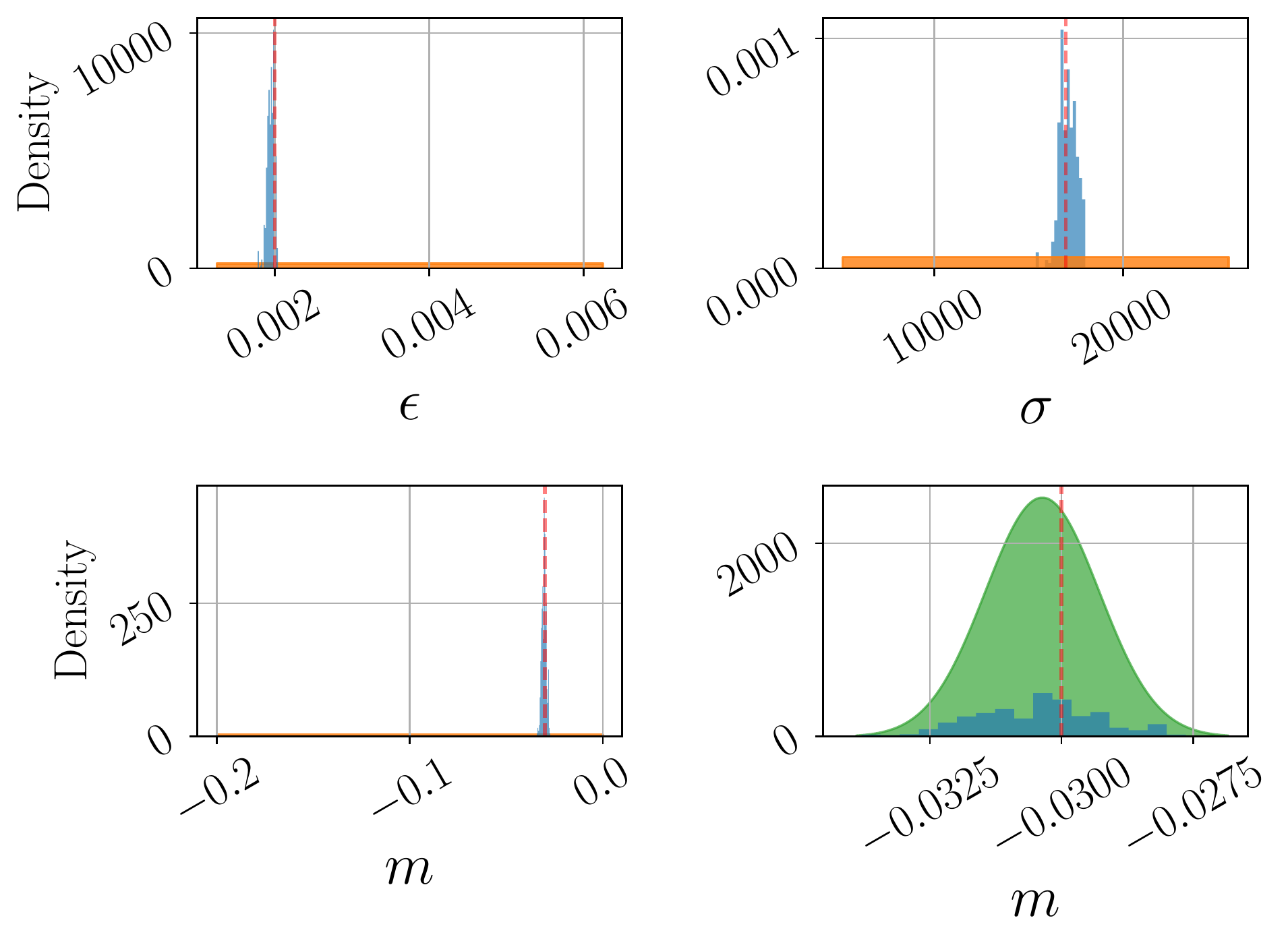} &  \includegraphics[width = 0.49\linewidth]{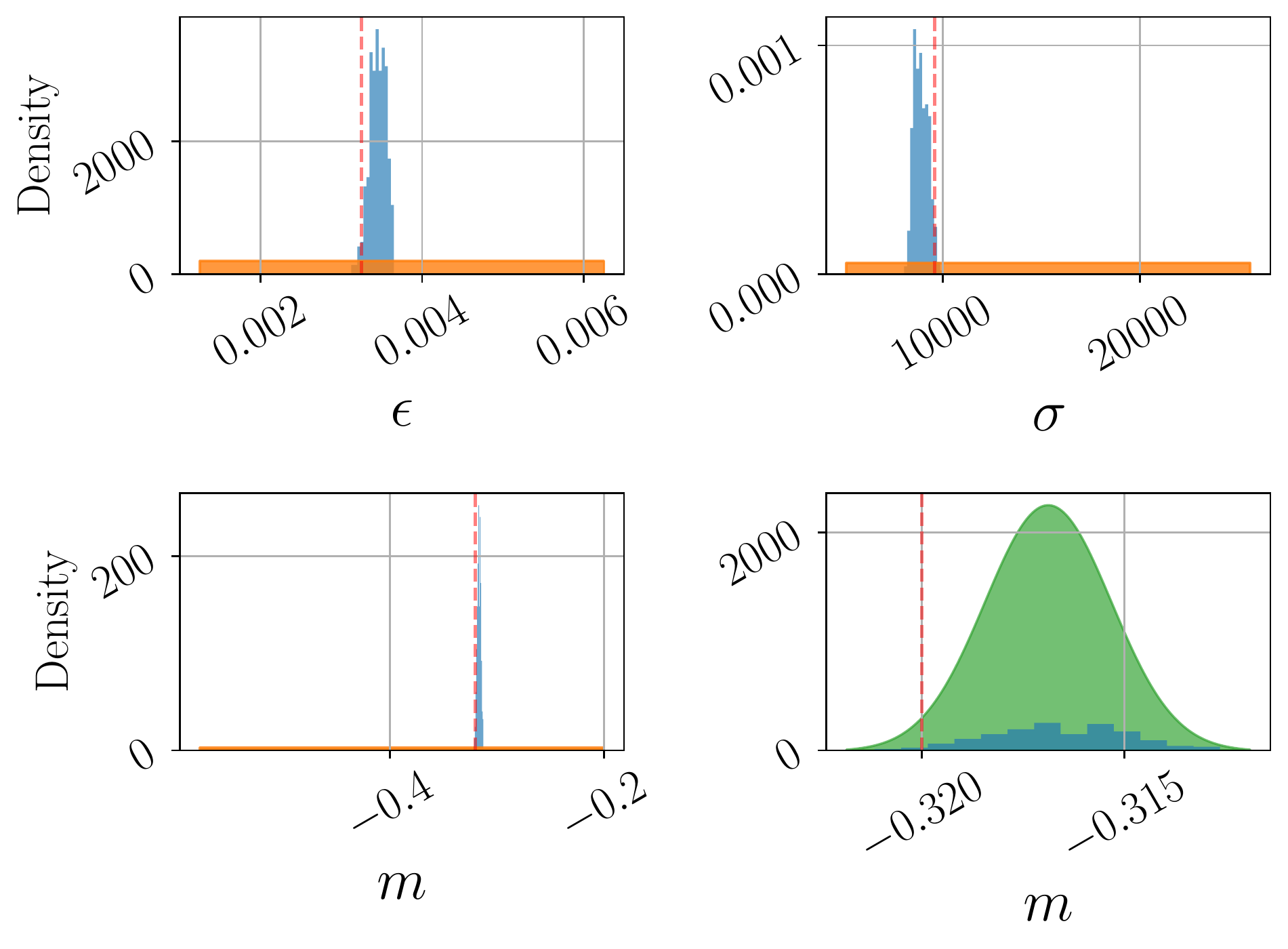}
    \end{tabular}
    \begin{tabular}{>{\centering\arraybackslash}p{0.24\linewidth} >{\centering\arraybackslash}p{0.24\linewidth} >{\centering\arraybackslash}p{0.24\linewidth} >{\centering\arraybackslash}p{0.24\linewidth} }
        Prior & Phase-informed prior & Posterior & Reference\\
        \includegraphics[width = 0.15\linewidth]{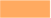} $\pi_{\bx}$ & \includegraphics[width = 0.15\linewidth]{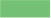} $\pi_{M|\langle\boldsymbol{D}\rangle}$ & \includegraphics[width = 0.15\linewidth]{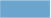} $\pi_{\bx|\by, \langle\boldsymbol{D}\rangle}$ & \includegraphics[width = 0.15\linewidth]{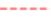} $\x^*$
    \end{tabular}
    \addtolength{\tabcolsep}{4pt} 
    \caption{The marginal prior densities, the phase-informed prior conditioned on $\langle\boldsymbol{D}\rangle = \langle\boldsymbol{d}_1^*\rangle$ or $\langle\boldsymbol{D}\rangle = \langle\boldsymbol{d}_2^*\rangle$, and sample-based estimates of the marginal posterior densities conditioned on $\by = \y_1^*$ (\emph{left, the stripe pattern in Figure~\ref{fig:synthetic_data}}) or $\by = \y_2^*$ (\emph{right, the spot pattern in Figure~\ref{fig:synthetic_data}}). The posterior sampling is performed using the pseudo-marginal method and solving the OK model for conditional likelihood evaluations. The red dashed line indicates the parameter values that generated the synthetic images $\x^*$. The marginal posterior distributions for $E$ and $M$ are extremely concentrated and may be difficult to visualize together with their marginal prior distributions.}
    \label{fig:marginal_posterior_known}
\end{figure}

\begin{figure}[!ht]
    \centering
        \addtolength{\tabcolsep}{-4pt} 
    \begin{tabular}{cc}
        The stripe pattern in Figures \ref{fig:synthetic_data} and \ref{fig:synthetic_data_spectrum} & The spot pattern in Figures~\ref{fig:synthetic_data} and \ref{fig:synthetic_data_spectrum}\\
        $\pi_{\bx|\by,\langle \boldsymbol{D}\rangle}\left(\cdot| \y_1^*,\langle\data_1^*\rangle\right)$ & $\pi_{\bx|\by,\langle \boldsymbol{D}\rangle}\left(\cdot| \y_2^*,\langle\data_2^*\rangle\right)$ \\
        \includegraphics[width =0.49\linewidth]{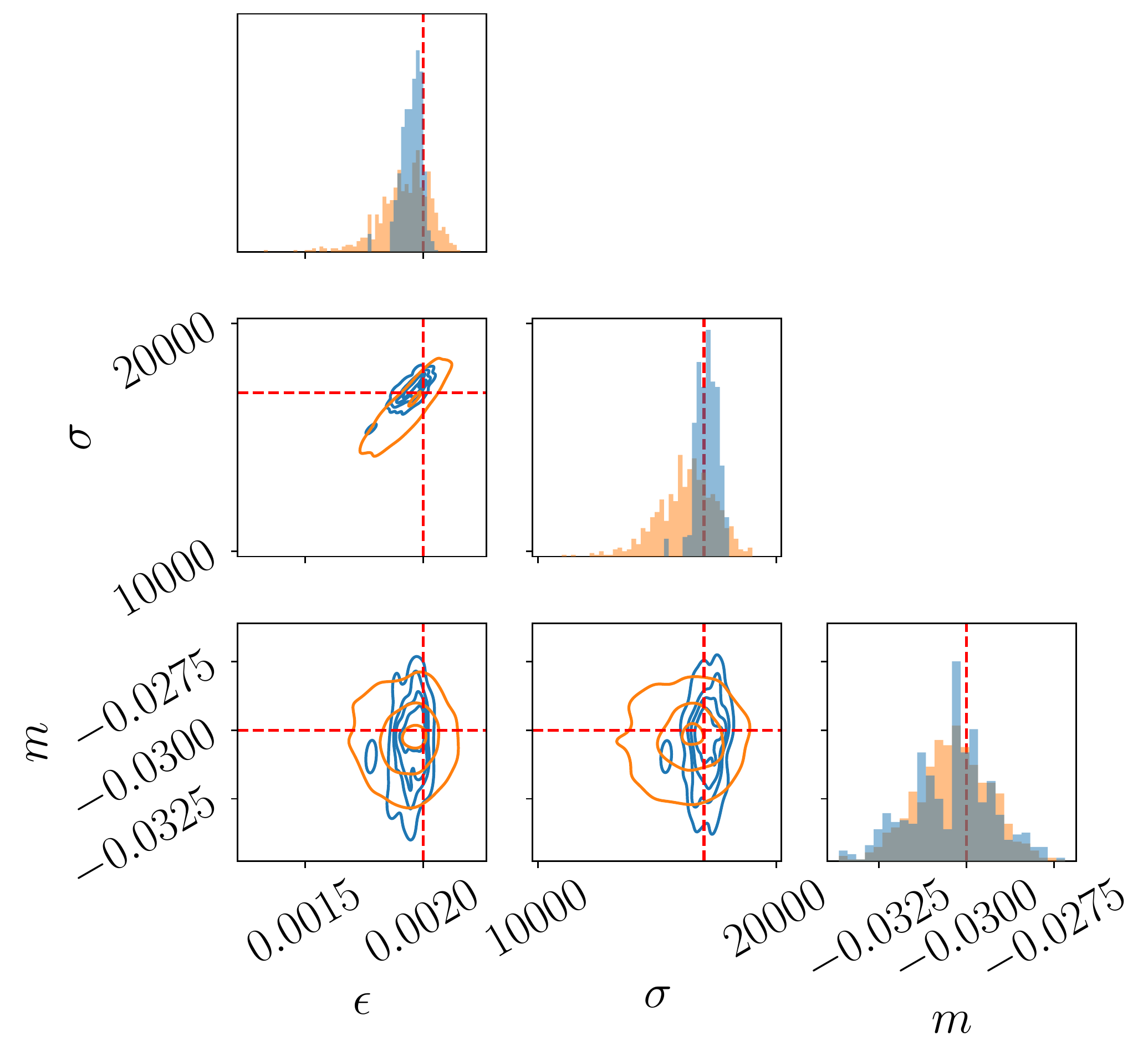} & \includegraphics[width =0.49\linewidth]{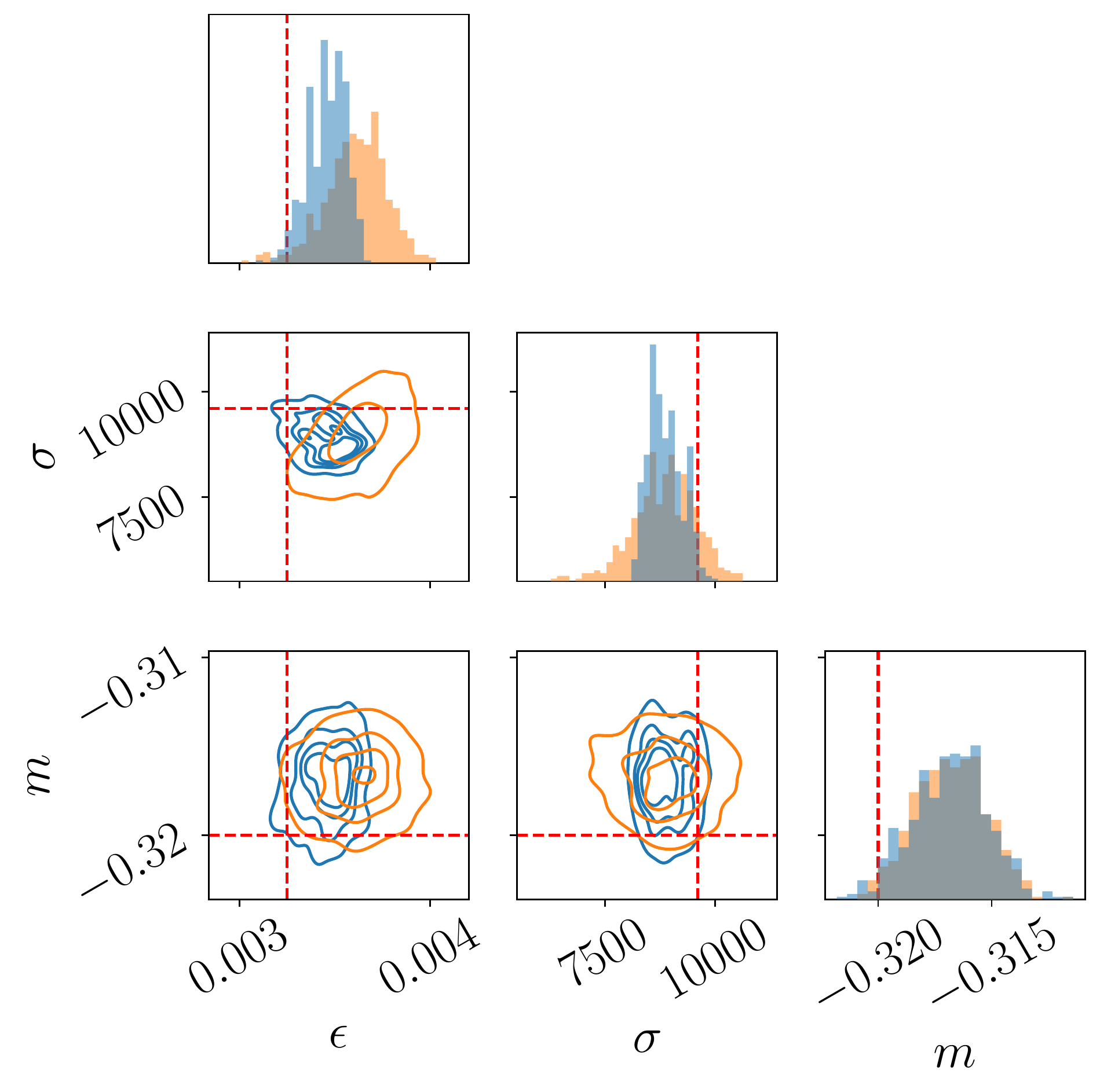}\\
    \end{tabular}
    \begin{tabular}{>{\centering\arraybackslash}p{0.3\linewidth} >{\centering\arraybackslash}p{0.3\linewidth} >{\centering\arraybackslash}p{0.3\linewidth}}
        \includegraphics[width = 0.1\linewidth]{num_results_c0.png} Model & \includegraphics[width = 0.1\linewidth]{num_results_c1.png} Neural network surrogate  &\includegraphics[width=0.1\linewidth]{num_results_red.png} Reference
    \end{tabular}
        \addtolength{\tabcolsep}{4pt} 
    \caption{The corner plots of the sample-based estimates of the posterior distribution $\pi_{\bx|\by, \langle\boldsymbol{D}\rangle}$ with $\by = \y_1^*$ and $\langle\boldsymbol{D}\rangle = \langle\boldsymbol{d}_1^*\rangle$ (\emph{left, the stripe pattern in Figure~\ref{fig:synthetic_data}}) or $\by = \y_2^*$ and $\langle\boldsymbol{D}\rangle = \langle\boldsymbol{d}_2^*\rangle$  (\emph{right, the spot pattern in Figure~\ref{fig:synthetic_data}}) generated via the pseudo-marginal method and simulating the OK model (\emph{blue}) and surrogate-based integrated likelihood evaluations (\emph{orange}). The red dashed line indicates the parameter values that generated the synthetic images.}
    \label{fig:joint_posterior_known}
\end{figure}

In Figures~\ref{fig:marginal_posterior_known} and \ref{fig:joint_posterior_known}, we visualize sample-based estimates of the posterior distributions generated via solving the OK model for conditional likelihood evaluations. We observe that the first stage inference results in a significant decrease in the uncertainty in $M$, and the second stage inference results in the same for $E$ and $\Sigma$. When comparing the marginal posterior distributions for the two sets of synthetic images, we see that the proposed Bayesian model calibration procedure leads to much more reduced uncertainty in $E$ for the stripe pattern with a lower signal-to-noise ratio. We note that $\epsilon$ scales the interfacial energy (the $H^1$-seminorm of order parameters) in the OK model~\eqref{eq:ok_energy}, and it has a large influence on the formation of phase interfaces in the simulated Di-BCP patterns. We expect the mutual information between the image data and $E$ to be low when the observed phase interfaces are polluted by white noise. Furthermore, the reference parameter values that generated the synthetic images $\x_j^*$, $j = 1,2$, are recovered well. The phase-informed prior dominates the inference of the marginal parameter $M$, for which we observe little change when conditioned on $\y^*_j$, $j= 1,2$.

\begin{figure}[!hbt]
    \centering
    \addtolength{\tabcolsep}{-6pt} 
    \begin{tabular}{D D D C D D D C D D D E}
        \multicolumn{3}{c}{\makecell{Prior\\predictive samples}} &  &\multicolumn{3}{c}{\makecell{Phase-informed prior\\predictive samples}}&  & \multicolumn{3}{c}{\makecell{Posterior\\ predictive samples}} & \\
        \includegraphics[width = \linewidth]{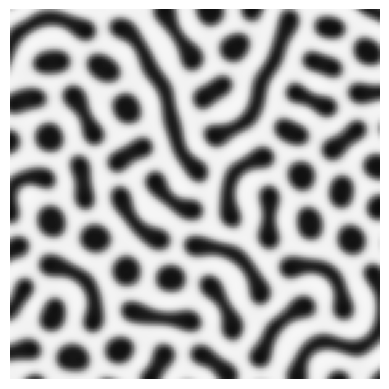} & \includegraphics[width = \linewidth]{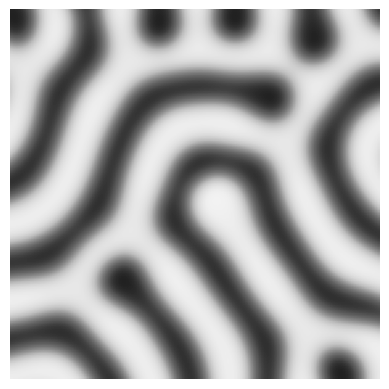} & \includegraphics[width = \linewidth]{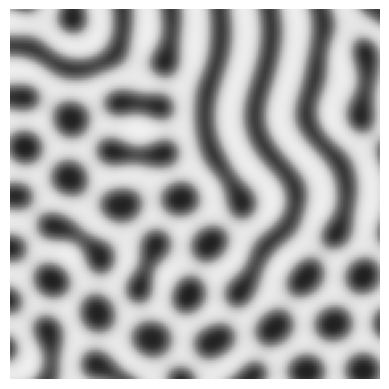} & \scalebox{1.2}{$\xrightarrow{\langle\data_1^*\rangle}$} & \includegraphics[width = \linewidth]{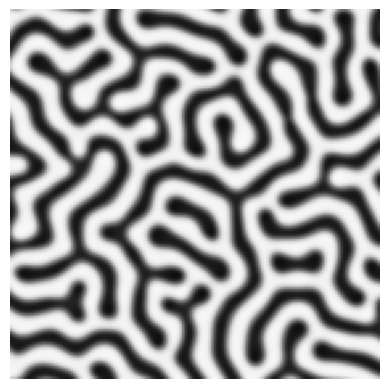} & \includegraphics[width = \linewidth]{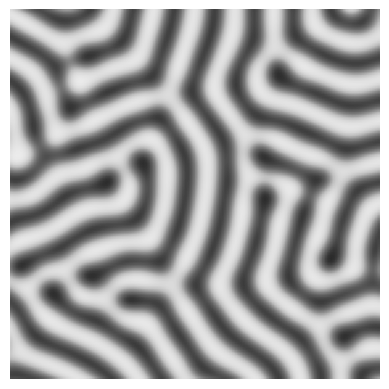} & \includegraphics[width = \linewidth]{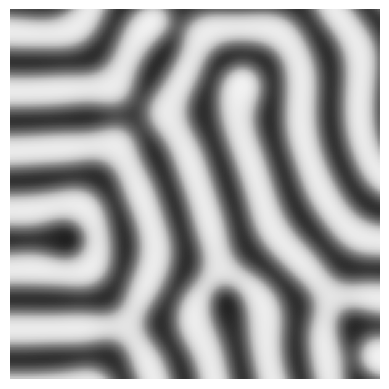} & \scalebox{1.2}{$\xrightarrow{\y_1^*}$} & \includegraphics[width = \linewidth]{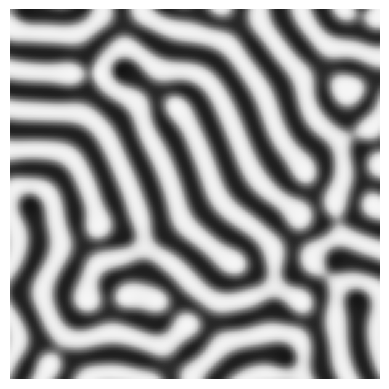}  & \includegraphics[width = \linewidth]{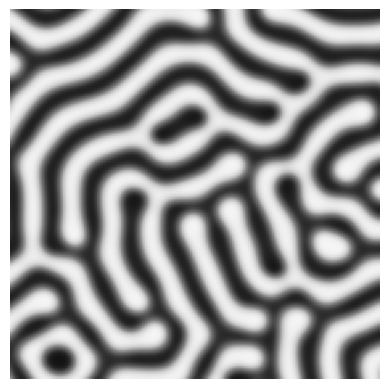} & \includegraphics[width = \linewidth]{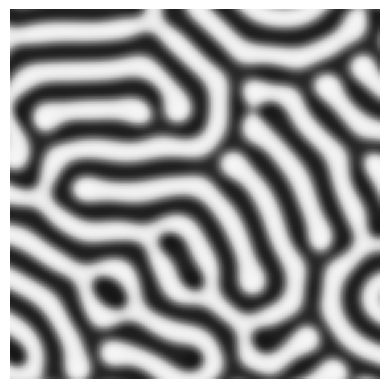} & \includegraphics[width = \linewidth]{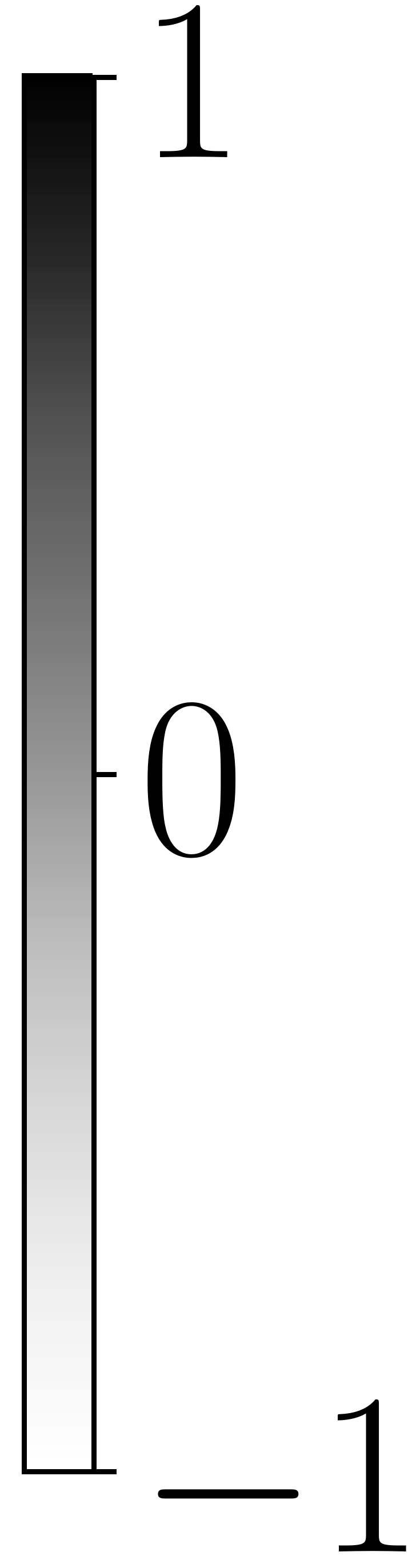}\\
        \includegraphics[width = \linewidth]{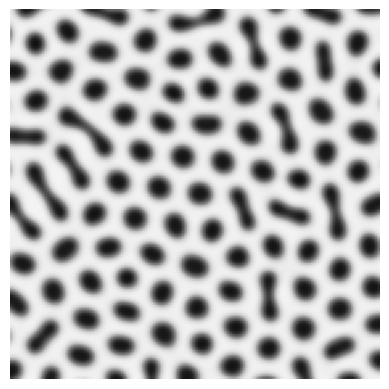} & \includegraphics[width = \linewidth]{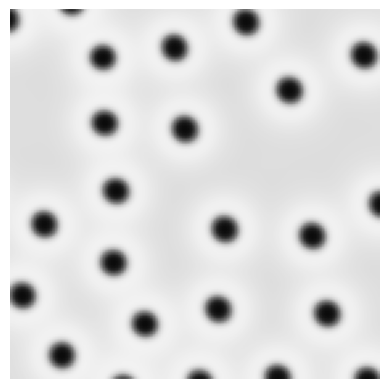} & \includegraphics[width = \linewidth]{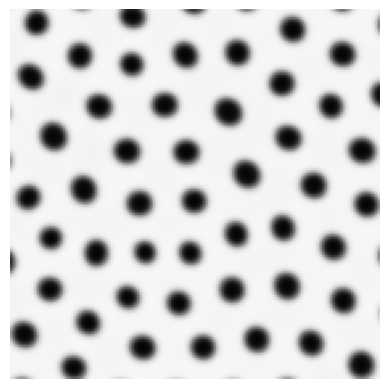} & \scalebox{1.2}{$\xrightarrow{\langle\data_2^*\rangle}$} & \includegraphics[width = \linewidth]{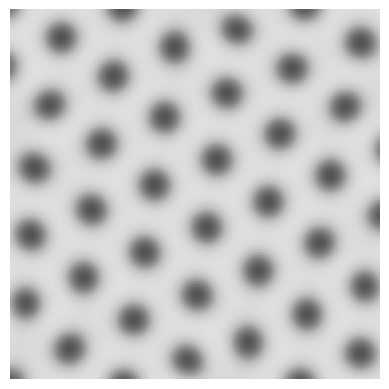} & \includegraphics[width = \linewidth]{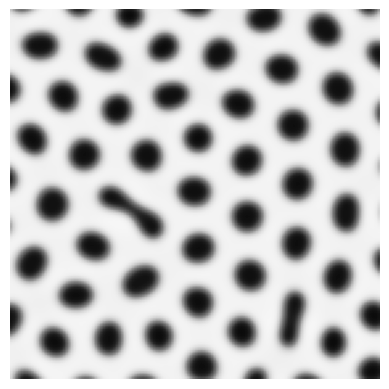} & \includegraphics[width = \linewidth]{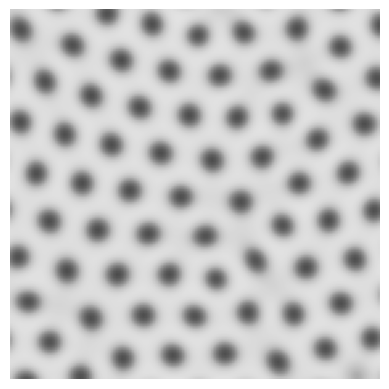} & \scalebox{1.2}{$\xrightarrow{\y_2^*}$} & \includegraphics[width = \linewidth]{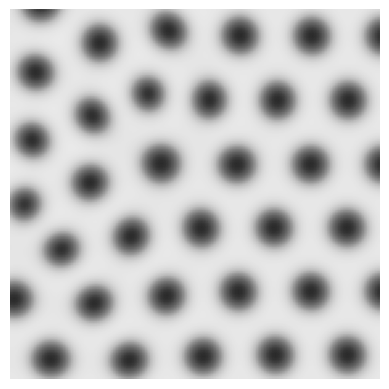} & \includegraphics[width = \linewidth]{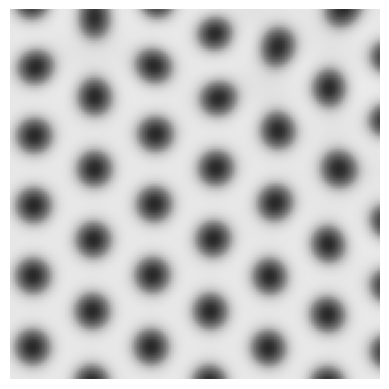} &\includegraphics[width = \linewidth]{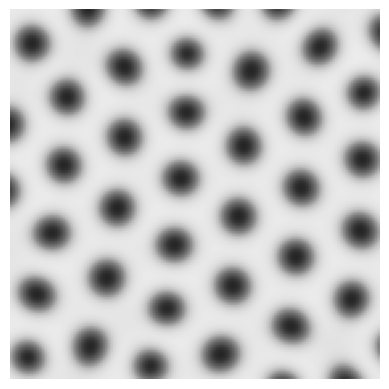} & \includegraphics[width = \linewidth]{colorbar_colorbar_op_36.pdf}
    \end{tabular}
        \addtolength{\tabcolsep}{6pt} 
    \caption{Predictive samples generated via simulating the OK model at parameter samples from the prior distributions (\emph{left}), the phase-informed prior distributions (\emph{middle}), and the posterior distributions (\emph{right}). The samples from the posterior distributions are generated by the pseudo-marginal method and simulating the OK model for conditional likelihood evaluations. These distributions are visualized in Figures \ref{fig:marginal_posterior_known} and \ref{fig:joint_posterior_known}. The predictive samples are generated on a domain size consistent with the sub-image size of $100\times 100$ shown in Figure~\ref{fig:synthetic_data_spectrum}.}
    \label{fig:predictive_samples_known}
\end{figure}

In Figure \ref{fig:predictive_samples_known}, we plot samples of the prior predictive distributions, the phase-informed prior distributions, and the posterior predictive samples generated on a domain size consistent with the sub-image size. After assimilating $\langle \data_j^*\rangle$, $j=1,2$, we observe a significant reduction of the uncertainty in predicting the ratio of stripes and spots (\emph{top row and middle column}) and the density of the spots (\emph{bottom row and middle column}). After assimilating the magnitude values of the selected smooth AAPS entries $\y^*_j$, $j=1,2$, we observe a significant reduction of uncertainty in predicting the periodicity length and feature size (e.g., the width of the stripes and radius of the spots). These observations agree with our analysis in Section~\ref{sec:bayesian_model_calibration}.

\subsubsection{Posterior sampling via surrogate-based integrated likelihood evaluations}

We present posterior sampling results using the surrogate parameter-to-spectrum map and the data $\y^*_j$, $j=1,2$, described in Section~\ref{subsubsec:sampling_model_known}. The integrated likelihood is now defined with the surrogate parameter-to-spectrum map as in \eqref{eq:likelihood_integration} and \eqref{eq:surrogate_likelihood}. The integrated likelihood is estimated using $40$ Gauss--Hermite quadrature points. 

In Figure~\ref{fig:joint_posterior_known}, we visualize sample-based estimates of posterior distributions generated using the surrogate-based integrated likelihood evaluations (\emph{orange}) along with the reference posterior distribution estimated using the pseudo-marginal method and model simulations (\emph{blue}). The reference parameter values that generated the synthetic images $\x_j^*$, $j = 1,2$, are recovered well via the surrogate-based sampling. The anisotropy of the marginal posterior distribution for $(E, \Sigma)$ inferred from the stripe pattern (\emph{left}) is also captured. However, we observe decreased uncertainty reduction for the length scale--related parameter $E$ and $\Sigma$ for both numerical examples. This phenomenon is likely caused by the normal approximation with independent components introduced in~\eqref{eq:surrogate}.

We note that the $200,000$ model simulations for characterizing a single posterior distribution in the numerical examples are $5$ times the $40,000$ model simulations for generating samples that are used to construct the surrogate parameter-to-spectrum map; see Section~\ref{subsec:surrogate_results}. In practice, however, one may require fewer posterior samples than needed to finely resolve features of the posterior, in which case the discrepancy in computational cost may not be as dramatic. Nevertheless, this comparison in computational cost demonstrates the necessity of the surrogate, as it transfers the online computational burden of repeated model simulations for each calibration task to one-time offline sample generation.

\subsection{Bayesian model calibration with uncertain nuisance parameters}
\begin{figure}[!hbt]
    \centering
    \addtolength{\tabcolsep}{-4pt} 
    \begin{tabular}{c c}
    The stripe pattern in Figures \ref{fig:synthetic_data} & The spot pattern in Figures~\ref{fig:synthetic_data}\\
    $\pi_{M,C_1,C_2|\langle \boldsymbol{D}\rangle}\left(\cdot| \langle\data_1^*\rangle\right)$ & $\pi_{M,C_1,C_2|\langle \boldsymbol{D}\rangle}\left(\cdot| \langle\data_2^*\rangle\right)$ \\
    \includegraphics[width =0.49\linewidth]{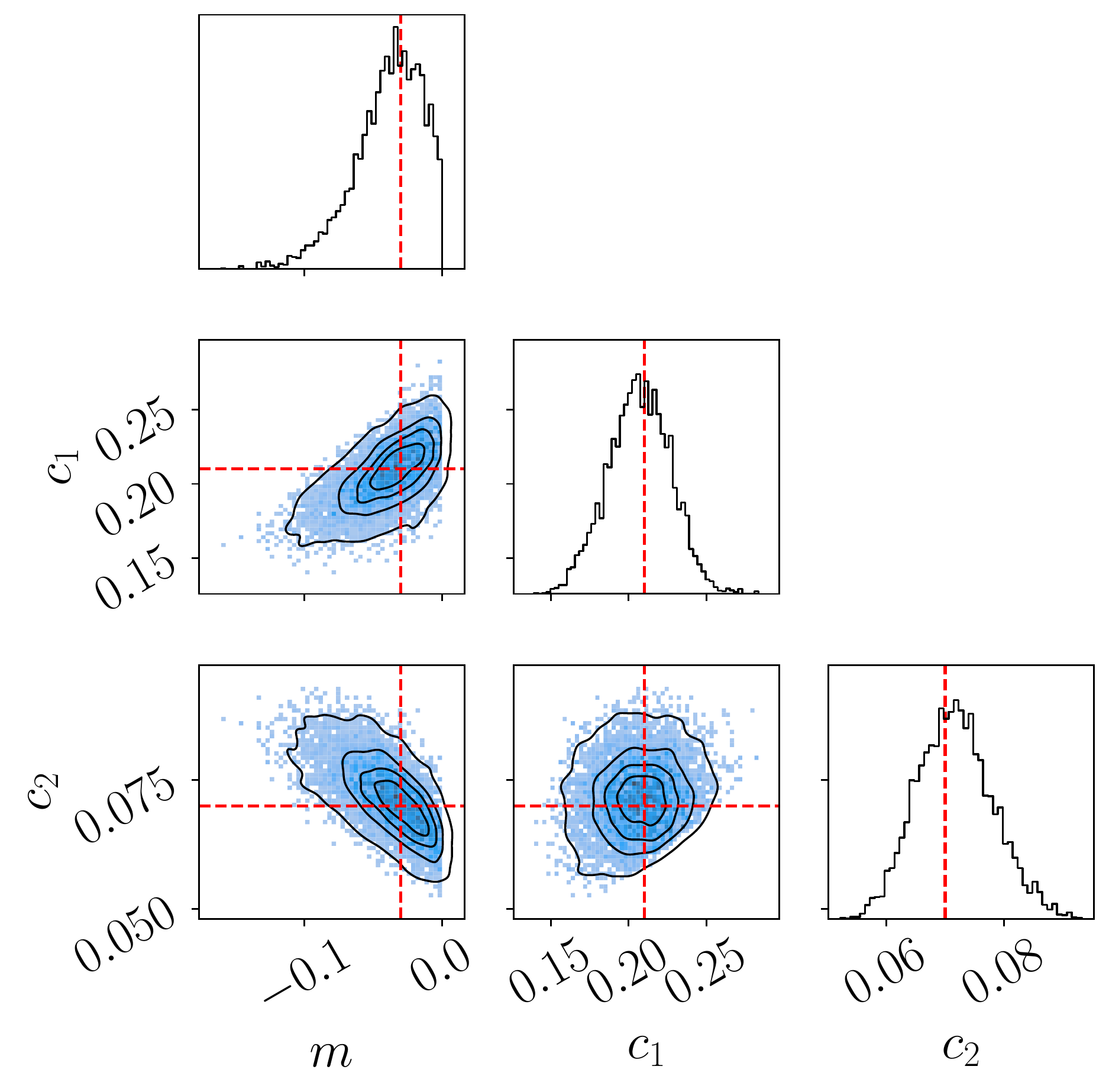} &
    \includegraphics[width =0.49\linewidth]{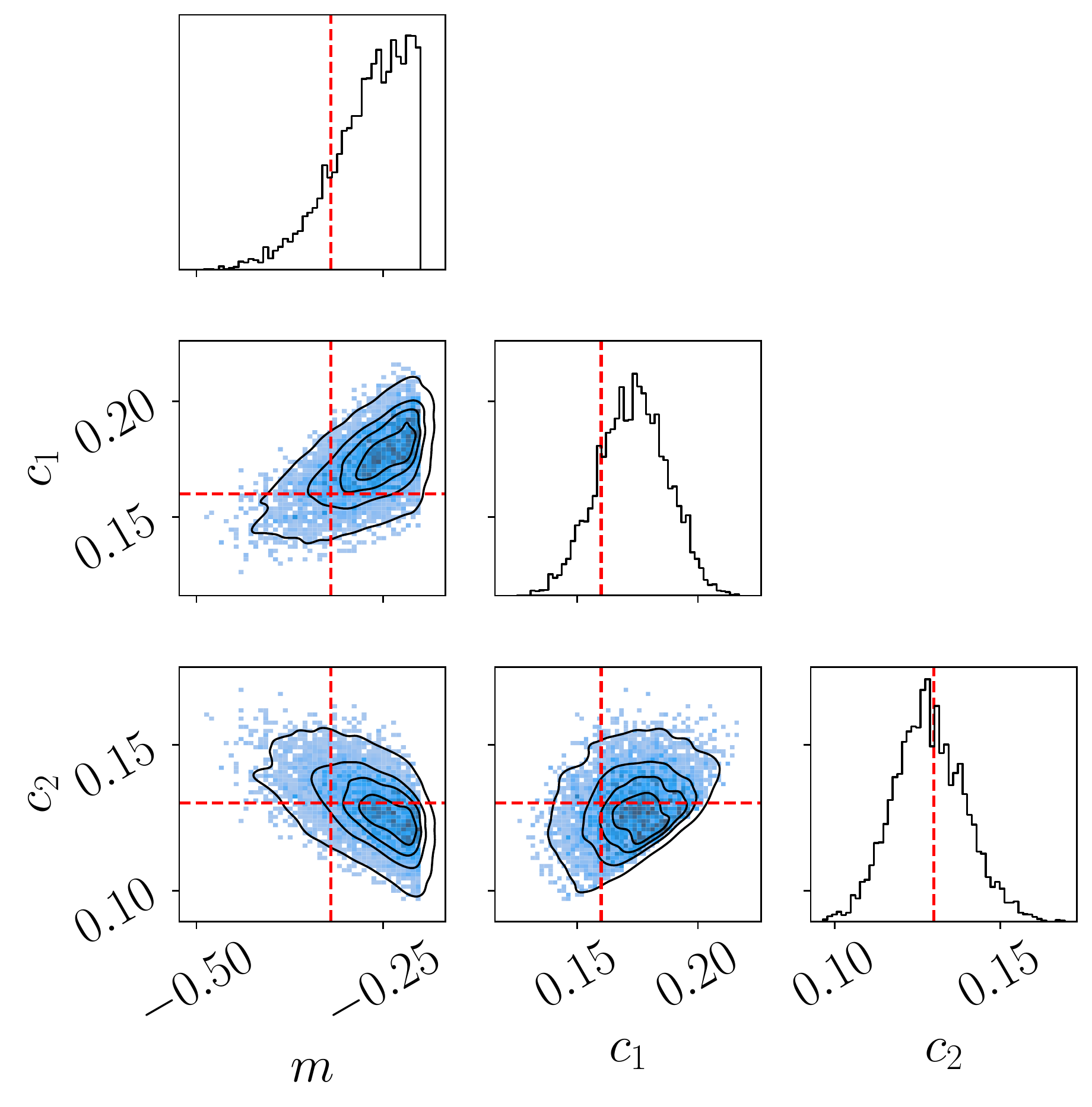}
    \end{tabular}
        \addtolength{\tabcolsep}{4pt} 
    \caption{The corner plots of the sample-based estimates of the phase-informed prior $\pi_{M,C_1,C_2|\langle\boldsymbol{D}\rangle}$ with $\langle\boldsymbol{D}\rangle = \langle\boldsymbol{d}_1^*\rangle$ (\emph{left, the stripe pattern in Figure~\ref{fig:synthetic_data}}) or $\langle\boldsymbol{D}\rangle = \langle\boldsymbol{d}_2^*\rangle$  (\emph{right, the spot pattern in Figure~\ref{fig:synthetic_data}}). The red dashed line indicates the parameter values that generated the synthetic images.}
    \label{fig:joint_phase-informed_prior_uncertain}
\end{figure}

Now we assume the nuisance parameters associated with contrast scaling ($c_1$), brightness shift ($c_2$), and blurring level $(\sigma_b)$ are uncertain, while the noise level ($\sigma_n$) is known\footnote{The noise level can be determined from the AAPS entries with high radial frequency as shown in Figure~\ref{fig:synthetic_data_spectrum}. In particular, we have $(\overline{\aaps}(\boldsymbol{d}^*))_j \approx M_1 M_2\sigma_n^2$ for high radial frequency AAPS entires.}. In particular, we assume the uncertain variables are independent and normally distributed:
\begin{equation*}
    \nu_{\bw} = \nu_{C_1}\otimes\nu_{C_2}\otimes\nu_{\Sigma_b}\otimes\delta_{\sigma_n^*}\,.
\end{equation*}
We use the truncated Gaussian distribution to construct a marginal prior distribution for each variable. The mean values of these distributions are randomly selected to fall within $5\%$ deviation from the reference $\w_j^*$, $j=1,2$. Moreover, we set a standard deviation of $10\%$ of the reference $\w_j^*$, $j=1,2$, for these marginal prior distributions. The goal of constructing such uncertainty is to mimic the experimental scenario where these parameters are only known up to limited confidence via instrument calibration. In Figure~\ref{fig:joint_phase-informed_prior_uncertain}, we present the sample-based estimates of the phase-informed prior distributions. We observe that the uncertainty in material contrast $C_1$ and brightness shift $C_2$ leads to a significant decrease in the uncertainty reduction of the marginal parameter $M$ compared to the case where these nuisance parameters are assumed to be known \eqref{eq:phase_informed_prior_known}; see the phase-informed prior distribution (\emph{green}) in Figure~\ref{fig:marginal_posterior_known} for a comparison. In particular, the uncertainty in the nuisance parameters leads to difficulty in recovering the reference value $m^*$ for the spot pattern with a low signal-to-noise ratio. The samples of $C_2$ are discarded to form the phase-informed prior to be used for the second stage inference using the smooth AAPS as data:
\begin{equation*}
    \pi_{\bx,C_1,\Sigma_b|\langle \boldsymbol{D}\rangle}(\x, c_1, \sigma_b|\langle \data^*\rangle) = \pi_{E}(\epsilon)\pi_{\Sigma}(\sigma)\pi_{M,C_1|\langle \boldsymbol{D}\rangle}(m,c_1|\langle\data^*\rangle)\pi_{\Sigma_b}(\sigma_b)\,.
\end{equation*}

\subsubsection{Posterior sampling via evaluating the parameter-to-spectrum surrogate}\label{subsec:surrogate_uncertain}
We present posterior sampling results using surrogate-based integrated likelihood evaluations and the phase-informed prior shown in Figure~\ref{fig:joint_phase-informed_prior_uncertain}. In particular, we randomly extract $10$ sub-images in each of $\data_j^*$, $j=1,2$, and compute the magnitude values for the first $25$ entries of the smooth AAPS for these sub-images as the data $\y^*_j\in\R_+^{250}$, $j = 1,2$, for Bayesian inference. These entries match the full output of the neural network surrogate described in Section~\ref{subsec:surrogate_results}. We formulate the inference problem using~\eqref{eq:bayes_rule_joint} and collect samples from the joint posterior $\pi_{\bx,C_1,\Sigma_b|\by,\left\langle D\right\rangle}(\cdot|\y^*_j, \langle\data_j^*\rangle)$, $j=1,2$.

\begin{figure}[!hbt]
    \centering
    \addtolength{\tabcolsep}{-4pt} 
    \begin{tabular}{c c}
    The stripe pattern in Figures \ref{fig:synthetic_data} & The spot pattern in Figures~\ref{fig:synthetic_data}\\
        \includegraphics[width = 0.49\linewidth]{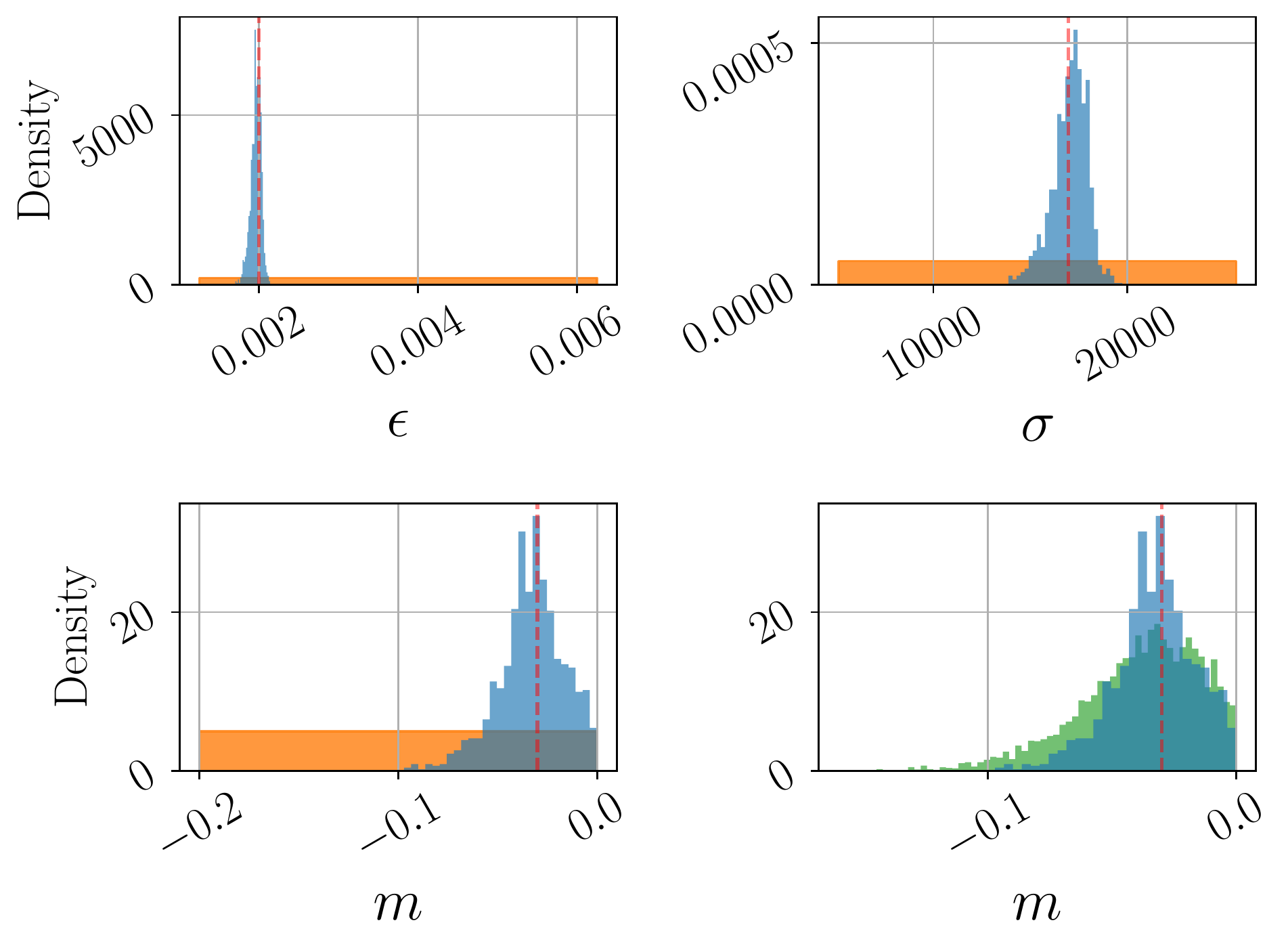} &  \includegraphics[width = 0.49\linewidth]{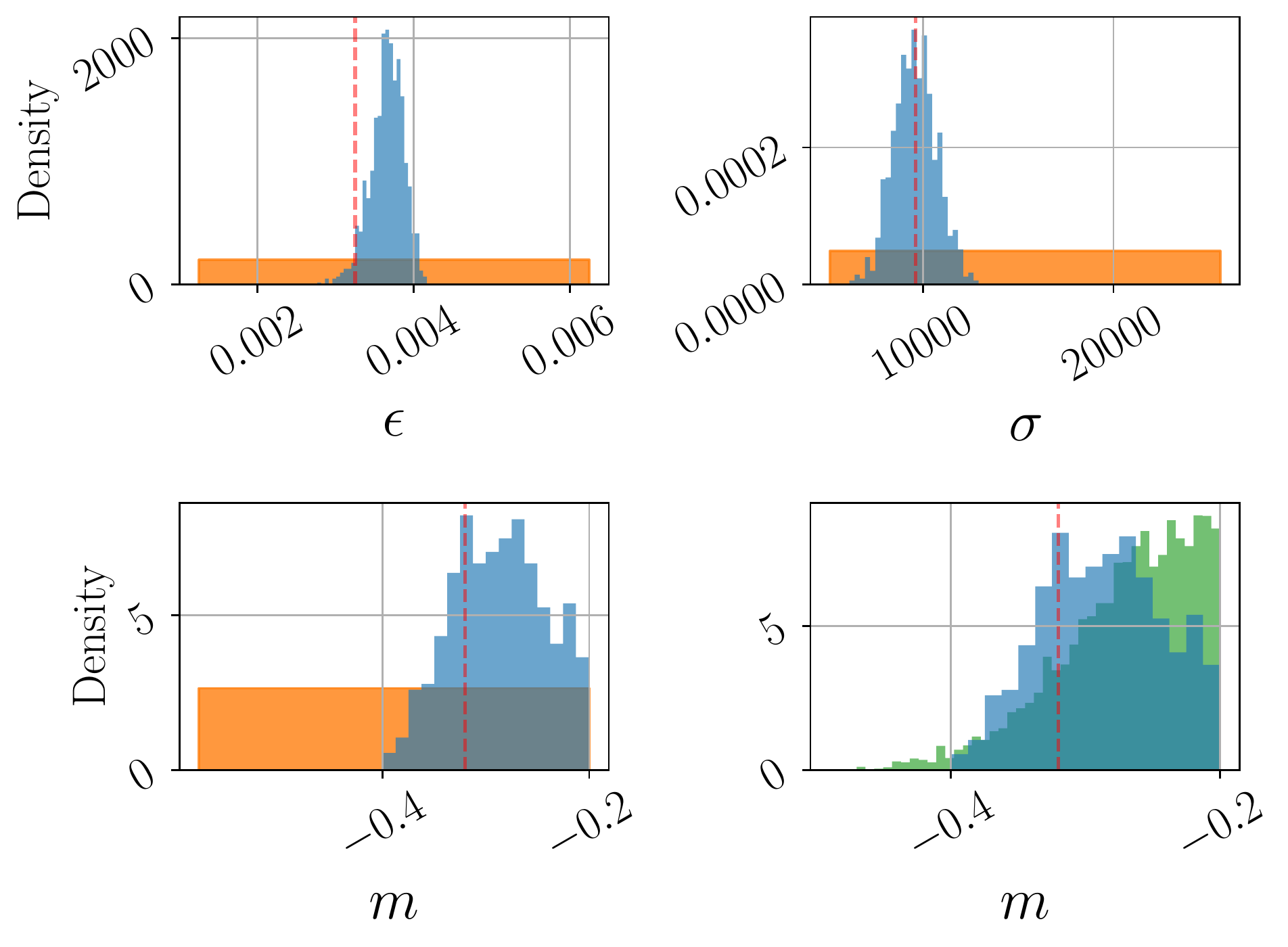}
    \end{tabular}
    \begin{tabular}{>{\centering\arraybackslash}p{0.24\linewidth} >{\centering\arraybackslash}p{0.24\linewidth} >{\centering\arraybackslash}p{0.24\linewidth} >{\centering\arraybackslash}p{0.24\linewidth} }
        Prior & Phase-informed prior & Posterior & Reference\\
        \includegraphics[width = 0.15\linewidth]{num_results_c1.png} $\pi_{\bx}$ & \includegraphics[width = 0.15\linewidth]{num_results_c2.png} $\pi_{M|\langle\boldsymbol{D}\rangle}$ & \includegraphics[width = 0.15\linewidth]{num_results_c0.png} $\pi_{\bx|\by, \langle\boldsymbol{D}\rangle}$ & \includegraphics[width = 0.15\linewidth]{num_results_red.png} $\x^*$
    \end{tabular}
    \addtolength{\tabcolsep}{4pt} 
    \caption{The marginal prior densities, the phase-informed prior conditioned on $\langle\boldsymbol{D}\rangle = \langle\boldsymbol{d}_1^*\rangle$ or $\langle\boldsymbol{D}\rangle = \langle\boldsymbol{d}_2^*\rangle$, and sample-based estimates of the marginal posterior densities conditioned on the first $25$ entries of the smooth AAPS for $10$ randomly selected sub-images of size $100\times 100$ in $\data_1^*$ (\emph{left, the stripe pattern in Figure~\ref{fig:synthetic_data}}) or in $\data_2^*$ (\emph{right, the spot pattern in Figure~\ref{fig:synthetic_data}}). The posterior sampling is performed using surrogate-based integrated likelihood evaluations. The red dashed line indicates the parameter values that generated the synthetic images $\x^*$.}
    \label{fig:marginal_posterior_uncertain}
\end{figure}

\begin{figure}[!hbt]
    \centering
        \addtolength{\tabcolsep}{-4pt} 
    \begin{tabular}{c c}
    The stripe pattern in Figures \ref{fig:synthetic_data} & The spot pattern in Figures~\ref{fig:synthetic_data}\\
        $\pi_{\bx,C_1,\Sigma_b|\by,\langle \boldsymbol{D}\rangle}\left(\cdot| \y_1^*,\langle\data_1^*\rangle\right)$ & $\pi_{\bx,C_1,\Sigma_b|\by,\langle \boldsymbol{D}\rangle}\left(\cdot| \y_2^*,\langle\data_2^*\rangle\right)$ \\
    \includegraphics[width = 0.49\linewidth]{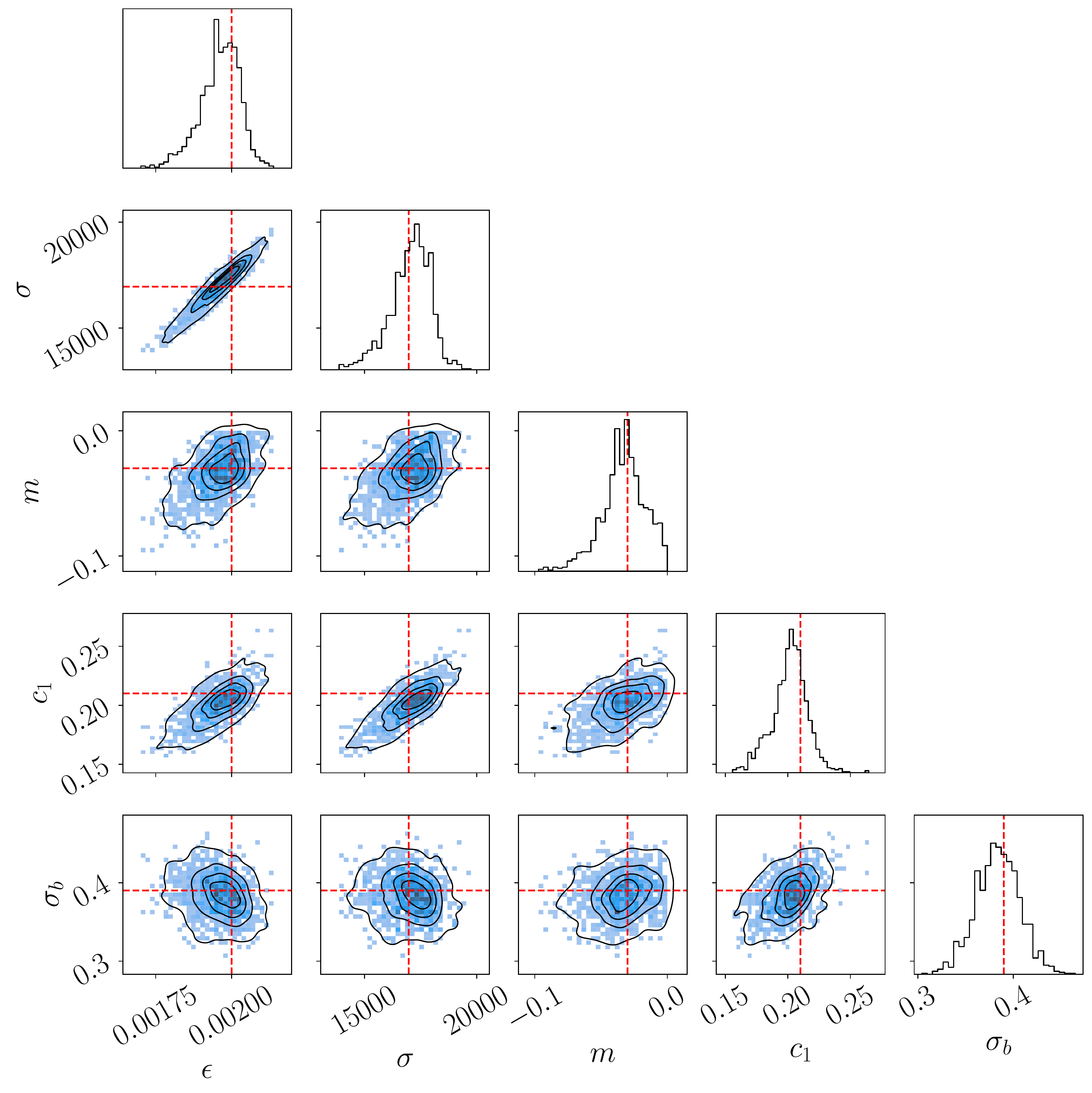} &\includegraphics[width = 0.49\linewidth]{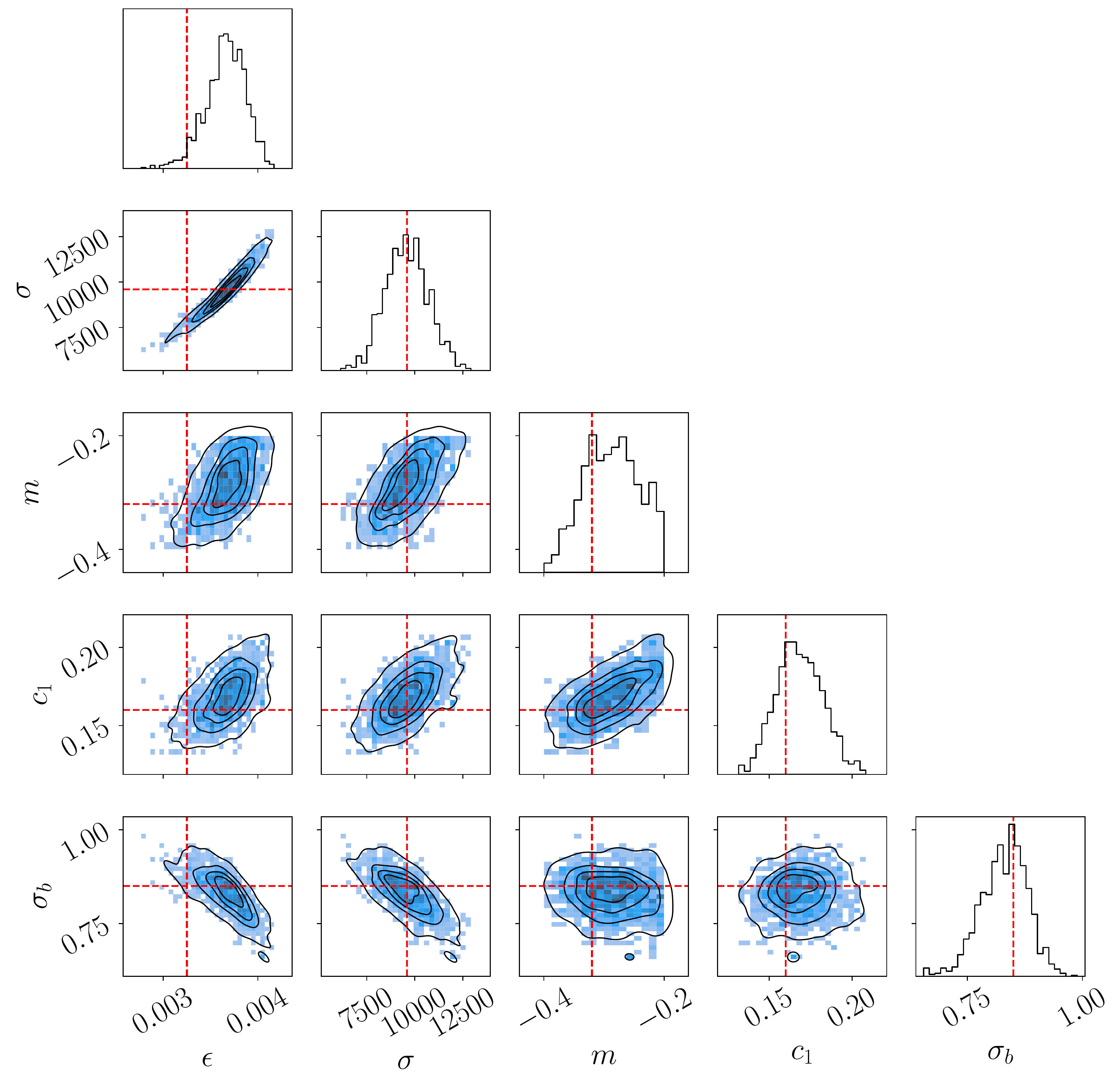}
    \end{tabular}{c c}
            \addtolength{\tabcolsep}{4pt} 
    \caption{The corner plots of the sample-based estimates of the joint posterior distribution $\pi_{\bx, C_1,\Sigma_b|\by, \langle\boldsymbol{D}\rangle}$ conditioned on the first $25$ entries of the smooth AAPS for $10$ randomly selected sub-images of size $100\times 100$ in $\data_1^*$ (\emph{left, the stripe pattern in Figure~\ref{fig:synthetic_data}}) or in $\data_2^*$ (\emph{right, the spot pattern in Figure~\ref{fig:synthetic_data}}). The posterior sampling is performed using surrogate-based integrated likelihood evaluations. The red dashed line indicates the parameter values that generated the synthetic images $\x^*$.}
    \label{fig:joint_posterior_uncertain}
\end{figure}

In Figures~\ref{fig:marginal_posterior_uncertain} and \ref{fig:joint_posterior_uncertain}, we visualize sample-based estimates of the posterior distributions generated via surrogate-based integrated likelihood evaluations. We again observe anisotropy in the marginal posterior distribution of $(E, \Sigma)$ for both the stripe and spot patterns. Such a consistent characteristic of the posterior distribution is possibly explained by the fact that the periodic length of the Di-BCP film pattern scales like $(\epsilon/\sigma)^{1/3}$~\cite{choksi2001scaling}, and an increase in the $\epsilon$ value leads to an increase in the $\sigma$ value when the periodic length is fixed. Moreover, we observe that the second stage inference using the smooth AAPS leads to a slight reduction of uncertainty of the marginal parameter $M$. Compared to the posterior distribution conditioned on $\y^*_j\in\R^{5}$ with known nuisance parameters, we see that the uncertainty reduction with uncertain nuisance parameters is mild despite assimilating many smooth AAPS entries. This is likely caused by both the uncertainty in the nuisance parameter and the entry-wise independent normal approximation in~\eqref{eq:surrogate} for the surrogate-based integrated likelihood. Nevertheless, the reference parameter value $\x^*$ that generates the synthetic images are recovered well. 

\begin{figure}[!hbt]
    \centering
    \addtolength{\tabcolsep}{-6pt} 
    \begin{tabular}{D D D C D D D C D D D E}
        \multicolumn{3}{c}{\makecell{Prior\\predictive samples}} &  &\multicolumn{3}{c}{\makecell{Phase-informed prior\\predictive samples}}&  & \multicolumn{3}{c}{\makecell{Posterior\\ predictive samples}} & \\
        \includegraphics[width = \linewidth]{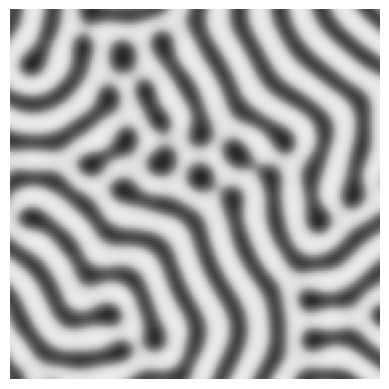} & \includegraphics[width = \linewidth]{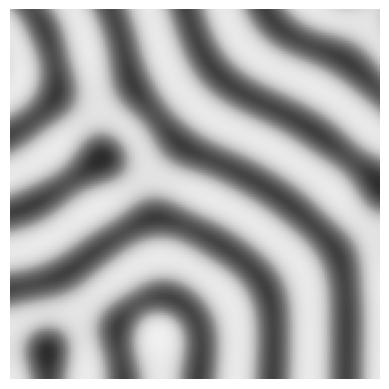} & \includegraphics[width = \linewidth]{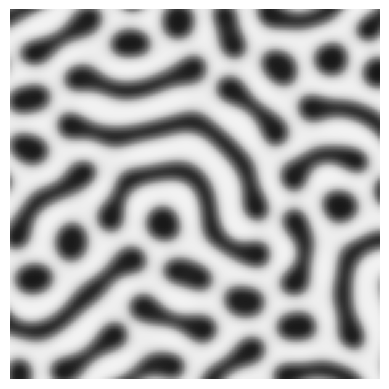} & \scalebox{1.2}{$\xrightarrow{\langle\data_1^*\rangle}$} & \includegraphics[width = \linewidth]{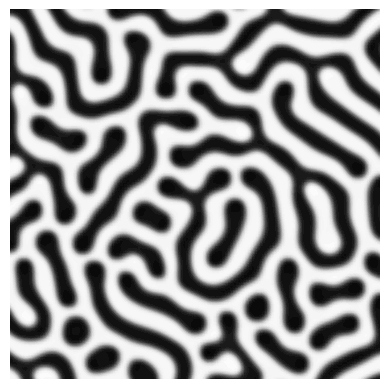} & \includegraphics[width = \linewidth]{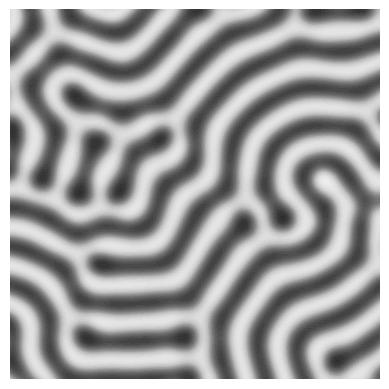} & \includegraphics[width = \linewidth]{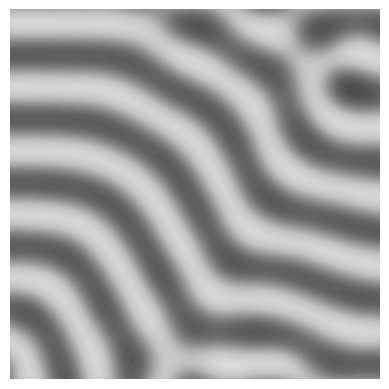} & \scalebox{1.2}{$\xrightarrow{\y_1^*}$} & \includegraphics[width = \linewidth]{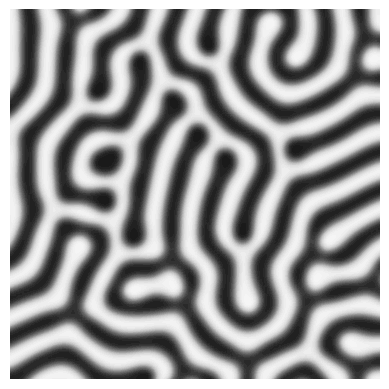}  & \includegraphics[width = \linewidth]{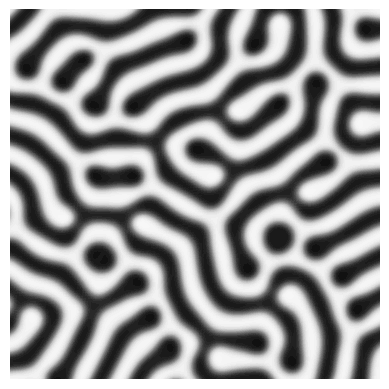} & \includegraphics[width = \linewidth]{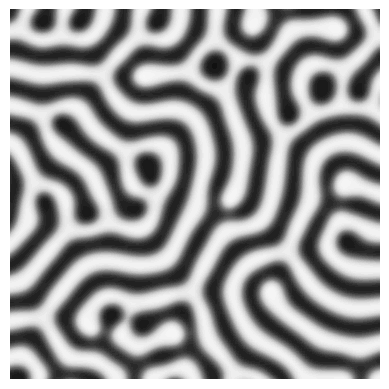} & \includegraphics[width = \linewidth]{colorbar_colorbar_op_36.pdf}\\
        \includegraphics[width = \linewidth]{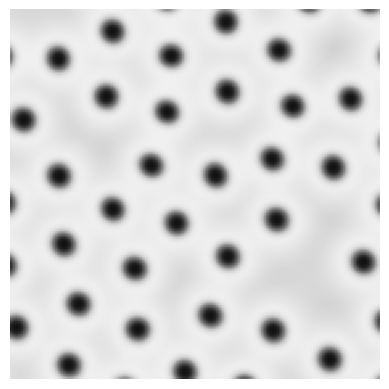} & \includegraphics[width = \linewidth]{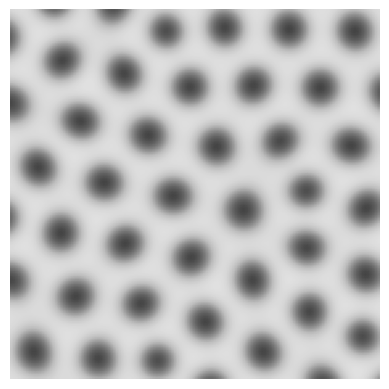} & \includegraphics[width = \linewidth]{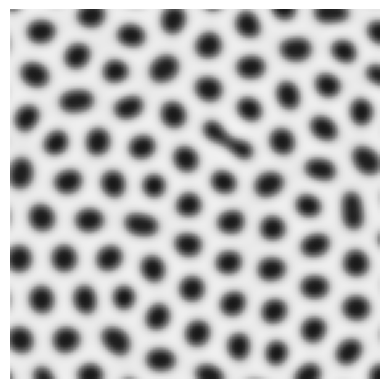} & \scalebox{1.2}{$\xrightarrow{\langle\data_2^*\rangle}$} & \includegraphics[width = \linewidth]{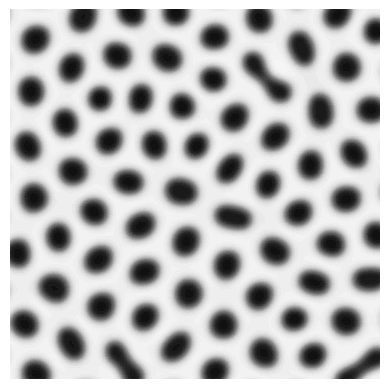} & \includegraphics[width = \linewidth]{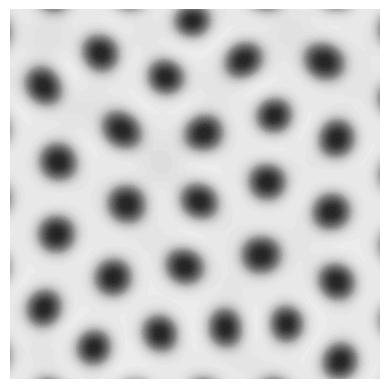} & \includegraphics[width = \linewidth]{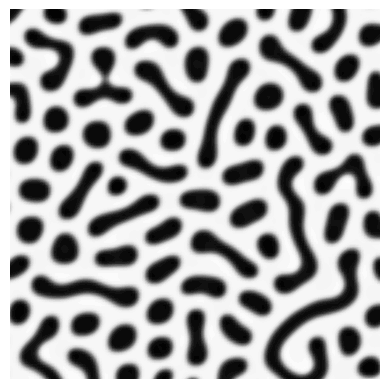} & \scalebox{1.2}{$\xrightarrow{\y_2^*}$} & \includegraphics[width = \linewidth]{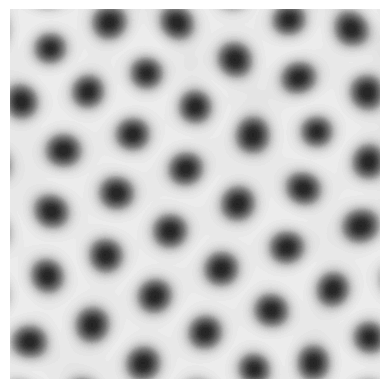} & \includegraphics[width = \linewidth]{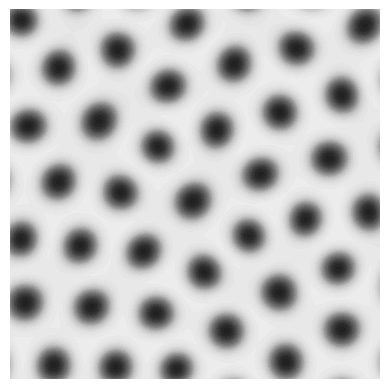} &\includegraphics[width = \linewidth]{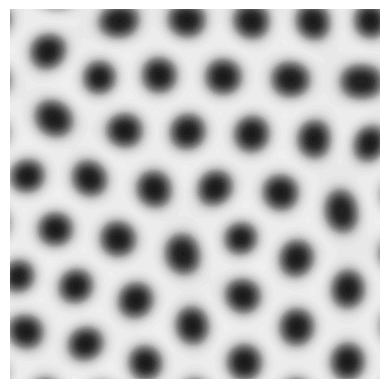} & \includegraphics[width = \linewidth]{colorbar_colorbar_op_36.pdf}
    \end{tabular}
        \addtolength{\tabcolsep}{6pt} 
    \caption{Predictive samples generated by simulating the OK model at parameter samples from the prior distributions (\emph{left}), the phase-informed prior distributions (\emph{middle}), and the posterior distributions (\emph{right}). The samples from the posterior distributions are generated by the surrogate parameter-to-spectrum map for integrated likelihood evaluations. These distributions are visualized in Figures \ref{fig:marginal_posterior_uncertain} and \ref{fig:joint_posterior_uncertain}. The predictive samples are generated on a domain size consistent with the sub-image size of $100\times 100$ in Figure~\ref{fig:synthetic_data_spectrum}.}
    \label{fig:predictive_samples_uncertain}
\end{figure}

In Figure~\ref{fig:predictive_samples_uncertain}, we show predictive samples for the prior, phase-informed prior, and posterior distributions generated via the surrogate-based integrated likelihood evaluations. These samples qualitatively demonstrate a significant reduction in predicted pattern variation after performing the Bayesian model calibration procedure.

\section{Conclusion}\label{sec:conclusion}

We propose an approach for Bayesian calibration of models predicting the morphological pattern of Di-BCP thin film self-assembly using top-down microscopy image data. The proposed approach extracts the smooth AAPS from image data to define a conditional likelihood for parameter inference through the pseudo-marginal method. The conditional likelihood function is given by the probability density of the non-central chi-squared distribution, and its normal approximation is derived. Through numerical experiments on the sensitivity of the conditional likelihood, we demonstrate that the smooth AAPS of image data is suitable for inferring model parameters in the presence of aleatoric uncertainties represented by the random long-range disorder (metastability) in top surface patterns of Di-BCP thin films. Additionally, we derive a prior distribution for the Bayesian model calibration procedure using the mean pixel value of image data. Such a prior distribution represents the information in the image about the morphological phase of the latent Di-BCP pattern. Such a prior distribution is complementary to the conditional likelihood based on the smooth AAPS that represents the information in the image about the important length scales of the latent Di-BCP pattern. The effectiveness of the proposed Bayesian model calibration procedure is confirmed by numerical experiments.

Furthermore, we formulate and train a neural network surrogate of the parameter-to-spectrum map. The surrogate assumes an entry-wise independent truncated normal approximation of the power spectrum response to the random long-range disorder in the top surface patterns of Di-BCP thin films. We train two MLPs to approximate the logarithmic mean and standard deviation of the smooth AAPS entries predicted by the OK model over a wide range of input parameters. When deployed in Bayesian model calibration, the surrogate-based integrated likelihood evaluations lead to posterior distributions with decreased uncertainty reduction. Such a disadvantage can be compensated by efficiently introducing more data into the proposed Bayesian calibration procedure.

\section*{Acknowledgement}
The work was partially supported by the U.S. Department of Energy, Office of Science, Office of Advanced Scientific Computing Research under awards DE-SC0019303 and DE-SC0023171, and by the U.S. National Science Foundation Division of
Mathematical Sciences under award 2245674. This work benefited from discussions with Joshua Chen on the pseudo-marginal method and deriving conditional likelihoods, Dingcheng Luo on implementing parallel solves for evaluating the Monte Carlo estimator, and Danial Faghihi on the treatment of hyper-parameters in Bayesian inference. Some parts of this work have appeared in Lianghao Cao's Ph.D. dissertation \cite{cao2022thesis}.

%-------------
% References
%-------------
\addcontentsline{toc}{section}{References}
\bibliographystyle{model1-num-names}
\biboptions{sort,numbers,comma,compress}                 
%\bibliography{main.bib}

\end{document}